\documentclass[letterreport, 11pt]{article}

\usepackage{geometry}
\usepackage{amsmath,amsthm,amssymb}
\usepackage{tikz}
\usepackage{enumitem}
\usepackage[ruled,vlined]{algorithm2e}
\usepackage{multicol, multirow}
\usepackage{booktabs}
\usepackage{subcaption}
\usepackage{mdframed}
\usepackage{ulem}
\usepackage[numbers]{natbib}
\usepackage{tabularx}
\usepackage[colorlinks, allcolors=blue]{hyperref}

\tikzstyle{dot}=[circle,fill,inner sep=1.5pt]

\providecommand{\comment}[1]{}
\providecommand{\nitin}[1]{}

\newtheorem{assumption}{Assumption}
\newtheorem{definition}{Definition}
\newtheorem{theorem}{Theorem}

\newtheorem{lemma}{Lemma}

\providecommand{\iprod}[2]{\ensuremath{\left\langle #1,\,#2  \right\rangle}}
\providecommand{\norm}[1]{\ensuremath{\left\lVert#1\right\rVert }}
\providecommand{\mnorm}[1]{\ensuremath{\left\lvert#1\right\rvert}}
\providecommand{\dist}[2]{\ensuremath{\mathrm{dist}\left( #1,\,#2 \right)}}

\def\R{\mathbb{R}}
\def\Z{\mathbb{Z}}
\def\H{\mathcal{H}}
\def\B{\mathcal{B}}

\def\O{\mathcal{O}}

\def\S{\mathcal{S}}

\def\D{\mathsf{D}}

\def\M{\mathsf{M}}
\def\W{\mathcal{W}}
\def\E{\mathbb{E}}
\def\g{\boldsymbol{g}}
\def\gf{\mathsf{GradFilter}}
\def\ga{\mathsf{GradAgg}}

\def\z{\boldsymbol{z}}

\begin{document}

\title{Impact of redundancy in resilient distributed optimization and learning\thanks{This is a full version of the paper of the same title in ICDCN 2023 \cite{icdcn}. This version updates the results in Section~\ref{sec:stochastic} and their proofs in Appendix~\ref{part:3}, among other minor fixings.}}

\author{Shuo Liu \thanks{Georgetown University. Email: {\tt sl1539@georgetown.edu}.} \hspace{0.5in} Nirupam Gupta \thanks{École Polytechnique Fédérale de Lausanne (EPFL). Email: {\tt nirupam.gupta@epfl.ch}.} \hspace{0.5in} Nitin H. Vaidya \thanks{Georgetown University. Email: {\tt nitin.vaidya@georgetown.edu}.}
}
\date{}
\maketitle

\begin{abstract}
    This paper considers the problem of resilient distributed optimization and stochastic learning in a server-based architecture. The system comprises a server and multiple agents, where each agent has its own \textit{local} cost function. The agents collaborate with the server to find a minimum of the aggregate of the local cost functions. In the context of stochastic learning, the local cost of an agent is the \textit{loss function} computed over the data at that agent. In this paper, we consider this problem in a system wherein some of the agents may be Byzantine faulty and some of the agents may be slow (also called \textit{stragglers}). In this setting, we investigate the conditions under which it is possible to obtain an ``approximate'' solution to the above problem. In particular, we 
    introduce the notion of \textit{$(f, r; \epsilon)$-resilience} to characterize how well the true solution is approximated in the presence of up to $f$ Byzantine faulty agents, and up to $r$ slow agents (or stragglers) -- smaller $\epsilon$ represents a better approximation. We also introduce a measure named \textit{$(f, r; \epsilon)$-redundancy} to characterize the \textit{redundancy} in the cost functions of the agents. Greater redundancy allows {for} a better approximation when solving the problem of aggregate cost minimization.
    
    
    In this paper, we 
    constructively show (both theoretically and empirically) that \textit{$(f, r; \O(\epsilon))$-resilience} can indeed be achieved in practice, given that the local cost functions are sufficiently redundant.

\end{abstract}



\newpage
\tableofcontents


\section{Introduction}
\label{sec:intro}

With the rapid growth in the computational power of modern computer systems and the scale of optimization tasks, e.g., training of deep neural networks~\citep{otter2020survey}, the problem of distributed optimization in a multi-agent system has gained significant attention in recent years. 
This paper considers the problem of \textit{resilient} distributed optimization and stochastic learning in a server-based architecture.

The system under consideration consists of a \textit{trusted} server and multiple agents, where each agent has its own ``local'' cost function. The agents collaborate with the server to find a minimum of the aggregate cost functions (i.e., the aggregate of the local cost functions)~\citep{boyd2011distributed}.
Specifically, suppose that there are $n$ agents in the system where each agent $i$ has a cost function $Q_i: \W\subset\R^d \to \R$. The goal then is to enable the agents to compute a global minimum $x^*$ such that 
\begin{equation}
     x^*\in\arg\min_{x\in\W}\sum_{i=1}^nQ_i(x), 
     \label{eqn:goal}
\end{equation}
where $\W\subset\R^d$ is a compact set where the problem is defined on.
As a simple example, consider a group of $n$ people who want to pick a place for a meeting. Suppose that function $Q_i(x)$ represents the cost for the $i$-th person (agent) to travel to location $x$, {$x$ being its coordinates}. Then, $x^*\in\arg\min_{x\in\R^d}\sum_{i=1}^nQ_i(x)$ is a location such that the travel cost of meeting at $x^*$, when aggregated over all the agents, is the smallest possible.
In the context of training of a deep neural network, $Q_i(x)$ represents the \textit{loss function} corresponding to the data at agent $i$
when using machine parameter vector $x$. Thus, our results are relevant in the context of distributed machine learning~\citep{boyd2011distributed}. The above multi-agent optimization problem has many other applications as well, including distributed sensing \citep{rabbat2004distributed}, and swarm robotics \citep{raffard2004distributed}.


Such distributed multi-agent systems may encounter some challenges in practice. We consider two challenges in this paper:
\begin{itemize}
    \item \textit{Byzantine faulty agents}: A Byzantine faulty  \cite{lamport1982byzantine} agent may share wrong information with the server, or not send any information at all. The Byzantine fault model  \cite{lamport1982byzantine} does not impose any constraints on the behavior of a faulty agent. An agent's faulty behavior may occur either due to software/hardware failure or due to a security compromise of the agent. Prior work has shown that even a single faulty agent can compromise the entire distributed optimization process \citep{su2016fault}. 
    \item \textit{Slow agents (Stragglers)}: Stragglers are agents that operate much more slowly than the other agents. Synchronous algorithms that require each agent to communicate with the server in each ``round'' of the computation can perform poorly, since the stragglers will slow down the progress of the computation \citep{hannah2017more, leblond2018asynchronous, assran2020advances}. 
\end{itemize}
The related work is discussed in more detail in Section~\ref{sec:related}. {Recall the travel and meeting example in the previous paragraph. A ``faulty'' person may deliberately send a wrong cost function so that the minimum would be his desired place; while a ``straggler'' may be unresponsive, delaying everyone from reaching a decision.} The above two challenges have been considered independently in the past work \citep{chen2018draco, blanchard2017machine, liu2021approximate, tandon2017gradient, halbawi2018improving, karakus2017encoded, niu2011hogwild}. This paper makes contributions that further our understanding of distributed optimization algorithms that are resilient to both the forms of adversities above.

\subsection{Contributions of this paper}

It is known that, under the above adverse conditions, it is not always possible to solve the problem of interest exactly. This paper addresses the interaction between these two forms of adversities.
In particular, we consider a multi-agent system with up to $f$ faulty agents and up to $r$ stragglers (or slow agents) out of $n$ agents. It is possible that the same agent may be both faulty and slow. Let $\H$ denote the set of non-faulty (or honest) agents in a given execution. Note that an agent in $\H$ may be slow (i.e., stragglers). In this paper, we consider the \textit{Resilient Distributed Optimization} (RDO) problem, with the goal of approximately computing
\begin{equation}
\arg\min_{x\in\W}\sum_{i\in\H}Q_i(x),
\label{eqn:goal-ft}
\end{equation}
where $\W\subset\R^d$ is a compact set where the problem is defined on,
in the presence of up to $f$ Byzantine faulty agents and up to $r$ slow agents.
For the problem to be solvable, we {assume} that $n>2f+r$.
{Previous research has established that Byzantine fault-tolerance problem is solvable when the non-faulty agents dominates the set of agents, or $n>2f$ \cite{liu2021approximate}; we also need enough agents to be synchronous for the problem to be solvable.}

We characterize how well the true solution of (\ref{eqn:goal-ft}) can be approximated, as a function of the level of ``redundancy'' in the cost functions. As a trivial example, if all the agents have an identical cost function, then (\ref{eqn:goal-ft}) can be obtained by simply taking a majority vote on the outcome of local cost function optimization performed by each agent separately. In general, such a high degree of redundancy may not be available, and only an approximate solution of  (\ref{eqn:goal-ft}) may be obtainable. We
    introduce the notion of \textit{$(f, r; \epsilon)$-resilience} to characterize how well the true solution is approximated in the presence of up to $f$ Byzantine faulty agents, and up to $r$ slow agents (or stragglers) -- smaller $\epsilon$ represents a better approximation. We also introduce a measure named \textit{$(f, r; \epsilon)$-redundancy} to characterize the \textit{redundancy} in the cost functions of the agents. Greater redundancy allows a better approximation when solving the problem of aggregate cost minimization.  
    
    We constructively show  (both theoretically and empirically) that \textit{$(f, r; \O(\epsilon))$-resilience} can indeed be achieved in practice, provided that the local cost functions are sufficiently redundant. Our empirical evaluation considers a distributed gradient descent (DGD)-based solution; in particular, for distributed learning in the presence of Byzantine and asynchronous agents, we evaluate a distributed stochastic gradient descent (D-SGD)-based algorithm.
    
\subsection{\textit{$(f, r; \epsilon)$-resilience} and \textit{$(f, r; \epsilon)$-redundancy}}

\subsubsection{Euclidean and Hausdorff Distance}

To help define the notions of \textit{$(f, r; \epsilon)$-resilience} and \textit{$(f, r; \epsilon)$-redundancy}, we will use Euclidean and Hausdorff distance measures. Let $\norm{\cdot}$ represents the Euclidean norm. Then Euclidean distance between points $x$ and $y$ in $\mathbb{R}^d$ equals $\|x-y\|$.
The Euclidean distance between a point $x$ and a set $Y$ in $\mathbb{R}^d$, denoted by $\dist{x}{Y}$, is defined as:
\begin{equation*}
     \dist{x}{Y}=\inf_{y\in Y}\dist{x}{y}=\inf_{y\in Y}\norm{x-y}.
\end{equation*}
The Hausdorff distance between two sets $X$ and $Y$ in $\mathbb{R}^d$, denoted by $\dist{X}{Y}$, is defined as: 
\begin{equation*}
    \dist{X}{Y}\triangleq\max\left\{\sup_{x\in X}\dist{x}{Y}, \sup_{y\in Y}\dist{y}{X}\right\}.
\end{equation*}
It is worth noting that Huasdorff distance is a metric on compact sets, and therefore satisfies triangle inequality \cite{conci2018distance}.

\subsubsection{Define \textit{$(f, r; \epsilon)$-resilience} and \textit{$(f, r; \epsilon)$-redundancy}}

We now define the notion of \textit{$(f,r;\epsilon)$-resilience} to characterize how closely (\ref{eqn:goal-ft}) is approximated by a given RDO algorithm (the definition below denotes by $\widehat{x}$ the output produced by the RDO algorithm).

\begin{definition}[$(f,r;\epsilon)$-resilience]
    \label{def:resilience}
    For $\epsilon \geq 0$, a distributed optimization algorithm is said to be $(f,r;\epsilon)$-resilient if its output $\widehat{x}$ satisfies
    \begin{equation*}
         \dist{\widehat{x}}{\arg\min_{x\in\W}\sum_{i \in\H}Q_i(x)}\leq\epsilon
    \end{equation*}
    for each set $\H$ of $n-f$ non-faulty agents, despite the presence of up to $f$ faulty agents and up to $r$ stragglers. 
\end{definition}

This resilience notion captures how well a given RDO algorithm performs in terms of approximating the solution of 
(\ref{eqn:goal-ft}). Observe that the above definition considers the distance of $\widehat{x}$ (i.e., the output of the given RDO algorithm) from the true solution for every sub-problem corresponding to $n-f$ non-faulty agents. Intuitively, the reason is as follows: Consider the case when the faulty agents behave badly in a manner that is \textit{not detectable} -- that is, by simply looking at the information received from the faulty agents, the trusted server cannot determine if they are faulty. Thus, it is difficult to know how many agents are faulty. In particular, it is possible that exactly $n-f$ agents are non-faulty, with the rest being faulty. Therefore, intuitively, we want $\widehat{x}$ to be within $\epsilon$ of the true solution that minimizes the aggregate cost over any $n-f$ non-faulty agents.

In this paper, we show how the \textit{redundancy} in agents' cost functions can be utilized to obtain $(f,r;\epsilon)$-resilience. Formally, we define the property as \textit{$(f,r;\epsilon)$-redundancy}, stated below. 

\begin{definition}[$(f,r;\epsilon)$-redundancy]
    \label{def:redundancy}
    For $\epsilon \geq 0$, the agents' local cost functions are said to satisfy the $(f,r;\epsilon)$-redundancy property if and only if for every pair of subsets of agents $S,\widehat{S}\subsetneq\{1,...,n\}$, where $\mnorm{S}=n-f$, $\mnorm{\widehat{S}}\geq n-r-2f$ and $\widehat{S}\subsetneq S$, 
    \begin{equation*}
         \dist{\arg\min_x\sum_{i\in S}Q_i(x)}{\arg\min_x\sum_{i\in\widehat{S}}Q_i(x)}\leq\epsilon.
    \end{equation*}
\end{definition}

\begin{figure}
    \centering
    \includegraphics[width=.5\textwidth]{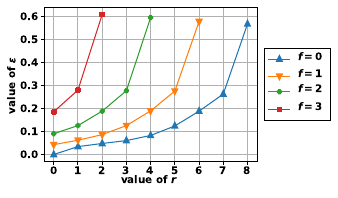}
    \caption{Trade-off between $\epsilon$, $f$ and $r$ for the numerical example in Section~\ref{sec:experiments}. The total number of agents $n=10$.}
    \label{fig:trade-off-main}
\end{figure}

The $(f,r;\epsilon)$-redundancy condition also characterizes the trade-off between parameters $f$, $r$ and $\epsilon$.
In particular, for any set of local cost functions (i.e., $Q_i(x)$'s), and for any $r< n$ and $f < (n - r)/2$, \textit{there exists} some  $\epsilon$ such that the agents' cost functions satisfy the $(f,r;\epsilon)$-redundancy property for that $\epsilon$ value. For a given set of local cost functions, 
\begin{itemize}
    \item With a fixed value of $f$, larger $r$ results in larger $\epsilon$.
    \item With a fixed value of $r$, larger $f$ results larger $\epsilon$.
\end{itemize}
%
Thus, the notion of $(f,r;\epsilon)$-redundancy is a \textit{description} of the redundancy in the cost functions. {Note that for a given set of functions, we always consider the smallest $\epsilon$ for which redundancy condition holds.} We will discuss a numerical example of distributed linear regression in Section~\ref{sec:experiments}. For this example, Figure~\ref{fig:trade-off-main} shows the relationship between $\epsilon$, $f$ and $r$. Each curve in Figure~\ref{fig:trade-off-main} corresponds to a fixed value of $f$. The horizontal axis varies parameter $r$. The corresponding value of $\epsilon$, plotted on the vertical axis, illustrates the trade-off discussed above.

\begin{table}
    \centering
    \small
    \begin{tabular}{c|c|c}
        {Deterministic} / Stochastic & Synchronous & Asynchronous \\
        \midrule
        Fault-free & - & {Sec.} \ref{ssub:async} / {Sec.} \ref{ssub:async-stoc} \\ 
        \midrule
        With Byzantine agents & \cite{gupta2020resilience, liu2021approximate} / {Sec.} \ref{ssub:byz-stoc} & {Sec.} \ref{sub:rdo} / {Sec.} \ref{sec:stochastic} \\
    \end{tabular}
    \caption{Problem space of solving RDO using resilience in cost functions.}
    \label{tab:prob-space}
\end{table}

In the context of solving RDO problems with resilience in cost functions defined above, there are several special cases that {have} been solved {in} previous work. The current status of the problem space is shown in Table~\ref{tab:prob-space}. There are three dimensions: deterministic vs. stochastic, synchronous vs. asynchronous, and fault-free vs. Byzantine agents in presence. 
We provide analysis from the redundancy-in-cost-function perspective for all remaining problems, as indicated in Table~\ref{tab:prob-space}, showing that redundancy can be utilized to solve RDO problems.

\subsection{Paper Outline}
The related work is discussed in more detail in Section~\ref{sec:related}.
In Section~\ref{sec:full-grad}, we propose a gradient-based algorithm framework that can be used to solve the RDO problems in practice, and analyze its resilience under various settings, while utilizing the redundancy property of the local cost functions. In Section~\ref{sec:stochastic}, we adapt the proposed framework to solve resilient distributed \textit{stochastic optimization problems}, which has relevance in distributed machine learning, and present convergence results for the stochastic algorithm. To validate our theoretical findings, we present empirical results in Section~\ref{sec:experiments}.

Table~\ref{tab:notations} in Appendix~\ref{appdx:0} provides a reference for some notions that appear in this paper. Proofs for theoretical results in Sections~\ref{sec:full-grad} and~\ref{sec:stochastic} are provided in Appendix~\ref{part:2} and Appendix~\ref{part:3}. 

\section{Related work}
\label{sec:related}

\paragraph{Byzantine fault-tolerant optimization} Byzantine agents make it difficult to achieve the goal of distributed optimization \eqref{eqn:goal} \cite{su2015byzantine1, blanchard2017machine}.
 Various methods have been proposed to solve Byzantine fault-tolerant optimization or learning problems \citep{liu2021survey}, including robust gradient aggregation \citep{blanchard2017machine, chen2017distributed}, gradient coding \citep{chen2018draco}, and other methods \citep{gupta2021byzantine,xie2018zeno, yin2018byzantine}. 
 
 In general, instead of solving the problem exactly, an approximate solution may be found. In some cases, however, exact solution is feasible. For instance, when every agent has the same (identical) cost function, it is easy to see that an exact solution may be obtained via majority vote on the output of the agents, so long as only a minority of the agents are faulty. Stochastic versions of distributed optimization algorithms are also proposed to exploit such forms of \textit{redundancy} in the context of distributed machine learning (e.g, \citep{blanchard2017machine, chen2017distributed}).
 \citep{gupta2020resilience} obtained conditions under which it is possible to tolerate $f$ Byzantine agents in a \textit{synchronous} system and yet obtain accurate solution for the optimization problem.  
 \citep{liu2021approximate} obtained a condition under which approximately correct outcome can be obtained in a \textit{synchronous system}. 
 
The contribution of this paper is to generalize the results in \cite{gupta2020resilience,liu2021approximate} to a system that allows some asynchrony in the form of stragglers -- allowing for stragglers makes the proposed approach more useful in practice; we also discuss deterministic and stochastic methods at the same time, closing all remaining gaps in the context of solving resilient distributed optimization problem through redundancy in cost functions.

\paragraph{Asynchronous optimization} For fault-free systems (i.e., in the absence of failures), there is past work addressing distributed optimization under asynchrony. 
Prior work shows that distributed optimization problems can be solved using \textit{stale} gradients with a constant delay (e.g., \citep{langford2009slow}) or bounded delay (e.g., \citep{agarwal2012distributed, feyzmahdavian2016asynchronous}).
Methods such as \textsc{Hogwild!} allow lock-free updates in shared memory \citep{niu2011hogwild}. Other works use variance reduction and incremental aggregation methods
to improve the convergence rate \citep{roux2012stochastic, johnson2013accelerating, defazio2014saga, shalev2013accelerated}. 

\textit{Coding} has been used to mitigate the effect of stragglers or failures \citep{lee2017speeding, karakus2017encoded, karakus2017straggler, yang2017coded}.
Tandon et al. \citep{tandon2017gradient} proposed a framework using maximum-distance separable coding across gradients to tolerate failures and stragglers. Similarly, Halbawi et al. \citep{halbawi2018improving} adopted coding to construct a coding scheme with a time-efficient online decoder.
Karakus et al. \citep{karakus2019redundancy} proposed an encoding distributed optimization framework with deterministic convergence guarantee.
Other replication- or repetition-based techniques involve either task-rescheduling or assigning the same tasks to multiple nodes \citep{ananthanarayanan2013effective, gardner2015reducing, shah2015redundant, wang2015using, yadwadkar2016multi}. These previous methods rely on algorithm-created redundancy of data or gradients to achieve robustness, while $(f,r;\epsilon)$-redundancy can be a property of the cost functions themselves, allowing us to exploit such redundancy without extra effort.
\section{Algorithmic framework for resilient distributed optimization problems}
\label{sec:full-grad}

 From now on, we use $[n]$ as a shorthand for the set $\{1,...,n\}$. In this section, we study the resilience of a class of
 distributed gradient descent (DGD)-based algorithms. In the algorithms considered here,  a server maintains an estimate for the optimum. In each iteration of the algorithm, the server sends its estimate to the agents, and the agents send the gradients of their local cost functions at that estimate to the server. The server uses these gradients to update the estimate.

Specifically, Algorithm~\ref{alg} is a framework for several instances of the RDO algorithms presented later in this paper. Different instances of the algorithm differ in the manner in which the iterative update \eqref{eqn:update} is performed at the server  -- different gradient aggregation rules are used in different instances of the RDO algorithm.

\begin{algorithm}[h]
    \SetAlgoLined
    \caption{Resilient distributed gradient descent under $(f,r;\epsilon)$-redundancy}
    \label{alg}
    \textbf{Input:} $n$, $f$, $r$, $\epsilon$. A convex compact set $\W$. Each agent $i$ has its cost function $Q_i(x)$.
    
    \vspace{2pt}
    The initial estimate of the optimum, $x^0${$\in\W$}, is chosen by the server. The new estimate $x^{t+1}$ is computed in iteration $t\geq 0$ as follows:
    \begin{description}[nosep]
        \item[Step 1:] The server requests each agent for the gradient of its local cost function at the current estimate $x^t$. Each agent $j$ is expected to send to the server the gradient (or stochastic gradient) with timestamp $t$. 
        Let us denote by $g_j^t$ sent by an agent $j$ in this iteration. If agent $j$ is non-faulty, then $g_j^t$ will be the gradient (or stochastic gradient) of $Q_j(x)$ at $x^t$. If agent $j$ is faulty, then it may send arbitrary $g_j^t$.
        
        \item[Step 2:] The server waits until it receives $n-r$ vectors with the timestamp of $t$. Suppose $S^t\subseteq\{1,...,n\}$ is the set of agents whose vectors are received by the server at step $t$ where $\mnorm{S^t}=n-r$. The server updates its estimate to
        \begin{equation}
            x^{t+1}=\left[x^t-\eta_t \, \ga\left(g_j^t|\,j\in S^t;n,f,r\right) \right]_\W \label{eqn:update}
        \end{equation}
        where $\eta_t\geq0$ is the step-size for each iteration $t$,
        and $[\,\cdot\,]_{\mathcal{W}}$ denotes a projection onto $\mathcal{W}$.
    \end{description}
\end{algorithm}

A \textit{gradient aggregation rule} (GAR) 
$$\ga(\cdot;n,f,r):\R^{d\times (n-r)}\rightarrow\R^d$$
is a function that takes $n-r$ vectors in $\R^d$ (gradients) and outputs a vector in $\R^d$ for the update, with the knowledge that there are up to $f$ Byzantine faulty agents and up to $r$ stragglers out of the $n$ agents in the system. As an example, the GAR used in synchronous distributed optimization with no faulty agents is \textit{averaging}, i.e., $\ga(g_j^t|j\in[n];n,0,0)=(1/n)\sum_{j=1}^ng_j^t$. The fact that only $n-r$ gradients {are} in use reflects the asynchrony of the optimization process caused by stragglers, while the GAR counters the effect of Byzantine agents and stragglers, together achieving resilience.


\subsection{Convergence of resilient distributed gradient descent}
\label{sub:rdo}
Recall the problem setting that in a $n$-agent system, there are up to $f$ Byzantine agents and up to $r$ stragglers. With the existence of Byzantine agents, the goal is to solve \eqref{eqn:goal-ft} (cf. Section~\ref{sec:intro}). First, we need to introduce some standard assumptions that are necessary for our analysis. Suppose $\H\subseteq[n]$ is {any} subset of non-faulty agents with $\mnorm{\H}=n-f$.

\begin{assumption}
    \label{assum:lipschitz}
    For each (non-faulty) agent $i$, the function $Q_i(x)$ is $\mu$-Lipschitz smooth, i.e., $\forall x, x'\in\W$, 
    \begin{equation}
        \norm{\nabla Q_i(x)-\nabla Q_i(x')}\leq\mu\norm{x-x'}.
    \end{equation}
\end{assumption}

\begin{assumption}
    \label{assum:strongly-convex-ft}
    For any set $S\subseteq\H$, we define the average cost function to be $Q_S(x)=\frac{1}{\mnorm{S}}\sum_{j\in S}Q_j(x)$. We assume that $Q_S(x)$ is $\gamma$-strongly convex for any $S$ subject to $\mnorm{S}\geq n-{2}f$, i.e., $\forall x, x'\in\W$, 
    \begin{equation}
        \iprod{\nabla Q_S(x)-\nabla Q_S(x')}{x-x'}\geq\gamma\norm{x-x'}^2.
    \end{equation}
\end{assumption}

We also need to assume that a solution to the problem exists. Specifically, for each subset of non-faulty agents $S$ with $\mnorm{S}\geq n-f$, we assume that there exists a point $x_S\in\arg\min_{x\in\R^d}\sum_{j\in S}Q_j(x)$ such that $x_S\in\W$. By Assumption~\ref{assum:strongly-convex-ft}, there exists a unique minimum point $x_\H\in \W$ that minimize the aggregate cost functions of agents in $\H$, i.e.,
\begin{equation}
    \textstyle \{x_\H\}=\W\cap\arg\min_{x\in\R^d}\sum_{j\in\H}Q_j(x).
    \label{eqn:existence-ft}
\end{equation}

Recall \eqref{eqn:update} in Step 2 of Algorithm~\ref{alg}. The GAR used for RDO problems with up to $f$ Byzantine agents and up to $r$ stragglers is
\begin{align}
    &\ga \left(g_j^t|j\in S^t;n,f,r\right)=\gf\left(g_j^t|j\in S^t;n-r,f\right)
    \label{eqn:aggregation-rule-async-ft}
\end{align}
for every iteration $t$, where $\gf$ is a \textit{robust GAR}, or \textit{gradient filter}, that enables us fault-tolerant capability \citep{liu2021approximate, damaskinos2019aggregathor, karimireddy2021learning}. Specifically, a gradient filter $\gf(\cdot;m,f):\R^{d\times m}\rightarrow\R^d$ is a function that takes $m$ vectors of $d$-dimension and outputs a $d$-dimension vector given that there are up to $f$ Byzantine agents, where $m>f\geq0$. Each agent $j$ sends a vector
\begin{equation}
    g_j^t=\left\{\begin{array}{cl}
        \nabla Q_j(x^t), & \textrm{ if the agent is non-faulty,} \\
        \textrm{arbitrary vector}, & \textrm{ if the agent is faulty} 
    \end{array}\right.
\end{equation}
to the server at iteration $t$. 

Following the above GAR in \eqref{eqn:aggregation-rule-async-ft}, the server receives the first $n-r$ vectors from the agents in the set $S^t$, and send{s} the vectors through a gradient filter. Now we present an asymptotic convergence property of Algorithm~\ref{alg}.

    

\begin{theorem}
    \label{thm:async-fault-toler}
    Suppose $x_\H$ is any point in $\W$. Assume that $\eta_t$ satisfy $\sum_{t=0}^\infty\eta_t=\infty~\textrm{ and }~\sum_{t=0}^\infty\eta_t^2<\infty$. Suppose that for the gradient aggregation rule \linebreak $\ga\left(g_j^t|\,j\in S^t;n,f,r\right)$ in  Algorithm~\ref{alg}, there exists $\M<\infty$ such that
    \begin{equation}
        \norm{\ga\left(g_j^t|\,j\in S^t;n,f,r\right)}\leq\M
    \end{equation} for all $t$. Algorithm~\ref{alg} satisfies the following:
    
    For the point $x_\H\in\W$, if there exists $\D^*\in\left[0,\max_{x\in\W}\norm{x-x_\H}\right)$ 
    and $\xi>0$ such that for each iteration $t$,
    \begin{equation}
        \label{eqn:thm-condition}
        \begin{split}
        \phi_t\triangleq\iprod{x^t-x_\H}{\ga\left(g_j^t\vert{j\in S^t};n,f,r\right)}\geq\xi \\
        \textrm{ when }\norm{x^t-x_\H}\geq\D^*,
        \end{split}
    \end{equation}
    then we have $\lim_{t\rightarrow\infty}\norm{x^t-x_\H}\leq\D^*$.
\end{theorem}
\noindent The proof of Theorem~\ref{thm:async-fault-toler} is presented in Appendix~\ref{appdx:proof-lemma-bound}.

Intuitively, so long as the point $x_\H$ and the gradient filter in use satisfy the desired properties in Theorem~\ref{thm:async-fault-toler}, the iterative estimate $x^t$ in Algorithm~\ref{alg} will eventually be within $\D^*$ distance to $x_\H$. If the point $x_\H$ is the minimum point of the aggregate cost functions of agents in $\H$, as defined in \eqref{eqn:existence-ft}, then the algorithm is $(f,r;\D^*)$-{resilient}. 

Gradient filters that {satisfy} condition \eqref{eqn:thm-condition} in Theorem~\ref{thm:async-fault-toler} include CGE \citep{liu2021approximate} and coordinate-wise trimmed mean \citep{yin2018byzantine, liu2021approximate}. 
As an example, for the CGE gradient filter\footnote{The definition of CGE is the following: in each iteration $t$, sort $m$ vectors $g_j^t$'s as
   $ \norm{g_{i_1}^t}\leq\norm{g_{i_2}^t}\leq...\leq\norm{g_{i_{m}}^t},$
then we have CGE gradient filter:
\begin{align}
     \gf_\textrm{CGE}(g_j^t|j\in S^t;m,f)=\sum_{l=1}^{m-f}g_{i_l}^t.
    \label{eqn:cge-footnote}
\end{align}}, 
we obtain the result that
\begin{align}
    \begin{gathered}
        \phi_t \geq \alpha (n-r) \gamma \D^* \left(\D^* - \D\right) > 0 \textrm{ when } 
        \norm{x^t-x_\H} \geq \D^*, \\
        \textrm{for all }\D^* > \D \triangleq \frac{4\mu (f+r)\epsilon}{\alpha\gamma}, \\
        \textrm{where }\alpha = \frac{n-f}{n-r} - \frac{2\mu}{\gamma}\cdot\frac{f+r}{n-r}>0, \\
    \end{gathered}
    \label{eqn:thm2-params}
\end{align}
and therefore, $\lim_{t\rightarrow\infty}\norm{x^t-x_\H}\leq\D^*$. The proof for these results in \eqref{eqn:thm2-params} is presented in Appendix~\ref{appdx:a-2}.

Intuitively, by applying CGE, the output of our new algorithm can converge to a $\epsilon$-dependent region centered by the true minimum point $x_\H$ of aggregate cost functions of non-faulty agents, with up to $f$ Byzantine agents and up to $r$ stragglers, i.e., Algorithm~\ref{alg} with CGE is $(f,r;\O(\epsilon))$-resilient. For other valid gradient filters, the parameters in \eqref{eqn:thm2-params} may vary.

\subsection{Special cases of RDO problems}
\label{sub:fg-special}

There are several meaningful special cases of RDO problems. When $f=r=0$, RDO degenerates to the synchronous distributed optimization problem with no faulty agents in the system, which can be solved by several existing methods including DGD \citep{boyd2011distributed, nedic2009distributed, shi2015extra, varagnolo2015newton}. When $r=0$, RDO becomes the Byzantine fault-tolerant distributed optimization problem we discussed in Section~\ref{sec:related}. $(f,0;\epsilon)$-redundancy is necessary to achieving $(f,0;\epsilon)$-resilience \cite{liu2021approximate}.

\subsubsection{Asynchronous distributed optimization}
\label{ssub:async}
When $f=0$, RDO becomes the fault-free asynchronous distributed optimization problem. As discussed in Section~\ref{sec:related}, prior research has not considered exploiting the existing redundancy in cost functions. When $f=0$, we  have $\H=[n]$. The goal \eqref{eqn:goal-ft} of distributed optimization becomes \eqref{eqn:goal}. Recall \eqref{eqn:update} in \textbf{Step 2} of Algorithm~\ref{alg}. With $f=0$, we define the GAR for asynchronous optimization to be 
\begin{equation}
     \ga \left(g_j^t|j\in S^t;n,0,r\right)=\sum_{j\in{S^t}} g_j^t
    \label{eqn:aggregation-rule}
\end{equation}
for every iteration $t$. That is, the algorithm updates the current estimates using the sum of the first $n-r$ gradients it receives. Note that $g_j^t=\nabla Q_j(x^t)$ is the full gradient. For this problem, it can be shown for Theorem~\ref{thm:async-fault-toler} that 
if $(0,r;\epsilon)$-redundancy is satisfied,
\begin{align}
    \begin{gathered}
    \phi_t\geq \alpha n\gamma\D^*(\D^*-\D)>0\textrm{ when }\norm{x^t-x_\H}\geq\D^*, \\
    \textrm{for all }\D^*>\D=\frac{2r\mu}{\alpha\gamma}\epsilon, \\
    \textrm{where } \alpha = 1-\frac{r}{n}\cdot\frac{\mu}{\gamma}>0, \\
    \end{gathered}
    \label{eqn:async-parameters}
\end{align}
and therefore, $\lim_{t\rightarrow\infty}\norm{x^t-x_\H}\leq\D^*$. By substituting \eqref{eqn:thm2-params} with $f=0$, we see that \eqref{eqn:async-parameters} is a tighter bound. Also, note that there could be other GARs for the asynchronous problem to achieve better efficiency or further reduce computational overhead. The proof for these results in \eqref{eqn:async-parameters} is presented in Appendix~\ref{appdx:a-3}.

For asynchronous distributed optimization problems, we can also utilize \textit{stale gradients}, i.e., gradients from previous iterations to further reduce waiting time. 
We define the following two sets: during iteration $t$, 
$T^{t;k}$ denotes the set of agents from whom the most recent gradient that the server \textit{has received} is computed using $x^k$.
Note that the definition implies $T^{t;t-i}\cap T^{t;t-j}=\varnothing$ for any $i\neq j$. Let us further define $T^t=\bigcup_{i=0}^\tau T^{t;t-i}$, where $\tau\geq0$ is a predefined \textit{straggler parameter}. To use stale gradients, we define the GAR in the iterative update \eqref{eqn:update} to be 
\begin{equation}
    \label{eqn:update-straggler}
     \ga \left(g_j^t|j\in S^t;n,f,r\right)=\sum_{i=0}^\tau\sum_{j\in T^{t;t-i}} g_j^{t-i}.
\end{equation}
Also, we change the waiting criteria in Step 2 of Algorithm~\ref{alg} to be ``The server waits until it receives $n-r$ vectors with timestamps between $t-\tau$ and $t$'', in other words, $\mnorm{T^t}\geq n-r$.
With the above GAR, suppose that there exists a $\tau\geq0$ such that $\mnorm{T^t}\geq n-r$ for all $t$, and the step sizes satisfy $\eta_t\geq\eta_{t+1}$ for all $t$, in addition to conditions in Theorem~\ref{thm:async-fault-toler}, then Theorem~\ref{thm:async-fault-toler} holds with the following parameters 
\begin{align}
    \begin{gathered}
    \phi_t\geq \alpha n\gamma\D^*(\D^*-\D)>0\textrm{ when }\norm{x^t-x_\H}\geq\D^*, \\
    \textrm{for all }\D^*>\D=\dfrac{\mu\left(2r + \tau\eta_0G\right)}{\alpha\gamma}\epsilon, \\
    \textrm{where } \alpha = 1-\frac{r}{n}\cdot\frac{\mu}{\gamma}>0, \,G=n\mu(2n\epsilon+\Gamma), \\
    \textrm{and }\Gamma = \max_{x\in\W}\norm{x-x_\H}. \\
    \end{gathered}
    \label{eqn:async-parameters-stale}
\end{align}
The proof for these results in \eqref{eqn:async-parameters-stale} is presented in Appendix~\ref{appdx:a-4}.

Intuitively, this method allows gradients of at most $\tau$-iteration stale. This result indicates that the algorithm would still be resilient even when using stale gradients. It can be expected that with less waiting time involved, the algorithm will be more efficient. Still, comparing \eqref{eqn:async-parameters-stale} to \eqref{eqn:async-parameters}, the bound on $\D$ indicates that the error range of the output is also affected, and the convergence rate 
will also be affected by $\tau$ when stale gradients are introduced.
\section{Resilient stochastic distributed optimization}
\label{sec:stochastic}

As mentioned briefly in Section~\ref{sec:intro}, stochastic optimization is useful when the computation of full gradients is too expensive \citep{bottou1998online, bottou2008tradeoffs, bottou2018optimization, sra2012optimization}, which is commonly used in various scenarios including machine learning \cite{sra2012optimization}. So it is worthwhile for us to also examine the potential of utilizing $(f,r;\epsilon)$-redundancy for solving resilient stochastic distributed optimization problems.



Consider a distributed stochastic optimization problem on a $d$ dimensional real-valued space $\R^d$. Each agent $i$ has a \textit{data generation distribution} $\mathcal{D}_i$ over $\R^m$. Each data point $z\in\R^m$ is a real-valued vector that incurs a \textit{loss} defined by a \textit{loss function} $\ell:(x;z)\mapsto\R$. The \textit{expected loss function} for each agent $i$ can be defined as
\begin{equation}
    Q_i(x)=\E_{z\sim\mathcal{D}_i}\ell(x;z),~\textrm{ for all }x\in\R^d. \label{eqn:learning-cost}
\end{equation}

With {this problem formulation} above, the gradient-based Algorithm~\ref{alg} proposed in Section~\ref{sec:full-grad} can also be adapted to a stochastic version for solving various problems including resilient distributed machine learning (RDML). 

Naturally, the goal of resilient distributed stochastic optimization is also \eqref{eqn:goal-ft}, the same as RDO. 
For machine learning problems, the machine learning model $\Pi$ can be parameterized as a $d$-dimensional vector $x\in\R^d$, which is the optimization target vector.

We first briefly revisit the computation of stochastic gradients in a distributed optimization system. To compute a stochastic gradient in iteration $t$, a (non-faulty) agent $i$ samples $k$ i.i.d. data points $z_{i_1}^t,...,z_{i_k}^t$ from its distribution $\mathcal{D}_i$ and computes
\begin{equation}
     g_i^t=\frac{1}{k}\sum_{i=1}^k\nabla\ell(x^t, z_{i_j}^t),
    \label{eqn:def-g-i-t}
\end{equation}
where $k$ is referred to as the \textit{batch size}. A faulty agent $i$ sends an arbitrary vector $g_i^t$.

We will also use the following notations in this section. Suppose faulty agents (if any) in the system are \textit{fixed} during a certain execution. For each non-faulty agent $i$, let $\z_i^t=\left\{z_{i_1}^t,...,z_{i_k}^t\right\}$ denote the collection of $k$ i.i.d. data points sampled by agent $i$ at iteration $t$. For each agent $i$ and iteration $t$, we define a random variable 
\begin{equation}
    \zeta_i^t=\begin{cases}
        \z_i^t, & \textrm{ agent $i$ is non-faulty}, \\
        g_i^t, & \textrm{ agent $i$ is faulty}.
    \end{cases}
    \label{eqn:zeta-i-t-cge}
\end{equation}
Recall that $g_i^t$ can be an arbitrary $d$-dimensional random variable for each Byzantine faulty agent $i$. For each iteration $t$, let $\zeta^t=\left\{\zeta_i^t,\,i=1,...,n\right\}$, and let $\E_t$ denote the expectation with respect to the collective random variables $\zeta^0,...,\zeta^t$, given the initial estimate $x^0$. Specifically, $\E_t(\cdot) \triangleq \E_{\zeta^0,...,\zeta^t}(\cdot)$.

Similar to Section~\ref{sec:full-grad}, we make standard Assumptions~\ref{assum:lipschitz} and \ref{assum:strongly-convex-ft}. We also need an extra assumption to bound the variance of stochastic gradients from all non-faulty agents. 
\begin{assumption}
     \label{assum:bound-grad}
     For each non-faulty agent $i$, assume that the variance of $g_i^t$ is bounded. Specifically, there exists a finite real value $\sigma$ such that for each non-faulty agent $i$,
     \begin{equation}
          \E_{\zeta_i^t}\norm{g_i^t-\E_{\zeta_i^t}\left(g_i^t\right)}^2\leq\sigma^2.
          \label{eqn:assum-bound-var-ft}
     \end{equation}
\end{assumption}
\noindent We make no assumption over the behavior of Byzantine agents. 

Suppose $\H\subseteq[n]$ is a subset of non-faulty agents with $\mnorm{\H}=n-f$, and a solution $x_\H$ exists in $\W$. The following theorem shows the general results for Resilient distributed stochastic optimization problems. Note that here Algorithm~\ref{alg} uses stochastic gradients instead of full gradients.
\begin{theorem}
    \label{thm:cge}
    Consider Algorithm~\ref{alg} with stochastic gradients, and the GAR in use is the CGE gradient filter \eqref{eqn:cge-footnote}. Suppose Assumptions~\ref{assum:lipschitz}, \ref{assum:strongly-convex-ft}, and \ref{assum:bound-grad} hold true, the expected cost functions of non-faulty agents satisfy $(f,r;\epsilon)$-redundancy, a resilience margin $\alpha>0$, and the step size in \eqref{eqn:update}, $\eta_t=\eta>0$ for all $t$. Let $\M$ denote an error-related margin.
    There exists an $\overline{\eta}$ such that, for $\eta<\overline{\eta}$, the following holds true:
    \begin{itemize}[nosep]
        \item The value of a convergence rate parameter $\rho$
        satisfies $0\leq\rho<1$, and 
        \item Given the initial estimate $x^0$ arbitrarily chosen from $\R^d$, for all $t\geq0$,
            \begin{align}
                \E_{t}\norm{x^{t+1}-x_\H}^2&\leq\rho^{t+1}\norm{x^0-x_\H}^2 + \frac{1-\rho^{t+1}}{1-\rho}\M.
                \label{eqn:expectation-bound-thm}
            \end{align}
    \end{itemize}
\end{theorem}
\noindent We have the following parameters when $f\geq0$ and $r\geq0$ for Theorem~\ref{thm:cge}:  
\begin{gather}
    \begin{gathered}
    \alpha = \frac{n-f}{n-r}-\frac{f+r}{n-r}\cdot\frac{2\mu}{\gamma},  \\
    \overline{\eta} = \frac{2(n-r)\gamma\alpha}{(n-f)^2\mu^2+2(n-r)^2\mu^2},  \\
    \rho = 1-2(n-f)\eta\gamma+4(f+r)\eta\mu+(n-f)^2\eta^2\mu^2+2(n-r)^2\eta^2\mu^2, \\
    \begin{aligned}
    \M =& 4\left(\left(2(f+r)+(n-f)^2\eta\mu\right)^2 +m^2(n-f)^2\eta^2\mu^2\right)\epsilon^2 \\
        &\,+\left(4\left(\frac{f+r}{m\mu}\right)^2\left(\sqrt{n-f-1}+1\right)^2 + (n-f)^2\eta^2\right)\sigma^2.
    \end{aligned} 
    \end{gathered}
    \label{eqn:stochastic-params-a}
\end{gather}
Note that we also need $n\geq 2f+3r$ to guarantee that $\rho\geq0$. The proof for these results in \eqref{eqn:stochastic-params-a} is presented in Appendix~\ref{appdx:b-2}.

\subsection{Special cases of resilient distributed stochastic optimization}
\label{sub:stochastic-special}

Similar to what we have in Section~\ref{sec:full-grad}, the stochastic version of resilient optimization also has several meaningful special cases. When $f=r=0$, the problem degenerates to the synchronous distributed stochastic optimization with no faulty agents in the system. The problem has also been solved by various methods without redundancy, including D-SGD with convexity assumptions which is commonly used in machine learning \citep{li2014communication}.

\subsubsection{Stochastic Byzantine optimization}
\label{ssub:byz-stoc}
When $r=0$, the problem becomes the Byzantine fault-tolerant problem. 
CGE gradient filter \eqref{eqn:cge-footnote} degenerates to averaging \eqref{eqn:aggregation-rule}. Theorem 2 holds with $(f,0;\epsilon)$-redundancy and the same parameters as in \eqref{eqn:stochastic-params-a}, substituting with $r=0$.

\subsubsection{Stochastic asynchronous optimization}
\label{ssub:async-stoc}
When $f=0$, our problem degenerates to the asynchronous problem. Without faulty agents, the goal becomes \eqref{eqn:goal}. We also have $\H=[n]$. Note that when $f=0$, the gradient filter CGE degenerates to the GAR \eqref{eqn:aggregation-rule}, i.e., plain summation of all received stochastic gradients. Theorem~\ref{thm:cge} holds with $(0,r;\epsilon)$-redundancy and the following parameters: 
\begin{align}
&\begin{gathered}
     \alpha = 1-\frac{r}{n}\cdot\frac{\mu}{\gamma}, \qquad \overline{\eta} = \frac{2n\gamma\alpha}{3n^2\mu^2}, \\
    \rho = 1-2(n\gamma-r\mu)\eta + 3n^2\eta^2\mu^2, \\
    \begin{aligned}
\M &= 4\left(\left(r +n^2\eta\mu\right)^2+n^4\eta^2\mu^2\right)\epsilon^2 \\
&\qquad +\left(n^2\eta^2+\left(\frac{r}{n\mu}\right)^2\left(\sqrt{n-1}+1\right)^2 \right)\sigma^2.
\end{aligned}
\end{gathered}
\label{eqn:stochastic-params-c}
\end{align}
Note that the requirement of $n\geq 2f+3r$ is not needed here. The proof for these results in \eqref{eqn:stochastic-params-c} is presented in Appendix~\ref{appdx:b-4}.

\subsection{Discussion}
\label{sub:stochastic-discuss}

The results provided above indicate that with $(f,r;\epsilon)$-redundancy, there exist algorithms to approximate the true solution to \eqref{eqn:goal} or \eqref{eqn:goal-ft} with D-SGD, where \textit{linear convergence} is achievable, and the error range of that approximation is{, in expectation,} proportional to $\epsilon$ and $\sigma$. Specifically, in \eqref{eqn:expectation-bound-thm} when $t\rightarrow\infty$, 
\begin{equation}
     \lim_{t\rightarrow\infty}\E_t\norm{x^{t+1}-x_\H}^2\leq\frac{1}{1-\rho}\M,
\end{equation}
where $\M$ changes monotonically as $\epsilon$ and $\sigma$. 

Note that \citet{gupta2021byzantine} showed a special case of the Byzantine fault-tolerant problem with redundancy, where all agents have the same data distribution. 
Our results can be applied to a broader range of problems, including heterogeneous problems like \textit{federated learning} \citep{konevcny2015federated}.  

\section{Empirical studies}
\label{sec:experiments}

In this section, we empirically show the effectiveness of our scheme in Algorithm~\ref{alg} on some 
benchmark distributed machine learning tasks. 



\subsection{Machine learning experiments}

Now we examine some more complex optimization problems: resilient distributed machine learning with deep neural networks. Here, we use stochastic version of our Algorithm~\ref{alg}, validating theoretical results in Section~\ref{sec:stochastic}. 
It is worth noting that even though the actual values of redundancy parameter $\epsilon$ are difficult to compute, through the following results we can still see that the said redundancy property exists in real-world scenarios and it supports our algorithm to achieve its applicability.

We simulate a server-based distributed learning system using multiple threads, one for the server, the rest for agents, with inter-thread communications handled by message passing interface. The simulator is built in Python with PyTorch \citep{paszke2019pytorch} and MPI4py \citep{dalcin2011parallel}, and deployed on a virtual machine with 14 vCPUs and 16 GB memory.

The experiments are conducted on three benchmark image-classification datasets: 
\begin{itemize}
    \item MNIST \citep{bottou1998online} of monochrome handwritten digits, 
    \item Fashion-MNIST \citep{xiao2017fashion} of grayscale images of clothes, and
    \item KMNIST \cite{clanuwat2018deep} of monochrome Japanese Hiragana characters. 
\end{itemize}
Each of the above datasets comprises of 60,000 training and 10,000 testing data points in 10 non-overlapping classes. 
For each dataset, we train a benchmark neural network LeNet \citep{lecun1998gradient} with 431,080 learnable parameters. 
Data points are divided among agents such that each agent gets 2 out of 10 classes, and each class appears in 4 agents; $\mathcal{D}_i$ for each agent $i$ is unique.

\begin{figure*}[tb]
    \centering
    \includegraphics[width=.75\linewidth]{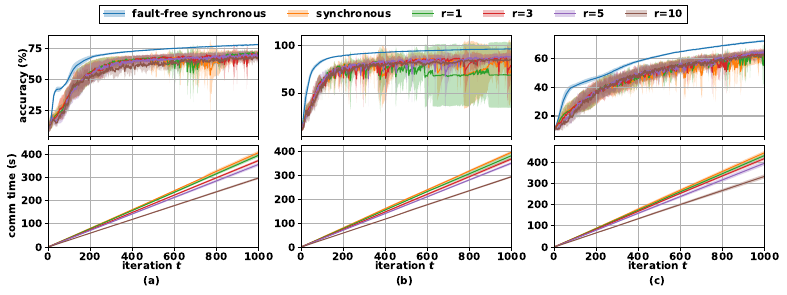}
    \includegraphics[width=.75\linewidth]{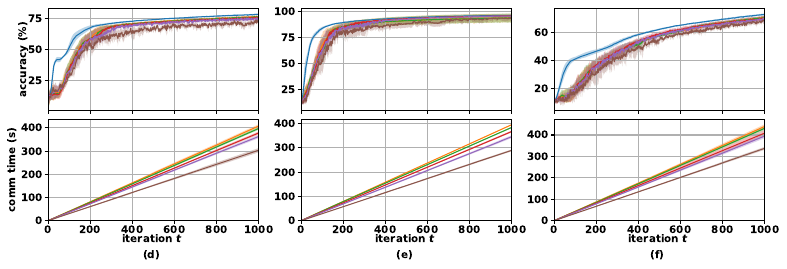}
    \caption{Results for resilient distributed learning with $n=20$ and $f=3$ using Algorithm~\ref{alg} with D-SGD and CGE gradient filter. Datasets: (a)(d) Fashion-MNIST, (b)(e) MNIST, (c)(f) KMNIST. Byzantine faults: (a-c) \textit{reverse-gradient}, (d-f) \textit{label-flipping}. Solid line for average; shade for standard deviation. Communication time for fault-free synchronous cases is omitted since the fault-free algorithm does not use CGE.}
    \label{fig:ds}
\end{figure*}

In each of our experiments, we simulate a distributed system of $n=20$ agents with $f=3$ and different values of $r=0,1,3,5,10$. Note that when $r=0$, the problem becomes synchronous Byzantine fault-tolerant learning. In each execution, we always designate the first $f$ agents to be Byzantine faulty. Stragglers are agents in each iteration whose gradients are the last $r$ ones received by the server. We also compare these results with the fault-free synchronous learning ($f=r=0$). We choose batch size $b=128$ for D-SGD, and fixed step size $\eta=0.01$. Performance of algorithms is measured by model accuracy at each step. We also document the cumulative communication time of each setting. Communication time of iteration $t$ is the time from the server's sending out of $x^t$ to its receiving of $n-r$ gradients. Experiments of each setting are run 4 times with different random seeds\footnote{Randomnesses exist in drawing of data points and stragglers in each iteration of each execution.}, and the averaged performance is reported. The results are shown in Figure~\ref{fig:ds}. We show the first 1,000 iterations as there is a clear trend of converging by the end of 1,000 iterations for both tasks in all four settings.

For Byzantine agents, we evaluate two types of faults: \textit{reverse-gradient} where faulty agents reverse the direction of its true gradients, and \textit{label-flipping} where faulty agents label data points of class $i$ as class $9-i$.

As is shown in the first row of Figure~\ref{fig:ds}, 
the learned model reaches comparable accuracy to the one learned by the synchronous algorithm at the same iteration; there is a gap between them and the fault-free case, echoing the error bound $\M$. In the second row of Figure~\ref{fig:ds}, we see that by dropping out $r$ stragglers, the communication time is gradually reduced when increasing the value of $r$. 

\subsection{Algorithm~\ref{alg} for practical use}

It is worth noting that $f$ and $r$ as parameters of Algorithm~\ref{alg} bear different meanings as $f$ and $r$ in the $(f,r;\epsilon)$-redundancy property. In the redundancy property, $f$ and $r$ together with $\epsilon$ describe how redundant the group of cost functions is. For the algorithm, $f$ and $r$ are hyperparameters set by users and indicate the numbers of faulty agents and stragglers we \textit{intend} to tolerate; these two values are not necessarily equal to the actual numbers in the system, which we may never know. Still, it is possible to estimate the value of $\epsilon$, and based on the theorems in Section~\ref{sec:full-grad} and \ref{sec:stochastic}, to estimate how close the output can be expected to be to the true solution. 

Like other hyper-parameters in optimization problems, there is no golden standard in choosing the values of $f$ and $r$. It would be better that $f$ and $r$ are close to the real numbers of faulty agents and stragglers in the system, which can be estimated by, for example, previous logs and statistics of the distributed system in use, to see how often are there failure of agents or how long of waiting can be qualified as stragglers.

Another interesting observation is that the algorithm does not require the Byzantine agents or stragglers to be ``\textit{fixed}''. In practice, it might be more common that the stragglers change from time to time during the training process, while malicious agents remains the same. Nonetheless, the gradient-based nature of our algorithm allows both of them to be changing; as long as the total number of them are bounded by $f$ and $r$, the algorithm remains valid.
\section{Summary}
\label{sec:summary}

We studied the impact of $(f,r;\epsilon)$-redundancy in cost functions on resilient distributed optimization and machine learning. Specifically, we presented an algorithm for resilient distributed optimization and learning, and analyzed its convergence when agents' cost functions have $(f,r;\epsilon)$-redundancy -- a generic characterization of redundancy in cost functions. We 
examined the resilient optimization and learning problem space. We showed that, under $(f,r;\epsilon)$-redundancy, Algorithm~\ref{alg} with DGD achieves $(f,r;\O(\epsilon))$-resilience for optimization, and Algorithm~\ref{alg} with D-SGD can solve resilient distributed stochastic optimization problems with error margins proportional to $\epsilon$.
We presented empirical results showing efficacy of Algorithm~\ref{alg} solving resilient distributed linear regression problems and  resilient distributed learning tasks. {Possible future work includes showing necessity of $(f,r;\epsilon)$-redundancy in solving RDO problems, analyzing the impact of redundancy in decentralized network, verifying our findings on larger, more complicated tasks, and so on.}

\textbf{Discussion on limitations } 
Note that our results in this paper are proved under strongly-convex assumptions. One may argue that such assumptions are too strong to be realistic. However, previous research has pointed out that although not a global property, cost functions of many machine learning problems are strongly-convex in the neighborhood of local minimizers \citep{bottou2018optimization}. Also, there is a research showing that the results on non-convex cost functions can be derived from those on strongly-convex cost functions \citep{allen2016optimal}, and therefore our results can be applied to a broader range of real-world problems. 
Our empirical results showing efficacy of our algorithm also concur with this argument.

It is also worth noting that the approximation bounds in optimization problems are linearly associated with the number of agents $n$, and the error margins in learning problems are related to $\Gamma$, the size of $\W$. These bounds can be loose when $n$ or $\Gamma$ is large. We do note that in practice $\W$ can be arbitrary, for example, a neighborhood of local minimizers mentioned above, making $\Gamma$ acceptably small. The value of $\epsilon$ can also be small in practice, as indicated by results in Section~\ref{sec:experiments}.

 \bibliographystyle{elsarticle-num-names} 
 \bibliography{bib}

\begin{thebibliography}{60}
\expandafter\ifx\csname natexlab\endcsname\relax\def\natexlab#1{#1}\fi
\providecommand{\url}[1]{\texttt{#1}}
\providecommand{\href}[2]{#2}
\providecommand{\path}[1]{#1}
\providecommand{\DOIprefix}{doi:}
\providecommand{\ArXivprefix}{arXiv:}
\providecommand{\URLprefix}{URL: }
\providecommand{\Pubmedprefix}{pmid:}
\providecommand{\doi}[1]{\href{http://dx.doi.org/#1}{\path{#1}}}
\providecommand{\Pubmed}[1]{\href{pmid:#1}{\path{#1}}}
\providecommand{\bibinfo}[2]{#2}
\ifx\xfnm\relax \def\xfnm[#1]{\unskip,\space#1}\fi
\bibitem[{Liu et~al.(2023)Liu, Gupta, and Vaidya}]{icdcn}
\bibinfo{author}{S.~Liu}, \bibinfo{author}{N.~Gupta}, \bibinfo{author}{N.~H. Vaidya},
\newblock \bibinfo{title}{Impact of redundancy on resilience in distributed optimization and learning},
\newblock in: \bibinfo{booktitle}{24th International Conference on Distributed Computing and Networking (ICDCN 2023)}, \bibinfo{organization}{ACM}, \bibinfo{year}{2023}.
\bibitem[{Otter et~al.(2020)Otter, Medina, and Kalita}]{otter2020survey}
\bibinfo{author}{D.~W. Otter}, \bibinfo{author}{J.~R. Medina}, \bibinfo{author}{J.~K. Kalita},
\newblock \bibinfo{title}{A survey of the usages of deep learning for natural language processing},
\newblock \bibinfo{journal}{IEEE Transactions on Neural Networks and Learning Systems} \bibinfo{volume}{32} (\bibinfo{year}{2020}) \bibinfo{pages}{604--624}.
\bibitem[{Boyd et~al.(2011)Boyd, Parikh, and Chu}]{boyd2011distributed}
\bibinfo{author}{S.~Boyd}, \bibinfo{author}{N.~Parikh}, \bibinfo{author}{E.~Chu}, \bibinfo{title}{Distributed optimization and statistical learning via the alternating direction method of multipliers}, \bibinfo{publisher}{Now Publishers Inc}, \bibinfo{year}{2011}.
\bibitem[{Rabbat and Nowak(2004)}]{rabbat2004distributed}
\bibinfo{author}{M.~Rabbat}, \bibinfo{author}{R.~Nowak},
\newblock \bibinfo{title}{Distributed optimization in sensor networks},
\newblock in: \bibinfo{booktitle}{Proceedings of the 3rd international symposium on Information processing in sensor networks}, \bibinfo{year}{2004}, pp. \bibinfo{pages}{20--27}.
\bibitem[{Raffard et~al.(2004)Raffard, Tomlin, and Boyd}]{raffard2004distributed}
\bibinfo{author}{R.~L. Raffard}, \bibinfo{author}{C.~J. Tomlin}, \bibinfo{author}{S.~P. Boyd},
\newblock \bibinfo{title}{Distributed optimization for cooperative agents: Application to formation flight},
\newblock in: \bibinfo{booktitle}{2004 43rd IEEE Conference on Decision and Control (CDC)(IEEE Cat. No. 04CH37601)}, volume~\bibinfo{volume}{3}, \bibinfo{organization}{IEEE}, \bibinfo{year}{2004}, pp. \bibinfo{pages}{2453--2459}.
\bibitem[{Lamport et~al.(1982)Lamport, Shostak, and Pease}]{lamport1982byzantine}
\bibinfo{author}{L.~Lamport}, \bibinfo{author}{R.~Shostak}, \bibinfo{author}{M.~Pease},
\newblock \bibinfo{title}{The byzantine generals problem},
\newblock \bibinfo{journal}{ACM Transactions on Programming Languages and Systems} \bibinfo{volume}{4} (\bibinfo{year}{1982}) \bibinfo{pages}{382--401}.
\bibitem[{Su and Vaidya(2016)}]{su2016fault}
\bibinfo{author}{L.~Su}, \bibinfo{author}{N.~H. Vaidya},
\newblock \bibinfo{title}{Fault-tolerant multi-agent optimization: optimal iterative distributed algorithms},
\newblock in: \bibinfo{booktitle}{Proceedings of the 2016 ACM symposium on principles of distributed computing}, \bibinfo{year}{2016}, pp. \bibinfo{pages}{425--434}.
\bibitem[{Hannah and Yin(2017)}]{hannah2017more}
\bibinfo{author}{R.~Hannah}, \bibinfo{author}{W.~Yin},
\newblock \bibinfo{title}{More iterations per second, same quality--why asynchronous algorithms may drastically outperform traditional ones},
\newblock \bibinfo{journal}{arXiv preprint arXiv:1708.05136}  (\bibinfo{year}{2017}).
\bibitem[{Leblond(2018)}]{leblond2018asynchronous}
\bibinfo{author}{R.~Leblond}, \bibinfo{title}{Asynchronous optimization for machine learning}, Ph.D. thesis, PSL Research University, \bibinfo{year}{2018}.
\bibitem[{Assran et~al.(2020)Assran, Aytekin, Feyzmahdavian, Johansson, and Rabbat}]{assran2020advances}
\bibinfo{author}{M.~Assran}, \bibinfo{author}{A.~Aytekin}, \bibinfo{author}{H.~R. Feyzmahdavian}, \bibinfo{author}{M.~Johansson}, \bibinfo{author}{M.~G. Rabbat},
\newblock \bibinfo{title}{Advances in asynchronous parallel and distributed optimization},
\newblock \bibinfo{journal}{Proceedings of the IEEE} \bibinfo{volume}{108} (\bibinfo{year}{2020}) \bibinfo{pages}{2013--2031}.
\bibitem[{Chen et~al.(2018)Chen, Wang, Charles, and Papailiopoulos}]{chen2018draco}
\bibinfo{author}{L.~Chen}, \bibinfo{author}{H.~Wang}, \bibinfo{author}{Z.~Charles}, \bibinfo{author}{D.~Papailiopoulos},
\newblock \bibinfo{title}{Draco: Byzantine-resilient distributed training via redundant gradients},
\newblock in: \bibinfo{booktitle}{International Conference on Machine Learning}, \bibinfo{organization}{PMLR}, \bibinfo{year}{2018}, pp. \bibinfo{pages}{903--912}.
\bibitem[{Blanchard et~al.(2017)Blanchard, El~Mhamdi, Guerraoui, and Stainer}]{blanchard2017machine}
\bibinfo{author}{P.~Blanchard}, \bibinfo{author}{E.~M. El~Mhamdi}, \bibinfo{author}{R.~Guerraoui}, \bibinfo{author}{J.~Stainer},
\newblock \bibinfo{title}{Machine learning with adversaries: Byzantine tolerant gradient descent},
\newblock in: \bibinfo{booktitle}{Proceedings of the 31st International Conference on Neural Information Processing Systems}, \bibinfo{year}{2017}, pp. \bibinfo{pages}{118--128}.
\bibitem[{Liu et~al.(2021)Liu, Gupta, and Vaidya}]{liu2021approximate}
\bibinfo{author}{S.~Liu}, \bibinfo{author}{N.~Gupta}, \bibinfo{author}{N.~H. Vaidya},
\newblock \bibinfo{title}{Approximate byzantine fault-tolerance in distributed optimization},
\newblock in: \bibinfo{booktitle}{Proceedings of the 2021 ACM Symposium on Principles of Distributed Computing}, PODC'21, \bibinfo{publisher}{Association for Computing Machinery}, \bibinfo{address}{New York, NY, USA}, \bibinfo{year}{2021}, p. \bibinfo{pages}{379–389}.
\bibitem[{Tandon et~al.(2017)Tandon, Lei, Dimakis, and Karampatziakis}]{tandon2017gradient}
\bibinfo{author}{R.~Tandon}, \bibinfo{author}{Q.~Lei}, \bibinfo{author}{A.~G. Dimakis}, \bibinfo{author}{N.~Karampatziakis},
\newblock \bibinfo{title}{Gradient coding: Avoiding stragglers in distributed learning},
\newblock in: \bibinfo{booktitle}{International Conference on Machine Learning}, \bibinfo{organization}{PMLR}, \bibinfo{year}{2017}, pp. \bibinfo{pages}{3368--3376}.
\bibitem[{Halbawi et~al.(2018)Halbawi, Azizan, Salehi, and Hassibi}]{halbawi2018improving}
\bibinfo{author}{W.~Halbawi}, \bibinfo{author}{N.~Azizan}, \bibinfo{author}{F.~Salehi}, \bibinfo{author}{B.~Hassibi},
\newblock \bibinfo{title}{Improving distributed gradient descent using reed-solomon codes},
\newblock in: \bibinfo{booktitle}{2018 IEEE International Symposium on Information Theory (ISIT)}, \bibinfo{organization}{IEEE}, \bibinfo{year}{2018}, pp. \bibinfo{pages}{2027--2031}.
\bibitem[{Karakus et~al.(2017)Karakus, Sun, and Diggavi}]{karakus2017encoded}
\bibinfo{author}{C.~Karakus}, \bibinfo{author}{Y.~Sun}, \bibinfo{author}{S.~Diggavi},
\newblock \bibinfo{title}{Encoded distributed optimization},
\newblock in: \bibinfo{booktitle}{2017 IEEE International Symposium on Information Theory (ISIT)}, \bibinfo{organization}{IEEE}, \bibinfo{year}{2017}, pp. \bibinfo{pages}{2890--2894}.
\bibitem[{Niu et~al.(2011)Niu, Recht, R{\'e}, and Wright}]{niu2011hogwild}
\bibinfo{author}{F.~Niu}, \bibinfo{author}{B.~Recht}, \bibinfo{author}{C.~R{\'e}}, \bibinfo{author}{S.~J. Wright},
\newblock \bibinfo{title}{Hogwild!: A lock-free approach to parallelizing stochastic gradient descent},
\newblock \bibinfo{journal}{arXiv preprint arXiv:1106.5730}  (\bibinfo{year}{2011}).
\bibitem[{Conci and Kubrusly(2018)}]{conci2018distance}
\bibinfo{author}{A.~Conci}, \bibinfo{author}{C.~S. Kubrusly}, \bibinfo{title}{Distance between sets - a survey}, \bibinfo{year}{2018}. \href{http://arxiv.org/abs/1808.02574}{{\tt arXiv:1808.02574}}.
\bibitem[{Gupta and Vaidya(2020)}]{gupta2020resilience}
\bibinfo{author}{N.~Gupta}, \bibinfo{author}{N.~H. Vaidya},
\newblock \bibinfo{title}{Resilience in collaborative optimization: redundant and independent cost functions},
\newblock \bibinfo{journal}{arXiv preprint arXiv:2003.09675}  (\bibinfo{year}{2020}).
\bibitem[{Su and Vaidya(2015)}]{su2015byzantine1}
\bibinfo{author}{L.~Su}, \bibinfo{author}{N.~Vaidya},
\newblock \bibinfo{title}{{Byzantine} multi-agent optimization: Part {I}},
\newblock \bibinfo{journal}{arXiv preprint arXiv:1506.04681}  (\bibinfo{year}{2015}).
\bibitem[{Liu(2021)}]{liu2021survey}
\bibinfo{author}{S.~Liu},
\newblock \bibinfo{title}{A survey on fault-tolerance in distributed optimization and machine learning},
\newblock \bibinfo{journal}{arXiv preprint arXiv:2106.08545}  (\bibinfo{year}{2021}).
\bibitem[{Chen et~al.(2017)Chen, Su, and Xu}]{chen2017distributed}
\bibinfo{author}{Y.~Chen}, \bibinfo{author}{L.~Su}, \bibinfo{author}{J.~Xu},
\newblock \bibinfo{title}{Distributed statistical machine learning in adversarial settings: Byzantine gradient descent},
\newblock \bibinfo{journal}{Proceedings of the ACM on Measurement and Analysis of Computing Systems} \bibinfo{volume}{1} (\bibinfo{year}{2017}) \bibinfo{pages}{1--25}.
\bibitem[{Gupta et~al.(2021)Gupta, Liu, and Vaidya}]{gupta2021byzantine}
\bibinfo{author}{N.~Gupta}, \bibinfo{author}{S.~Liu}, \bibinfo{author}{N.~Vaidya},
\newblock \bibinfo{title}{Byzantine fault-tolerant distributed machine learning with norm-based comparative gradient elimination},
\newblock in: \bibinfo{booktitle}{2021 51st Annual IEEE/IFIP International Conference on Dependable Systems and Networks Workshops (DSN-W)}, \bibinfo{organization}{IEEE}, \bibinfo{year}{2021}, pp. \bibinfo{pages}{175--181}.
\bibitem[{Xie et~al.(2018)Xie, Koyejo, and Gupta}]{xie2018zeno}
\bibinfo{author}{C.~Xie}, \bibinfo{author}{O.~Koyejo}, \bibinfo{author}{I.~Gupta},
\newblock \bibinfo{title}{Zeno: Byzantine-suspicious stochastic gradient descent},
\newblock \bibinfo{journal}{arXiv preprint arXiv:1805.10032} \bibinfo{volume}{24} (\bibinfo{year}{2018}).
\bibitem[{Yin et~al.(2018)Yin, Chen, Kannan, and Bartlett}]{yin2018byzantine}
\bibinfo{author}{D.~Yin}, \bibinfo{author}{Y.~Chen}, \bibinfo{author}{R.~Kannan}, \bibinfo{author}{P.~Bartlett},
\newblock \bibinfo{title}{Byzantine-robust distributed learning: Towards optimal statistical rates},
\newblock in: \bibinfo{booktitle}{International Conference on Machine Learning}, \bibinfo{organization}{PMLR}, \bibinfo{year}{2018}, pp. \bibinfo{pages}{5650--5659}.
\bibitem[{Langford et~al.(2009)Langford, Smola, and Zinkevich}]{langford2009slow}
\bibinfo{author}{J.~Langford}, \bibinfo{author}{A.~Smola}, \bibinfo{author}{M.~Zinkevich},
\newblock \bibinfo{title}{Slow learners are fast},
\newblock \bibinfo{journal}{arXiv preprint arXiv:0911.0491}  (\bibinfo{year}{2009}).
\bibitem[{Agarwal and Duchi(2012)}]{agarwal2012distributed}
\bibinfo{author}{A.~Agarwal}, \bibinfo{author}{J.~C. Duchi},
\newblock \bibinfo{title}{Distributed delayed stochastic optimization},
\newblock in: \bibinfo{booktitle}{51st IEEE Conference on Decision and Control}, \bibinfo{publisher}{IEEE}, \bibinfo{year}{2012}, pp. \bibinfo{pages}{5451--5452}.
\bibitem[{Feyzmahdavian et~al.(2016)Feyzmahdavian, Aytekin, and Johansson}]{feyzmahdavian2016asynchronous}
\bibinfo{author}{H.~R. Feyzmahdavian}, \bibinfo{author}{A.~Aytekin}, \bibinfo{author}{M.~Johansson},
\newblock \bibinfo{title}{An asynchronous mini-batch algorithm for regularized stochastic optimization},
\newblock \bibinfo{journal}{IEEE Transactions on Automatic Control} \bibinfo{volume}{61} (\bibinfo{year}{2016}) \bibinfo{pages}{3740--3754}.
\bibitem[{Roux et~al.(2012)Roux, Schmidt, and Bach}]{roux2012stochastic}
\bibinfo{author}{N.~L. Roux}, \bibinfo{author}{M.~Schmidt}, \bibinfo{author}{F.~Bach},
\newblock \bibinfo{title}{A stochastic gradient method with an exponential convergence rate for finite training sets},
\newblock \bibinfo{journal}{arXiv preprint arXiv:1202.6258}  (\bibinfo{year}{2012}).
\bibitem[{Johnson and Zhang(2013)}]{johnson2013accelerating}
\bibinfo{author}{R.~Johnson}, \bibinfo{author}{T.~Zhang},
\newblock \bibinfo{title}{Accelerating stochastic gradient descent using predictive variance reduction},
\newblock \bibinfo{journal}{Advances in neural information processing systems} \bibinfo{volume}{26} (\bibinfo{year}{2013}) \bibinfo{pages}{315--323}.
\bibitem[{Defazio et~al.(2014)Defazio, Bach, and Lacoste-Julien}]{defazio2014saga}
\bibinfo{author}{A.~Defazio}, \bibinfo{author}{F.~Bach}, \bibinfo{author}{S.~Lacoste-Julien},
\newblock \bibinfo{title}{Saga: A fast incremental gradient method with support for non-strongly convex composite objectives},
\newblock \bibinfo{journal}{arXiv preprint arXiv:1407.0202}  (\bibinfo{year}{2014}).
\bibitem[{Shalev-Shwartz and Zhang(2013)}]{shalev2013accelerated}
\bibinfo{author}{S.~Shalev-Shwartz}, \bibinfo{author}{T.~Zhang},
\newblock \bibinfo{title}{Accelerated mini-batch stochastic dual coordinate ascent},
\newblock \bibinfo{journal}{arXiv preprint arXiv:1305.2581}  (\bibinfo{year}{2013}).
\bibitem[{Lee et~al.(2017)Lee, Lam, Pedarsani, Papailiopoulos, and Ramchandran}]{lee2017speeding}
\bibinfo{author}{K.~Lee}, \bibinfo{author}{M.~Lam}, \bibinfo{author}{R.~Pedarsani}, \bibinfo{author}{D.~Papailiopoulos}, \bibinfo{author}{K.~Ramchandran},
\newblock \bibinfo{title}{Speeding up distributed machine learning using codes},
\newblock \bibinfo{journal}{IEEE Transactions on Information Theory} \bibinfo{volume}{64} (\bibinfo{year}{2017}) \bibinfo{pages}{1514--1529}.
\bibitem[{Karakus et~al.(2017)Karakus, Sun, Diggavi, and Yin}]{karakus2017straggler}
\bibinfo{author}{C.~Karakus}, \bibinfo{author}{Y.~Sun}, \bibinfo{author}{S.~Diggavi}, \bibinfo{author}{W.~Yin},
\newblock \bibinfo{title}{Straggler mitigation in distributed optimization through data encoding},
\newblock \bibinfo{journal}{Advances in Neural Information Processing Systems} \bibinfo{volume}{30} (\bibinfo{year}{2017}) \bibinfo{pages}{5434--5442}.
\bibitem[{Yang et~al.(2017)Yang, Grover, and Kar}]{yang2017coded}
\bibinfo{author}{Y.~Yang}, \bibinfo{author}{P.~Grover}, \bibinfo{author}{S.~Kar},
\newblock \bibinfo{title}{Coded distributed computing for inverse problems},
\newblock in: \bibinfo{booktitle}{Proceedings of the 31st International Conference on Neural Information Processing Systems}, \bibinfo{year}{2017}, pp. \bibinfo{pages}{709--719}.
\bibitem[{Karakus et~al.(2019)Karakus, Sun, Diggavi, and Yin}]{karakus2019redundancy}
\bibinfo{author}{C.~Karakus}, \bibinfo{author}{Y.~Sun}, \bibinfo{author}{S.~Diggavi}, \bibinfo{author}{W.~Yin},
\newblock \bibinfo{title}{Redundancy techniques for straggler mitigation in distributed optimization and learning},
\newblock \bibinfo{journal}{The Journal of Machine Learning Research} \bibinfo{volume}{20} (\bibinfo{year}{2019}) \bibinfo{pages}{2619--2665}.
\bibitem[{Ananthanarayanan et~al.(2013)Ananthanarayanan, Ghodsi, Shenker, and Stoica}]{ananthanarayanan2013effective}
\bibinfo{author}{G.~Ananthanarayanan}, \bibinfo{author}{A.~Ghodsi}, \bibinfo{author}{S.~Shenker}, \bibinfo{author}{I.~Stoica},
\newblock \bibinfo{title}{Effective straggler mitigation: Attack of the clones},
\newblock in: \bibinfo{booktitle}{10th {USENIX} Symposium on Networked Systems Design and Implementation ({NSDI} 13)}, \bibinfo{year}{2013}, pp. \bibinfo{pages}{185--198}.
\bibitem[{Gardner et~al.(2015)Gardner, Zbarsky, Doroudi, Harchol-Balter, and Hyytia}]{gardner2015reducing}
\bibinfo{author}{K.~Gardner}, \bibinfo{author}{S.~Zbarsky}, \bibinfo{author}{S.~Doroudi}, \bibinfo{author}{M.~Harchol-Balter}, \bibinfo{author}{E.~Hyytia},
\newblock \bibinfo{title}{Reducing latency via redundant requests: Exact analysis},
\newblock \bibinfo{journal}{ACM SIGMETRICS Performance Evaluation Review} \bibinfo{volume}{43} (\bibinfo{year}{2015}) \bibinfo{pages}{347--360}.
\bibitem[{Shah et~al.(2015)Shah, Lee, and Ramchandran}]{shah2015redundant}
\bibinfo{author}{N.~B. Shah}, \bibinfo{author}{K.~Lee}, \bibinfo{author}{K.~Ramchandran},
\newblock \bibinfo{title}{When do redundant requests reduce latency?},
\newblock \bibinfo{journal}{IEEE Transactions on Communications} \bibinfo{volume}{64} (\bibinfo{year}{2015}) \bibinfo{pages}{715--722}.
\bibitem[{Wang et~al.(2015)Wang, Joshi, and Wornell}]{wang2015using}
\bibinfo{author}{D.~Wang}, \bibinfo{author}{G.~Joshi}, \bibinfo{author}{G.~Wornell},
\newblock \bibinfo{title}{Using straggler replication to reduce latency in large-scale parallel computing},
\newblock \bibinfo{journal}{ACM SIGMETRICS Performance Evaluation Review} \bibinfo{volume}{43} (\bibinfo{year}{2015}) \bibinfo{pages}{7--11}.
\bibitem[{Yadwadkar et~al.(2016)Yadwadkar, Hariharan, Gonzalez, and Katz}]{yadwadkar2016multi}
\bibinfo{author}{N.~J. Yadwadkar}, \bibinfo{author}{B.~Hariharan}, \bibinfo{author}{J.~E. Gonzalez}, \bibinfo{author}{R.~Katz},
\newblock \bibinfo{title}{Multi-task learning for straggler avoiding predictive job scheduling},
\newblock \bibinfo{journal}{The Journal of Machine Learning Research} \bibinfo{volume}{17} (\bibinfo{year}{2016}) \bibinfo{pages}{3692--3728}.
\bibitem[{Damaskinos et~al.(2019)Damaskinos, El~Mhamdi, Guerraoui, Guirguis, and Rouault}]{damaskinos2019aggregathor}
\bibinfo{author}{G.~Damaskinos}, \bibinfo{author}{E.~M. El~Mhamdi}, \bibinfo{author}{R.~Guerraoui}, \bibinfo{author}{A.~H.~A. Guirguis}, \bibinfo{author}{S.~L.~A. Rouault},
\newblock \bibinfo{title}{Aggregathor: Byzantine machine learning via robust gradient aggregation},
\newblock in: \bibinfo{booktitle}{The Conference on Systems and Machine Learning (SysML), 2019}, \bibinfo{number}{CONF}, \bibinfo{year}{2019}.
\bibitem[{Karimireddy et~al.(2021)Karimireddy, He, and Jaggi}]{karimireddy2021learning}
\bibinfo{author}{S.~P. Karimireddy}, \bibinfo{author}{L.~He}, \bibinfo{author}{M.~Jaggi},
\newblock \bibinfo{title}{Learning from history for byzantine robust optimization},
\newblock in: \bibinfo{booktitle}{International Conference on Machine Learning}, \bibinfo{organization}{PMLR}, \bibinfo{year}{2021}, pp. \bibinfo{pages}{5311--5319}.
\bibitem[{Nedic and Ozdaglar(2009)}]{nedic2009distributed}
\bibinfo{author}{A.~Nedic}, \bibinfo{author}{A.~Ozdaglar},
\newblock \bibinfo{title}{Distributed subgradient methods for multi-agent optimization},
\newblock \bibinfo{journal}{IEEE Transactions on Automatic Control} \bibinfo{volume}{54} (\bibinfo{year}{2009}) \bibinfo{pages}{48--61}.
\bibitem[{Shi et~al.(2015)Shi, Ling, Wu, and Yin}]{shi2015extra}
\bibinfo{author}{W.~Shi}, \bibinfo{author}{Q.~Ling}, \bibinfo{author}{G.~Wu}, \bibinfo{author}{W.~Yin},
\newblock \bibinfo{title}{Extra: An exact first-order algorithm for decentralized consensus optimization},
\newblock \bibinfo{journal}{SIAM Journal on Optimization} \bibinfo{volume}{25} (\bibinfo{year}{2015}) \bibinfo{pages}{944--966}.
\bibitem[{Varagnolo et~al.(2015)Varagnolo, Zanella, Cenedese, Pillonetto, and Schenato}]{varagnolo2015newton}
\bibinfo{author}{D.~Varagnolo}, \bibinfo{author}{F.~Zanella}, \bibinfo{author}{A.~Cenedese}, \bibinfo{author}{G.~Pillonetto}, \bibinfo{author}{L.~Schenato},
\newblock \bibinfo{title}{Newton-raphson consensus for distributed convex optimization},
\newblock \bibinfo{journal}{IEEE Transactions on Automatic Control} \bibinfo{volume}{61} (\bibinfo{year}{2015}) \bibinfo{pages}{994--1009}.
\bibitem[{Bottou(1998)}]{bottou1998online}
\bibinfo{author}{L.~Bottou},
\newblock \bibinfo{title}{Online learning and stochastic approximations},
\newblock \bibinfo{journal}{On-line learning in neural networks} \bibinfo{volume}{17} (\bibinfo{year}{1998}) \bibinfo{pages}{142}.
\bibitem[{Bottou and Bousquet(2008)}]{bottou2008tradeoffs}
\bibinfo{author}{L.~Bottou}, \bibinfo{author}{O.~Bousquet},
\newblock \bibinfo{title}{The tradeoffs of large scale learning},
\newblock in: \bibinfo{booktitle}{Advances in Neural Information Processing Systems 20 (NIPS 2007)}, \bibinfo{publisher}{NIPS Foundation}, \bibinfo{year}{2008}, pp. \bibinfo{pages}{161--168}.
\bibitem[{Bottou et~al.(2018)Bottou, Curtis, and Nocedal}]{bottou2018optimization}
\bibinfo{author}{L.~Bottou}, \bibinfo{author}{F.~E. Curtis}, \bibinfo{author}{J.~Nocedal},
\newblock \bibinfo{title}{Optimization methods for large-scale machine learning},
\newblock \bibinfo{journal}{Siam Review} \bibinfo{volume}{60} (\bibinfo{year}{2018}) \bibinfo{pages}{223--311}.
\bibitem[{Sra et~al.(2012)Sra, Nowozin, and Wright}]{sra2012optimization}
\bibinfo{author}{S.~Sra}, \bibinfo{author}{S.~Nowozin}, \bibinfo{author}{S.~Wright}, \bibinfo{title}{Optimization for Machine Learning}, Neural information processing series, \bibinfo{publisher}{MIT Press}, \bibinfo{year}{2012}.
\bibitem[{Li et~al.(2014)Li, Andersen, Smola, and Yu}]{li2014communication}
\bibinfo{author}{M.~Li}, \bibinfo{author}{D.~G. Andersen}, \bibinfo{author}{A.~J. Smola}, \bibinfo{author}{K.~Yu},
\newblock \bibinfo{title}{Communication efficient distributed machine learning with the parameter server},
\newblock \bibinfo{journal}{Advances in Neural Information Processing Systems} \bibinfo{volume}{27} (\bibinfo{year}{2014}) \bibinfo{pages}{19--27}.
\bibitem[{Kone{\v{c}}n{\`y} et~al.(2015)Kone{\v{c}}n{\`y}, McMahan, and Ramage}]{konevcny2015federated}
\bibinfo{author}{J.~Kone{\v{c}}n{\`y}}, \bibinfo{author}{B.~McMahan}, \bibinfo{author}{D.~Ramage},
\newblock \bibinfo{title}{Federated optimization: Distributed optimization beyond the datacenter},
\newblock \bibinfo{journal}{arXiv preprint arXiv:1511.03575}  (\bibinfo{year}{2015}).
\bibitem[{Paszke et~al.(2019)Paszke, Gross, Massa, Lerer, Bradbury, Chanan, Killeen, Lin, Gimelshein, Antiga et~al.}]{paszke2019pytorch}
\bibinfo{author}{A.~Paszke}, \bibinfo{author}{S.~Gross}, \bibinfo{author}{F.~Massa}, \bibinfo{author}{A.~Lerer}, \bibinfo{author}{J.~Bradbury}, \bibinfo{author}{G.~Chanan}, \bibinfo{author}{T.~Killeen}, \bibinfo{author}{Z.~Lin}, \bibinfo{author}{N.~Gimelshein}, \bibinfo{author}{L.~Antiga}, et~al.,
\newblock \bibinfo{title}{Pytorch: An imperative style, high-performance deep learning library},
\newblock \bibinfo{journal}{arXiv preprint arXiv:1912.01703}  (\bibinfo{year}{2019}).
\bibitem[{Dalcin et~al.(2011)Dalcin, Paz, Kler, and Cosimo}]{dalcin2011parallel}
\bibinfo{author}{L.~D. Dalcin}, \bibinfo{author}{R.~R. Paz}, \bibinfo{author}{P.~A. Kler}, \bibinfo{author}{A.~Cosimo},
\newblock \bibinfo{title}{Parallel distributed computing using python},
\newblock \bibinfo{journal}{Advances in Water Resources} \bibinfo{volume}{34} (\bibinfo{year}{2011}) \bibinfo{pages}{1124--1139}.
\bibitem[{Xiao et~al.(2017)Xiao, Rasul, and Vollgraf}]{xiao2017fashion}
\bibinfo{author}{H.~Xiao}, \bibinfo{author}{K.~Rasul}, \bibinfo{author}{R.~Vollgraf},
\newblock \bibinfo{title}{Fashion-mnist: a novel image dataset for benchmarking machine learning algorithms},
\newblock \bibinfo{journal}{arXiv preprint arXiv:1708.07747}  (\bibinfo{year}{2017}).
\bibitem[{Clanuwat et~al.(2018)Clanuwat, Bober-Irizar, Kitamoto, Lamb, Yamamoto, and Ha}]{clanuwat2018deep}
\bibinfo{author}{T.~Clanuwat}, \bibinfo{author}{M.~Bober-Irizar}, \bibinfo{author}{A.~Kitamoto}, \bibinfo{author}{A.~Lamb}, \bibinfo{author}{K.~Yamamoto}, \bibinfo{author}{D.~Ha},
\newblock \bibinfo{title}{Deep learning for classical japanese literature},
\newblock \bibinfo{journal}{arXiv preprint arXiv:1812.01718}  (\bibinfo{year}{2018}).
\bibitem[{LeCun et~al.(1998)LeCun, Bottou, Bengio, and Haffner}]{lecun1998gradient}
\bibinfo{author}{Y.~LeCun}, \bibinfo{author}{L.~Bottou}, \bibinfo{author}{Y.~Bengio}, \bibinfo{author}{P.~Haffner},
\newblock \bibinfo{title}{Gradient-based learning applied to document recognition},
\newblock \bibinfo{journal}{Proceedings of the IEEE} \bibinfo{volume}{86} (\bibinfo{year}{1998}) \bibinfo{pages}{2278--2324}.
\bibitem[{Allen-Zhu and Hazan(2016)}]{allen2016optimal}
\bibinfo{author}{Z.~Allen-Zhu}, \bibinfo{author}{E.~Hazan},
\newblock \bibinfo{title}{Optimal black-box reductions between optimization objectives},
\newblock \bibinfo{journal}{arXiv preprint arXiv:1603.05642}  (\bibinfo{year}{2016}).
\bibitem[{Arnold and Groeneveld(1979)}]{arnold1979bounds}
\bibinfo{author}{B.~C. Arnold}, \bibinfo{author}{R.~A. Groeneveld},
\newblock \bibinfo{title}{Bounds on expectations of linear systematic statistics based on dependent samples},
\newblock \bibinfo{journal}{The Annals of Statistics}  (\bibinfo{year}{1979}) \bibinfo{pages}{220--223}.
\bibitem[{Liu et~al.(2021)Liu, Gupta, and Vaidya}]{liu2021approximate-full}
\bibinfo{author}{S.~Liu}, \bibinfo{author}{N.~Gupta}, \bibinfo{author}{N.~H. Vaidya},
\newblock \bibinfo{title}{Approximate byzantine fault-tolerance in distributed optimization},
\newblock \bibinfo{journal}{CoRR} \bibinfo{volume}{abs/2101.09337} (\bibinfo{year}{2021}). \href{http://arxiv.org/abs/2101.09337}{{\tt arXiv:2101.09337}}.

\end{thebibliography}


\appendix
\section{Table of notations in this paper}
\label{appdx:0}

For reference purposes, Table~\ref{tab:notations} summarizes common notations that have appeared in this paper.

\begin{table}[h]
    \centering
    \footnotesize
    \begin{tabularx}{\textwidth}{c|X}
        \toprule
        Notation & Meaning \\
        \midrule
        $n$ & The total number of agents in a distributed system \\
        $\W$ & A compact set in $\R^d$ where the optimization problem is defined \\
        $f$ & Upper bound on the number of faulty agents \\
        $r$ & Upper bound on the number of stragglers \\
        $\epsilon$ & The approximation parameter in the redundancy notation \\
        $\H$ & A set of $n-f$ non-faulty agents \\
        $\B$ & A set of faulty agents, see context for details \\
        $S^t$ & The set of agents whose gradients are received in iteration $t$ \\
        $\mu$ & Lipschitz constant in Assumption~\ref{assum:lipschitz} \\
        $\gamma$ & Strong-convexity constant in Assumption~\ref{assum:strongly-convex-ft} \\
        \midrule
        $\z_{i_k}^t$ & The $k$-th data point sampled by agent $i$ in iteration $t$ \\
        $\ell(x;z)$ & Loss incurred by data point $z$ with model parameter $x$ \\
        $\z_{i}^t$ & A batch of $k$ data points sampled by agent $i$ in iteration $t$ \\
        $\sigma$ & Upper bound on the variance of the norm of a non-faulty agent's stochastic gradient \\
        \midrule
        $\norm{\cdot}$ & Euclidean norm of a vector \\
        $\mnorm{\cdot}$ & Size of a set, or absolute value of a value \\
        \dist{\cdot}{\cdot} & Hausdorff distance between two points or sets \\
        $[k]$ & Shorthand for the set $\{1,...,k\}$ \\
        $\ga\left(\cdot;n,f,r\right)$ & A gradient aggregation rule (GAR) \\
        \bottomrule
    \end{tabularx}
    \caption{Common notations in this paper and their meanings}
    \label{tab:notations}
\end{table}

\section{Proofs of theorems in Section~\ref{sec:full-grad}}
\label{part:2}

\setcounter{equation}{23}

In this section, we present the detailed proof of the asymptotic convergence results presented in Section~\ref{sec:full-grad}, 
i.e., Theorem~\ref{thm:async-fault-toler}, with parameter settings \eqref{eqn:thm2-params} or \eqref{eqn:async-parameters}. For each of them, we restate the theorem with the parameters substituted in in full, then proceed with the proofs of the theorems.

\subsection{Proof of Theorem~\ref{thm:async-fault-toler}}
\label{appdx:proof-lemma-bound}

The proof of this theorem uses the following sufficient criterion for the convergence of non-negative sequences:
\begin{lemma}[Ref. \citep{bottou1998online}]
    \label{lemma:converge}
    Consider a sequence of real values $\{u_t\}$, $t=0,1,\dots$. If $u_t\geq0$, $\forall t$, 
    \begin{align*}
        &\sum_{t=0}^\infty(u_{t+1}-u_t)_+<\infty ~\textrm{ implies }~\left\{\begin{array}{l}
            u_t\xrightarrow[t\rightarrow\infty]{}u_\infty<\infty, \\
            \sum_{t=0}^\infty(u_{t+1}-u_t)_->-\infty,
        \end{array}\right.
    \end{align*}
    where the operators $(\cdot)_+$ and $(\cdot)_-$ are defined as follows for a real value scalar $x$:
    \begin{align*}
        (x)_+=\left\{\begin{array}{ll}
            x, & x>0, \\
            0, & \textrm{otherwise},
        \end{array}\right.~\textrm{ and }~
        (x)_-=\left\{\begin{array}{ll}
            0, & x>0, \\
            x, & \textrm{otherwise}.
        \end{array}\right.
    \end{align*}\qed
\end{lemma}

Recall that we defined in Section~\ref{sec:full-grad} that for any non-faulty agents $\H$ with $\mnorm{\H}=n-f$ in an execution, $x_\H\in\W$ is a unique minimum point of the aggregate cost function of agents in $\H$.

\begin{mdframed}[default]
\textsc{Theorem~\ref{thm:async-fault-toler}}. 
    \textit{Suppose $x_\H$ is any point in $\W$. Assume that $\eta_t$ satisfy $\sum_{t=0}^\infty\eta_t=\infty~\textrm{ and }~\sum_{t=0}^\infty\eta_t^2<\infty$. Suppose that for the gradient aggregation rule $\ga\left(g_j^t|\,j\in S^t;n,f,r\right)$ in  Algorithm~\ref{alg}, there exists $\M<\infty$ such that
    \begin{equation}
        \norm{\ga\left(g_j^t|\,j\in S^t;n,f,r\right)}\leq\M
        \label{eqn:bounded-aggregator-norm}
    \end{equation} for all $t$. Algorithm~\ref{alg} satisfies the following:}
    
    \textit{For the point $x_\H\in\W$, if there exists $\D^*\in\left[0,\max_{x\in\W}\norm{x-x_\H}\right)$ 
    and $\xi>0$ such that for each iteration $t$,
    \begin{equation}
        \phi_t\triangleq\iprod{x^t-x_\H}{\ga\left(g_j^t|\,j\in S^t;n,f,r\right)}\geq\xi\textrm{ when }\norm{x^t-x_\H}\geq\D^*,
        \label{eqn:phi-def}
    \end{equation}
    then we have $\lim_{t\rightarrow\infty}\norm{x^t-x_\H}\leq\D^*$.}
\end{mdframed}

Consider the iterative process \eqref{eqn:update}. From now on, we use $\mathsf{GradAgg}[t]$ as a shorthand for the output of the gradient aggregation rule at iteration $t$, i.e. 
\begin{equation}
    \label{eqn:def-gradagg-t}
    \mathsf{GradAgg}[t]\triangleq\ga\left(g_j^t|\,j\in S^t;n,f,r\right).
\end{equation}

\hrule

\begin{proof}
Let $e_t$ denote $\norm{x^t-x_\H}$. Define a scalar function $\psi$,
\begin{equation}
    \psi(y)=\left\{\begin{array}{ll}
        0, & y<\left(\D^*\right)^2, \\
        \left(y-\left(\D^*\right)^2\right)^2, & \textrm{otherwise}.
    \end{array}\right.
    \label{eqn:psi_def}
\end{equation}
Let $\psi'(y)$ denote the derivative of $\psi$ at $y$. Then (cf. \cite{bottou1998online})
\begin{equation}
    \label{eqn:psi_bnd}
    \psi(z)-\psi(y)\leq(z-y)\psi'(y)+(z-y)^2,~\forall y,z\in\mathbb{R}_{\geq0}.
\end{equation}
Note,
\begin{equation}
    \label{eqn:psi_prime}
    \psi'(y)=\max\left\{0,2\left(y-\left(\D^*\right)^2\right)\right\}.
\end{equation}

Now, define
\begin{equation}
    \label{eqn:ht_def}
    h_t\triangleq\psi(e_t^2).
\end{equation}
From \eqref{eqn:psi_bnd} and \eqref{eqn:ht_def},
\begin{equation}
    h_{t+1}-h_t=\psi\left(e_{t+1}^2\right)-\psi\left(e_t^2\right)\leq\left(e_{t+1}^2-e_t^2\right)\psi'\left(e_t^2\right)+\left(e_{t+1}^2-e_t^2\right)^2,~\forall t\in\mathbb{Z}_{\geq0}.
\end{equation}
From now on, we use $\psi_t'$ as the shorthand for $\psi'\left(e_t^2\right)$, i.e.,
\begin{equation}
    \label{eqn:def-psi-t}
    \psi_t'\triangleq\psi'\left(e_t^2\right).
\end{equation}
From above, for all $t\geq0$,
\begin{equation}
    \label{eqn:ht_1}
    h_{t+1}-h_t\leq\left(e_{t+1}^2-e_t^2\right)\psi'_t+\left(e_{t+1}^2-e_t^2\right)^2.
\end{equation}
Recall the iterative process \eqref{eqn:update}. Using the non-expansion property of Euclidean projection onto a closed convex set,
\begin{equation}
    \norm{x^{t+1}-x_\H}= \norm{\left[x^t-\eta_t\mathsf{GradAgg}[t]\right]_\W-x_\H} \leq\norm{x^t-\eta_t\mathsf{GradAgg}[t]-x_\H}.
\end{equation}
Taking square on both sides, and recalling that $e_t\triangleq\norm{x^t-x_\H}$,
\begin{equation*}
    e_{t+1}^2\leq e_t^2-2\eta_t\iprod{x_t-x_\H}{\mathsf{GradAgg}[t]}+\eta_t^2\norm{\mathsf{GradAgg}[t]}^2.
\end{equation*}
Recall from \eqref{eqn:phi-def} that $\iprod{x_t-x_\H}{\mathsf{GradAgg}[t]}=\phi_t$, therefore,
\begin{equation}
    \label{eqn:proj_bound}
    e_{t+1}^2\leq e_t^2-2\eta_t\phi_t+\eta_t^2\norm{\mathsf{GradAgg}[t]}^2,~\forall t\geq0.
\end{equation}

As $\psi'_t\geq0,~\forall t\in\mathbb{Z}_{\geq0}$, combining \eqref{eqn:ht_1} and \eqref{eqn:proj_bound},
\begin{equation}
    \label{eqn:ht_2}
    h_{t+1}-h_t\leq\left(-2\eta_t\phi_t+\eta_t^2\norm{\mathsf{GradAgg}[t]}^2\right)\psi'_t+\left(e_{t+1}^2-e_t^2\right)^2,~\forall t\geq0.
\end{equation}
Note that 
\begin{equation}
    \mnorm{e_{t+1}^2-e_t^2}=(e_{t+1}+e_t)\mnorm{e_{t+1}-e_t}.
\end{equation}
As $\W$ is assumed compact, there exists
\begin{equation}
    \Gamma=\max_{x\in\W}\norm{x-x_\H}\leq\infty.
    \label{eqn:def-gamma}
\end{equation}
Let $\Gamma>0$, since otherwise $\W=\{x_\H\}$ only contains one point, and the problem becomes trivial. As $x^t\in\W$, $\forall t\geq0$,
\begin{equation}
    \label{eqn:e_t_bound}
    e_t\leq\Gamma,
\end{equation}
which implies
\begin{equation}
    e_{t+1}+e_t\leq2\Gamma.
\end{equation}
Therefore,
\begin{equation}
    \mnorm{e_{t+1}^2-e_t^2}\leq2\Gamma\mnorm{e_{t+1}-e_t},~\forall t\geq0.
    \label{eqn:e2t-bound-1}
\end{equation}
By triangle inequality,
\begin{equation}
    \mnorm{e_{t+1}-e_t}=\mnorm{\norm{x^{t+1}-x_\H}-\norm{x^t-x_\H}}\leq\norm{x^{t+1}-x^t}.
    \label{eqn:e2t-bound-2}
\end{equation}
From \eqref{eqn:update} and the non-expansion property of Euclidean projection onto a closed convex set,
\begin{equation}
    \norm{x^{t+1}-x^t}=\norm{\left[x^t-\eta_t\mathsf{GradAgg}[t]\right]_\W-x^t}\leq\eta_t\norm{\mathsf{GradAgg}[t]}.
    \label{eqn:e2t-bound-3}
\end{equation}
So from \eqref{eqn:e2t-bound-1}, \eqref{eqn:e2t-bound-2}, and \eqref{eqn:e2t-bound-3},
\begin{align}
    &\mnorm{e_{t+1}^2-e_t^2}\leq2\eta_t\Gamma\norm{\mathsf{GradAgg}[t]}, \nonumber \\
    \Longrightarrow&\left(e_{t+1}^2-e_t^2\right)^2\leq4\eta_t^2\Gamma^2\norm{\mathsf{GradAgg}[t]}^2.
\end{align}
Substituting above in \eqref{eqn:ht_2},
\begin{align}
    \label{eqn:ht_3}
    h_{t+1}-h_t\leq&\left(-2\eta_t\phi_t+\eta_t^2\norm{\mathsf{GradAgg}[t]}^2\right)\psi'_t+4\eta_t^2\Gamma^2\norm{\mathsf{GradAgg}[t]}^2, \nonumber \\
    =&-2\eta_t\phi_t\psi'_t+(\psi'_t+4\Gamma^2)\eta_t^2\norm{\mathsf{GradAgg}[t]}^2,~\forall t\geq0.
\end{align}

Recall that the statement of Theorem~\ref{thm:async-fault-toler} assumes that $\D^*\in[0,\max_{x\in\W}\norm{x-x_\H})$, which indicates $\D^*<\Gamma$. When $e_t\geq \D^*$, using \eqref{eqn:psi_prime} and \eqref{eqn:e_t_bound}, we have
\begin{equation}
    0\leq\psi'_t\leq2\left(e_t^2-\left(\D^*\right)^2\right)\leq2\left(\Gamma^2-\left(\D^*\right)^2\right)\leq2\Gamma^2
    \label{eqn:psi_prime_t_bnd}
\end{equation}
Also, when $e_t\leq\D^*$, we have $\psi_t'=0\leq2\Gamma^2$.
Recall from \eqref{eqn:bounded-aggregator-norm} that  $\norm{\mathsf{GradAgg}[t]}\leq\M<\infty$ for all $t$. Substituting \eqref{eqn:psi_prime_t_bnd} in \eqref{eqn:ht_3},
\begin{align}
    \label{eqn:ht_4}
    h_{t+1}-h_t\leq&-2\eta_t\phi_t\psi'_t+(2\Gamma^2+4\Gamma^2)\eta_t^2\M^2=-2\eta_t\phi_t\psi'_t+6\Gamma^2\eta_t^2\M^2,~\forall t\geq0.
\end{align}

Now we use Lemma~\ref{lemma:converge} to show that $h_\infty=0$ as follows. For each iteration $t$, consider the following two cases:
\begin{description}
    \item[Case 1)] Suppose $e_t<\D^*$. In this case, $\psi'_t=0$. By Cauchy-Schwartz inequality,
        \begin{equation}
            \mnorm{\phi_t}=\mnorm{\iprod{x^t-x_\H}{\mathsf{GradAgg}[t]}}\leq e_t\norm{\mathsf{GradAgg}[t]}.
        \end{equation}
        By \eqref{eqn:bounded-aggregator-norm} and \eqref{eqn:e_t_bound}, this implies that
        \begin{equation}
            \mnorm{\phi_t}\leq\Gamma\M<\infty.
        \end{equation}
        Thus,
        \begin{equation}
            \label{eqn:phitpsit_1}
            \phi_t\psi'_t=0.
        \end{equation}
    \item[Case 2)] Suppose $e_t\geq\D^*$. Therefore, there exists $\delta\geq0$, $e_t=\D^*+\delta$. From \eqref{eqn:psi_prime}, we obtain that
    \begin{equation}
        \psi_t'=2\left(\left(\D^*+\delta\right)^2-\left(\D^*\right)^2\right)=2\delta\left(2\D^*+\delta\right).
    \end{equation}
    The statement of Theorem~\ref{thm:async-fault-toler} assumes that $\phi_t\geq\xi>0$ when $e_t\geq\D^*$, thus, 
        \begin{equation}
            \label{eqn:phitpsit_2}
            \phi_t\psi'_t\geq2\delta\xi\left(2\D^*+\delta\right)\geq0.
        \end{equation}
\end{description}
From \eqref{eqn:phitpsit_1} and \eqref{eqn:phitpsit_2}, for both cases,
\begin{equation}
    \phi_t\psi'_t\geq0, ~\forall t\geq0.
    \label{eqn:phi_psi_bound}
\end{equation}
Combining this with \eqref{eqn:ht_4},
\begin{equation}
    h_{t+1}-h_t\leq6\Gamma^2\eta_t^2\M^2.
\end{equation}
From above we have
\begin{equation}
    \left(h_{t+1}-h_t\right)_+\leq6\Gamma^2\eta_t^2\M^2.
\end{equation}
Since $\sum_{t=0}^\infty\eta_t^2<\infty$, $\Gamma,\M<\infty$,
\begin{equation}
    \sum_{t=0}^\infty\left(h_{t+1}-h_t\right)_+\leq6\Gamma^2\M^2\sum_{t=0}^\infty\eta_t^2<\infty.
\end{equation}
By the definition of $h_t$, we have $h_t\geq0,~\forall t$. Then Lemma~\ref{lemma:converge} implies that  
\begin{align}
    h_t\xrightarrow[t\rightarrow\infty]{}h_\infty<\infty,~\textrm{and} \label{eqn:upper_bound_h_infty}\\
    \sum_{t=0}^\infty\left(h_{t+1}-h_t\right)_->-\infty.
\end{align}
Note that $h_\infty-h_0=\sum_{t=0}^\infty(h_{t+1}-h_t)$. Thus, from \eqref{eqn:ht_4} we have 
\begin{equation}
    h_\infty-h_0\leq-2\sum_{t=0}^\infty\eta_t\phi_t\psi_t'+6\Gamma^2\M^2\sum_{t=0}^\infty\eta_t^2.
\end{equation}
Therefore, from above we obtain
\begin{equation}
    2\sum_{t=0}^\infty\eta_t\phi_t\psi_t'\leq h_0-h_\infty+6\Gamma^2\M^2\sum_{t=0}^\infty\eta_t^2.
    \label{eqn:bound_2_sum}
\end{equation}
By assumption, $\sum_{t=0}^\infty\eta_t^2<\infty$. Substituting from \eqref{eqn:e_t_bound} that $e_t<\infty$ in \eqref{eqn:ht_def}, we obtain that 
\begin{equation}
    h_0=\psi\left(e_0^2\right)<\infty.
\end{equation} 
Therefore, \eqref{eqn:bound_2_sum} implies that
\begin{equation}
    2\sum_{t=0}^\infty\eta_t\phi_t\psi_t'\leq h_0+6\Gamma^2\M^2\sum_{t=0}^\infty\eta_t^2<\infty.
\end{equation}
Or simply,
\begin{equation}
    \sum_{t=0}^\infty\eta_t\phi_t\psi_t'<\infty.
    \label{eqn:upper_bound_etatphitpsit}
\end{equation}

Finally, we reason below by contradiction that $h_\infty=0$. Note that for any $\zeta>0$, there exists a unique positive value $\beta$ such that $\zeta=2\beta\left(2\D^*+\sqrt{\beta}\right)^2$. Suppose that $h_\infty=2\beta\left(2\D^*+\sqrt{\beta}\right)^2$ for some positive value $\beta$. As the sequence $\{h_t\}_{t=0}^\infty$ converges to $h_\infty$ (see \eqref{eqn:upper_bound_h_infty}), there exists some finite $\tau\in\Z_{\geq0}$ such that for all $t\geq\tau$, 
\begin{align}
    &\mnorm{h_t-h_\infty}\leq\beta\left(2\D^*+\sqrt{\beta}\right)^2 \\
    \Longrightarrow & h_t\geq h_\infty-\beta\left(2\D^*+\sqrt{\beta}\right)^2.
\end{align}
As $h_\infty=2\beta\left(2\D^*+\sqrt{\beta}\right)^2$, the above implies that
\begin{equation}
    h_t\geq \beta\left(2\D^*+\sqrt{\beta}\right)^2, \forall t\geq\tau.
    \label{eqn:ht_lower_bound}
\end{equation}
Therefore (cf. \eqref{eqn:psi_def} and \eqref{eqn:ht_def}), for all $t\geq\tau$,
\begin{eqnarray*}
    \left(e_t^2-\left(\D^*\right)^2\right)^2\geq\beta\left(2\D^*+\sqrt{\beta}\right)^2, \textrm{ or} \\
    \mnorm{e_t^2-\left(\D^*\right)^2}\geq\sqrt{\beta}\left(2\D^*+\sqrt{\beta}\right).
\end{eqnarray*}
Thus, for each $t\geq\tau$, either
\begin{equation}
    e^2_t\geq\left(\D^*\right)^2+\sqrt{\beta}\left(2\D^*+\sqrt{\beta}\right)=\left(\D^*+\sqrt{\beta}\right)^2,
    \label{eqn:et_case_1}
\end{equation}
or
\begin{equation}
    e^2_t\leq\left(\D^*\right)^2-\sqrt{\beta}\left(2\D^*+\sqrt{\beta}\right)<\left(\D^*\right)^2.
    \label{eqn:et_case_2}
\end{equation}
If the latter, i.e., \eqref{eqn:et_case_2} holds true for some $t'\geq\tau$, 
\begin{equation}
    h_{t'}=\psi\left(e_{t'}^2\right)=0,
\end{equation}
which contradicts \eqref{eqn:ht_lower_bound}. Therefore, \eqref{eqn:ht_lower_bound} implies \eqref{eqn:et_case_1}.

From above we obtain that if $h_\infty=2\beta\left(2\D^*+\sqrt{\beta}\right)^2$, there exists $\tau<\infty$ such that for all $t\geq\tau$, 
\begin{equation}
    e_t\geq\D^*+\sqrt{\beta}.
\end{equation}
Thus, from \eqref{eqn:phitpsit_2}, with $\delta=\sqrt{\beta}$, we obtain that 
\begin{equation}
    \phi_t\psi_t'\geq2\xi\sqrt{\beta}\left(2\D^*+\sqrt{\beta}\right), \forall t\geq\tau.
\end{equation}
Therefore,
\begin{align}
    \sum_{t=\tau}^\infty\eta_t\phi_t\psi_t' &\geq2\xi\sqrt{\beta}\left(2\D^*+\sqrt{\beta}\right)\sum_{t=\tau}^\infty\eta_t.
\end{align}
Note that $\sum_{t=0}^{\tau-1}\eta_t$ is finite, and $\sum_{t=\tau}^\infty\eta_t = \sum_{t=0}^\infty\eta_t - \sum_{t=0}^{\tau-1}\eta_t$. Combining the assumption that $\sum_{t=0}^\infty\eta_t=\infty$ and above, we have
\begin{align}
    \sum_{t=\tau}^\infty\eta_t\phi_t\psi_t' &\geq2\xi\sqrt{\beta}\left(2\D^*+\sqrt{\beta}\right)\sum_{t=\tau}^\infty\eta_t=\infty.
\end{align}
This is a contradiction to \eqref{eqn:upper_bound_etatphitpsit}. Therefore, $h_\infty=0$, and by \eqref{eqn:ht_def}, the definition of $h_t$, 
\begin{equation}
    h_\infty=\lim_{t\rightarrow\infty}\psi\left(e_t^2\right)=0.
\end{equation}
Hence, by \eqref{eqn:psi_def}, the definition of $\psi(\cdot)$, 
\begin{equation}
    \lim_{t\rightarrow\infty}\norm{x^t-x_\H}\leq\D^*.
\end{equation}
\end{proof}

\hrule

\subsection{Derivation of \eqref{eqn:thm2-params} when using CGE}
\label{appdx:a-2}

Recall the CGE gradient filter defined in \eqref{eqn:cge-footnote}: in each iteration $t$, sort $m$ vectors $g_j^t$'s as
   $ \norm{g_{i_1}^t}\leq\norm{g_{i_2}^t}\leq...\leq\norm{g_{i_{m}}^t},$
then 
\begin{align}
     \gf(g_j^t|j\in S^t;m,f)=\sum_{l=1}^{m-f}g_{i_l}^t.
    \label{eqn:cge-filter-appdx}
\end{align}
Also recall the definition of $\phi_t$ in \eqref{eqn:phi-def}: 
\begin{align*}
    \phi_t\triangleq\iprod{x^t-x_\H}{\ga\left(g_j^t|\,j\in S^t;n,f,r\right)}.
\end{align*}
Recall that we defined in Section~\ref{sec:full-grad} that for any non-faulty agents $\H$ with $\mnorm{\H}=n-f$ in an execution, $x_\H\in\W$ is a unique minimum point of the aggregate cost function of agents in $\H$.

\begin{mdframed}
    \textsc{Theorem~\ref{thm:async-fault-toler}-CGE.}
    \textit{Let $\H$ be any set of $n-f$ non-faulty agents in the system. Let $x_\H = \arg\min_{x\in\R^d}\sum_{j\in\H}Q_j(x)$; $x_\H\in\W$ is unique. Suppose that Assumptions~\ref{assum:lipschitz} and \ref{assum:strongly-convex-ft} hold true, and the agents' cost functions satisfy $(f,r;\epsilon)$-redundancy. Assume that $\eta_t$ satisfy $\sum_{t=0}^\infty\eta_t=\infty$ and $\sum_{t=0}^\infty\eta_t^2<\infty$. Suppose the GAR in Algorithm~\ref{alg} is CGE, as defined in \eqref{eqn:cge-filter-appdx}. Assume that
    \begin{equation}
        \label{eqn:alpha-cge}
        \alpha \triangleq 1 - \frac{f-r}{n-r} - \frac{2\mu}{\gamma}\cdot\frac{f+r}{n-r}>0.
    \end{equation}
    It can be shown that for $\phi_t$ defined in \eqref{eqn:phi-def}, 
    \begin{align*}
        \phi_t \geq \alpha (n-r) \gamma \D^* (\D^* - \D) > 0 \textrm{ when } \norm{x^t-x_{\H}}\geq \D^*, \\
        \textrm{ for all }\D^*>\D \triangleq \frac{4\mu (f+r)\epsilon}{\alpha\gamma}.
    \end{align*}
    Therefore, Algorithm~\ref{alg} converges with
    $$\lim_{t\rightarrow\infty}\norm{x^t-x_\H}\leq\D^*.$$}
\end{mdframed}

\begin{proof}
Throughout, we assume $f > 0$ to ignore the trivial case of $f = 0$. In this proof, we use a shorthand $m=n-r$. 

\textbf{First}, we show that for CGE, $\norm{\gf\left(g_j^t|j\in S^t;n-r,f\right)}=\norm{\sum_{j=1}^{m-f}g_{i_j}^t} < \infty, ~ \forall t$. Consider a subset $S_1 \subset \H$ with $\mnorm{S_1} = m-2f$. From triangle inequality,
\begin{align*}
    \norm{\sum_{j\in S_1}\nabla Q_j(x)-\sum_{j\in S_1}\nabla Q_j(x_{\H})} \leq \sum_{j \in S_1}\norm{\nabla Q_j(x) - \nabla Q_j(x_{\H})}, \quad \forall x \in \R^d.
\end{align*}
Under Assumption~\ref{assum:lipschitz}, i.e., Lipschitz continuity of non-faulty gradients, for each non-faulty agent $j$, $\norm{\nabla Q_j(x) - \nabla Q_j(x_{\H})} \leq \mu \norm{x - x_{\H}}$. Substituting this above implies that 
\begin{equation}
    \norm{\sum_{j\in S_1}\nabla Q_j(x)-\sum_{j\in S_1}\nabla Q_j(x_{\H})} \leq \mnorm{S_1} \mu \, \norm{x-x_{\H}}.
    \label{eqn:lipschitz-distance}
\end{equation}
As $\mnorm{S_1} = m-2f$, the $(f,r;\epsilon)$-redundancy property defined in Definition~\ref{def:redundancy} implies that for all $x_1\in\arg\min_x\sum_{j\in S_1}Q_j(x)$,
\[\norm{x_1-x_{\H}} \leq \epsilon.\]
Substituting from above in~\eqref{eqn:lipschitz-distance} implies that, for all $x_1\in\arg\min_x\sum_{j\in S_1}Q_j(x)$,
\begin{equation}
    \norm{\sum_{j\in S_1}\nabla Q_j(x_1)-\sum_{j\in S_1}\nabla Q_j(x_{\H})} \leq \mnorm{S_1} \mu \, \norm{x_1-x_{\H}} \leq \mnorm{S_1} \mu \epsilon.
    \label{eqn:lipschitz-distance-2}
\end{equation}
For all $x_1 = \arg\min_x\sum_{j\in S_1}Q_j(x)$, $\nabla \sum_{j\in\S_1}Q_j(x_1) = 0$. Thus,~\eqref{eqn:lipschitz-distance-2} implies that
\begin{equation}
    \norm{\sum_{j\in S_1}\nabla Q_j(x_{\H})} \leq \mnorm{S_1} \mu \epsilon.
    \label{eqn:lipschitz-distance-S1}
\end{equation}
Now, consider an arbitrary non-faulty agent $i \in \H\setminus S_1$. Let $S_2=S_1\cup\{i\}$. 
Using similar arguments as above we obtain that under the $(f,r;\epsilon)$-redundancy property and Assumption~\ref{assum:lipschitz}, for all $x_2\in\arg\min_x\sum_{j\in S_2}Q_j(x)$,
\begin{align}
    \norm{\sum_{j\in S_2}\nabla Q_j(x_{\H})}=\norm{\sum_{j\in S_2}\nabla Q_j(x_2)-\sum_{j\in S_2}\nabla Q_j(x_{\H})}\leq\mnorm{S_2}\mu\epsilon.
\end{align}
Note that $\sum_{j\in S_2}\nabla Q_j(x)=\sum_{j\in S_1}\nabla Q_j(x)+\nabla Q_i(x)$. From triangle inequality,
\begin{equation}
    \norm{\nabla Q_i(x_{\H})}-\norm{\sum_{j\in S_1}\nabla Q_j(x_{\H})}\leq\norm{\sum_{j\in S_1}\nabla Q_j(x_{\H})+\nabla Q_i(x_{\H})}.
\end{equation}
Therefore, for each non-faulty agent $i\in\H$, 
\begin{align}
    \norm{\nabla Q_i(x_{\H})}&\leq\norm{\sum_{j\in S_1}\nabla Q_j(x_{\H})+\nabla Q_i(x_{\H})}+\norm{\sum_{j\in S_1}\nabla Q_j(x_{\H})} \leq\mnorm{S_2}\mu\epsilon+\mnorm{S_1}\mu\epsilon \nonumber\\
    & = (m-2f+1)\mu \epsilon + (m-2f) \mu \epsilon = (2m-4f+1)\mu\epsilon.
    \label{eqn:honest-norm-bound}
\end{align}
Now, for all $x$ and $i\in\H$, by Assumption \ref{assum:lipschitz},
\begin{equation*}
    \norm{\nabla Q_i(x)-\nabla Q_i(x_{\H})}\leq\mu\norm{x-x_{\H}}.
\end{equation*}
By triangle inequality,
\begin{equation*}
    \norm{\nabla Q_i(x)}\leq\norm{\nabla Q_i(x_{\H})}+\mu\norm{x-x_{\H}}.
\end{equation*}
Substituting from~\eqref{eqn:honest-norm-bound} above we obtain that
\begin{equation}
    \norm{\nabla Q_i(x)}\leq(2m-4f+1)\mu\epsilon+\mu\norm{x-x_{\H}}\leq2m\mu\epsilon+\mu\norm{x-x_{\H}}.
    \label{eqn:honest-bound-everywhere}
\end{equation}
We use the above inequality~\eqref{eqn:honest-bound-everywhere} to show below that $\norm{\sum_{j=1}^{m-f}g_{i_j}^t}$ is bounded for all $t$. Recall that when defining CGE we have, for each iteration $t$, 
\begin{equation*}
    \norm{g_{i_1}^t}\leq...\leq\norm{g_{i_{m-f}}^t}\leq\norm{g_{i_{m-f+1}}^t}\leq...\leq\norm{g_{i_m}^t}.
\end{equation*}
As there are at most $f$ Byzantine agents, for each $t$ there exists $\sigma_t\in\H$ such that
\begin{equation}
    \norm{g_{i_{m-f}}^t}\leq\norm{g_{i_{\sigma_t}}^t}.
    \label{eqn:honest-bound}
\end{equation}
As $g_j^t=\nabla Q_j(x^t)$ for all $j\in\H$, from~\eqref{eqn:honest-bound} we obtain that
\begin{equation*}
    \norm{g_{i_j}^t}\leq\norm{\nabla Q_{\sigma_t}(x^t)}, \quad \forall j \in \{1, \ldots, m-f\}, ~ t.
\end{equation*}
Substituting from~\eqref{eqn:honest-bound-everywhere} above we obtain that for every $j \in \{1, \ldots, m-f\}$,
\begin{equation*}
    \norm{g_{i_j}^t}\leq\norm{g_{i_{m-f}}^t}\leq2m\mu\epsilon+\mu\norm{x^t-x_{\H}}.
\end{equation*}
Therefore, from triangle inequality,
\begin{equation}
    \norm{\sum_{j=1}^{m-f}g_{i_j}^t}\leq\sum_{j=1}^{m-f}\norm{g_{i_j}^t}\leq(m-f)\left(2m\mu\epsilon+\mu\norm{x^t-x_{\H}}\right). \label{eqn:filtered-upperbound}
\end{equation}
Recall that $x_{\H} \in \W$. Also recall that we defined in \eqref{eqn:def-gamma} that $\Gamma = \max_{x \in \W} \norm{x - x_{\H}}$. As $\W$ is a compact set, $\Gamma < \infty$. Recall from the update rule~\eqref{eqn:update} that $x^t \in \W$ for all $t$. Thus, $\norm{x^t - x_{\H}} \leq \max_{x \in \W} \norm{x - x_{\H}} = \Gamma < \infty$. Substituting this in~\eqref{eqn:filtered-upperbound} implies that
\begin{equation}
    \norm{\sum_{j=1}^{m-f}g_{i_j}^t} \leq (m-f) \left( 2m \mu \epsilon + \mu \Gamma\right) < \infty. 
\end{equation}
Recall that in this particular case, $\sum_{j=1}^{m-f}g_{i_j}^t = \norm{\gf\left(g_j^t|j\in S^t;n-r,f\right)}$. Therefore, from above we obtain that
\begin{align}
    \norm{\gf\left(g_j^t|j\in S^t;n-r,f\right)} < \infty, \quad \forall t. \label{eqn:cge_bnd_grd}
\end{align}

\textbf{Second}, we show that for an arbitrary $\D^* > \D = \frac{4\mu (f+r)\epsilon}{\alpha\gamma}$,
\[\phi_t\triangleq\iprod{x^t-x_{\H}}{\sum_{j=1}^{m-f}g_{i_j}^t} \geq \alpha (n-r) \gamma \D^* (\D^* - \D) ~ \text{ when } ~ \norm{x^t-x_{\H}} \geq \D^*,\]
where $\alpha$ is defined by \eqref{eqn:alpha-cge}.

Consider an arbitrary iteration $t$. Note that, as $\mnorm{\H} = n-f$, there are at least $m-2f$ agents that are common to both sets $\H$ and $\{i_1,...,i_{m-f}\}$. We let $\H^t = \{i_1,...,i_{m-f}\} \cap \H$. The remaining set of agents $\B^t = \{i_1,...,i_{m-f}\} \setminus \H^t$ comprises of only faulty agents. Note that $\mnorm{\H^t} \geq m-2f $ and $\mnorm{\B^t} \leq f$. Therefore,
\begin{equation}
    \phi_t=\iprod{x^t-x_{\H}}{\sum_{j\in\H^t}g_j^t}+\iprod{x^t-x_{\H}}{\sum_{k\in\B^t}g_k^t}.
    \label{eqn:phi-t-two-parts}
\end{equation}
Consider the first term on the right-hand side of~\eqref{eqn:phi-t-two-parts}. Note that
\begin{align*}
    \iprod{x^t-x_{\H}}{\sum_{j\in\H^t}g_j^t}&=\iprod{x^t-x_{\H}}{\sum_{j\in\H^t}g_j^t+\sum_{j\in\H\backslash\H^t}g_j^t-\sum_{j\in\H\backslash\H^t}g_j^t} \\ 
    &=\iprod{x^t-x_{\H}}{\sum_{j\in\H}g_j^t}-\iprod{x^t-x_{\H}}{\sum_{j\in\H\backslash\H^t}g_j^t}.
\end{align*}
Recall that $g_j^t=\nabla Q_j(x^t)$, $\forall j\in\H$. Therefore,
\begin{equation}
    \iprod{x^t-x_{\H}}{\sum_{j\in\H^t}g_j^t}=\iprod{x^t-x_{\H}}{\sum_{j\in\H}\nabla Q_j(x^t)}-\iprod{x^t-x_{\H}}{\sum_{j\in\H\backslash\H^t}\nabla Q_j(x^t)}. \label{eqn:first_phi_1}
\end{equation}

\noindent Due to the strong convexity assumption (i.e., Assumption~\ref{assum:strongly-convex-ft}), for all $x, \, y \in \R^d$,
\begin{equation*}
    \iprod{x- y}{\nabla \sum_{j\in\H}Q_j(x)-\nabla\sum_{j\in\H} Q_j(y)} \geq \mnorm{\H}\, \gamma\norm{x-y}^2.
\end{equation*}
As $x_{\H}$ is minimum point of $\sum_{j \in \H}Q_j(x)$, $\nabla \sum_{j \in \H} Q_j(x_{\H}) = 0$. Thus, 
\begin{align}
    \iprod{x^t-x_{\H}}{\sum_{j\in\H}\nabla Q_j(x^t)} & = \iprod{x^t-x_{\H}}{\nabla\sum_{j\in\H}Q_j(x^t)-\nabla\sum_{j\in\H}Q_j(x_{\H})} \nonumber \\
    & \geq \mnorm{\H} \, \gamma\norm{x^t-x_{\H}}^2.
    \label{eqn:inner-prod-h}
\end{align}
Now, due to the Cauchy-Schwartz inequality, 
\begin{align}
    \iprod{x^t-x_{\H}}{\sum_{j\in\H\backslash\H^t}\nabla Q_j(x^t)}&=\sum_{j\in\H\backslash\H^t}\iprod{x^t-x_{\H}}{\nabla Q_j(x^t)} \nonumber\\
    &\leq\sum_{j\in\H\backslash\H^t}\norm{x^t-x_{\H}}\, \norm{\nabla Q_j(x^t)}.
    \label{eqn:inner-prod-h-ht}
\end{align}
Substituting from~\eqref{eqn:inner-prod-h} and~\eqref{eqn:inner-prod-h-ht} in~\eqref{eqn:first_phi_1} we obtain that
\begin{equation}
    \iprod{x^t-x_{\H}}{\sum_{j\in\H^t}g_j^t} \geq \gamma\mnorm{\H}\, \norm{x^t-x_{\H}}^2-\sum_{j\in\H\backslash\H^t}\norm{x^t-x_{\H}}\, \norm{\nabla Q_j(x^t)}.
    \label{eqn:phi-t-first-part}
\end{equation}

Next, we consider the second term on the right-hand side of~\eqref{eqn:phi-t-two-parts}. From the Cauchy-Schwartz inequality, 
\begin{equation*}
    \iprod{x^t-x_{\H}}{g_k^t}\geq-\norm{x^t-x_{\H}}\, \norm{g_k^t}.
\end{equation*}
Substituting from~\eqref{eqn:phi-t-first-part} and above in~\eqref{eqn:phi-t-two-parts} we obtain that
\begin{equation}
    \phi_t\geq\gamma\mnorm{\H}\, \norm{x^t-x_{\H}}^2-\sum_{j\in\H\backslash\H^t}\norm{x^t-x_{\H}}\, \norm{\nabla Q_j(x^t)}-\sum_{k\in\B^t}\norm{x^t-x_{\H}}\, \norm{g_k^t}.
    \label{eqn:phi-t-2}
\end{equation}

Recall that, due to the sorting of the gradients, for an arbitrary $k \in \B^t$ and an arbitrary $j \in \H\backslash\H^t$,
\begin{equation}
    \norm{g_k^t}\leq\norm{g_j^t}=\norm{\nabla Q_j(x^t)}. \label{eqn:k_B_j_H}
\end{equation}
Recall that $\B^t = \{i_1, \ldots, \, i_{m-f}\} \setminus \H^t$. Thus, $\mnorm{\B^t} = m-f-\mnorm{\H^t}$. Also, as $\mnorm{\H} = n-f$, $\mnorm{\H\backslash\H^t} = n- f - \mnorm{\H^t}$. That is, $\mnorm{\B^t} \leq \mnorm{\H\backslash\H^t}$. Therefore,~\eqref{eqn:k_B_j_H} implies that
\begin{align*}
    \sum_{k \in \B^t} \norm{g_k^t} \leq \sum_{j \in \H\backslash\H^t} \norm{\nabla Q_j(x^t)}.
\end{align*}
Substituting from above in~\eqref{eqn:phi-t-2}, we obtain that
\begin{align*}
    \phi_t&\geq\gamma\mnorm{\H}\, \norm{x^t-x_{\H}}^2-2\sum_{j\in\H\backslash\H^t}\norm{x^t-x_{\H}}\, \norm{\nabla Q_j(x^t)}.
\end{align*}
Substituting from \eqref{eqn:honest-bound-everywhere}, i.e., $\norm{\nabla Q_i(x)}\leq2m\mu\epsilon+\mu\norm{x-x_{\H}}$, in above we obtain that
\begin{align*}
    \phi_t&\geq\gamma\mnorm{\H}\, \norm{x^t-x_{\H}}^2 - 2\mnorm{\H\backslash\H^t}\, \norm{x^t-x_{\H}}\, (2m\mu\epsilon+\mu\norm{x^t-x_{\H}}) \nonumber\\
    & \geq \left(\gamma\mnorm{\H}-2\mu\mnorm{\H\backslash\H^t}\right)\norm{x^t-x_{\H}}^2 - 4m\mu\epsilon \mnorm{\H\backslash\H^t}\, \norm{x^t-x_{\H}}.
\end{align*}
As $\mnorm{\H} = n-f$ and $\mnorm{\H\backslash\H^t}\leq f+r$, the above implies that
\begin{align}
    \begin{split}
    \label{eqn:phi-t-pre-final}
        \phi_t & \geq\left(\gamma(n-f)-2\mu (f+r)\right)\norm{x^t-x_{\H}}^2-4m\mu\epsilon (f+r) \norm{x^t-x_{\H}} \\
            & = \left(\gamma(n-f)-2\mu (f+r)\right)\norm{x^t-x_{\H}} \left(\norm{x^t-x_{\H}}-\dfrac{4m\mu\epsilon (f+r)}{\gamma(n-f)-2\mu (f+r)} \right) \\
            & = \left(\gamma(n-r+r-f)-2\mu (f+r)\right)\norm{x^t-x_{\H}} \left(\norm{x^t-x_{\H}}-\dfrac{4m\mu\epsilon (f+r)}{\gamma(n-f)-2\mu (f+r)} \right) \\
            & = m \gamma \left( 1 - \frac{f-r}{m} - \frac{2\mu}{\gamma}\cdot\frac{f+r}{m}\right) \norm{x^t-x_{\H}}  \left(\norm{x^t-x_{\H}} - \frac{4 \mu (f+r) \epsilon}{\gamma\left( 1 - \frac{f-r}{m} + \frac{2\mu}{\gamma}\cdot\frac{f+r}{m}\right)} \right).
    \end{split}
\end{align}
Recall that we defined 
\[\alpha = 1 - \frac{f-r}{m} - \frac{2\mu}{\gamma}\cdot\frac{f+r}{m}. \]
Substituting from above in~\eqref{eqn:phi-t-pre-final} we obtain that
\begin{align}
    \phi_t \geq \alpha m \gamma \norm{x^t-x_{\H}} \left(\norm{x^t-x_{\H}}- \frac{4\mu (f+r)\epsilon}{\alpha\gamma} \right).
    \label{eqn:phi-t-final}
\end{align}
As it is assumed that $\alpha > 0$,~\eqref{eqn:phi-t-final} implies that for an arbitrary $\D^* > \D$,
\begin{align}
    \phi_t \geq \alpha (n-r) \gamma \D^* (\D^* - \D) > 0 ~ \text{ when } \norm{x^t-x_{\H}}\geq\D^*.
    \label{eqn:final-cge}
\end{align}
Combining \eqref{eqn:final-cge} with Theorem~\ref{thm:async-fault-toler}, we have $\lim_{t\rightarrow\infty}\norm{x^t - x_{\H}} \leq \D^*$.
\end{proof}

\hrule

\subsection{Derivation of \eqref{eqn:async-parameters}}
\label{appdx:a-3}

In this case of $f=0$ and $r\geq0$, we use the averaging GAR defined in \eqref{eqn:aggregation-rule}
\begin{equation}
    \mathsf{GradAgg}\left(g_j^t|j\in S^t;n,f,r\right)=\sum_{j\in{S^t}}\nabla Q_j(x^t).
    \label{eqn:aggregation-rule-appdx}
\end{equation}
Also recall the definition of $\phi_t$ in \eqref{eqn:phi-def}: 
\begin{align*}
    \phi_t\triangleq\iprod{x^t-x_\H}{\ga\left(g_j^t|\,j\in S^t;n,f,r\right)}.
\end{align*}
Recall that we defined in Section~\ref{sec:full-grad} that for any non-faulty agents $\H$ with $\mnorm{\H}=n-f$ in an execution, $x_\H\in\W$ is a unique minimum point of the aggregate cost function of agents in $\H$.
Note that $\H=[n]$, so $x_\H=\arg\min_{x\in\W}\sum_{j\in[n]}Q_j(x)$ is the unique minimum of the aggregate cost functions of \textit{all} agents, and the vector $g_j^t=\nabla Q_j(x^t)$ is the gradient of $Q_j$ at $x^t$. 

\begin{mdframed}[everyline=true]
    \textsc{Theorem~\ref{thm:async-fault-toler}-Async.}
    \textit{Let $\H=[n]$ be all agents in the system. Let $x_\H = \arg\min_{x\in\R^d}\sum_{j\in\H}Q_j(x)$; $x_\H\in\W$ is unique. Suppose that $f=0$, $r\geq0$, Assumptions~\ref{assum:lipschitz} and \ref{assum:strongly-convex-ft} hold true, and the agents' cost functions satisfy $(0,r;\epsilon)$-redundancy. Assume that $\eta_t$ satisfy $\sum_{t=0}^\infty\eta_t=\infty$ and $\sum_{t=0}^\infty\eta_t^2<\infty$. Suppose the GAR in Algorithm~\ref{alg} is \eqref{eqn:aggregation-rule-appdx}. Assume that     \begin{align*}
        \alpha\triangleq 1-\frac{r}{n}\cdot\frac{\mu}{\gamma}>0.
    \end{align*}
    It can be shown that 
    \begin{align*}
        \phi_t \geq \alpha n \gamma \D^*(\D^*-\D) > 0 \textrm{ when } \norm{x^t-x_{\H}}\geq\D^*,\\
        \textrm{ for all }\D^*>\D\triangleq\frac{2r\mu}{\alpha\gamma}\epsilon.
    \end{align*}
    And therefore, Algorithm~\ref{alg} converges with 
    \begin{equation}
        \lim_{t\rightarrow\infty}\norm{x^t-x_\H}\leq\D^*, 
    \end{equation}}
\end{mdframed}

\begin{proof}
\textbf{First}, we need to show that $\norm{\sum_{j\in S^t}\nabla Q_j(x^t)}$ is bounded for all $t$. By Assumption~\ref{assum:lipschitz}, for all $j\in[n]$, 
\begin{equation}
    \label{eqn:thm2-lip}
    \norm{\nabla Q_j(x)-\nabla Q_j(x_\H)}\leq\mu\norm{x-x_\H}.
\end{equation}

Let $x_S\in\arg\min_x\sum_{j\in S}Q_j(x)$ be a minimum point of the aggregated cost functions of a set $S$ of $n-r$ agents, i.e., $\mnorm{S}=n-r$. Note that $\sum_{j\in S}\nabla Q_j(x_S)=0$.
By triangle inequality,
\begin{equation}
    \label{eqn:thm2-triangle}
    \norm{\sum_{j\in S}\nabla Q_j(x_S)-\sum_{j\in S}\nabla Q_j(x_\H)}\leq\sum_{j\in S}\norm{\nabla Q_j(x_S)-\nabla Q_j(x_\H)}.
\end{equation}
Combining~\eqref{eqn:thm2-lip} through \eqref{eqn:thm2-triangle}, 
\begin{equation}
    \norm{\sum_{j\in S}\nabla Q_j(x_\H)}\leq\mnorm{S}\mu\norm{x_S-x_\H}.
\end{equation}
By Definition~\ref{def:redundancy} of $(0,r;\epsilon)$-redundancy, $\norm{x_S-x_\H}\leq\epsilon$.
Therefore,
\begin{equation}
    \norm{\sum_{j\in S}\nabla Q_j(x_\H)}\leq\mu\epsilon (n-r).
\end{equation}

Now, consider an arbitrary agent $i\in[n]\backslash S$. Let $T=S\cup\{i\}$. Using a similar argument as above, we obtain 
\begin{equation}
    \norm{\sum_{j\in T}\nabla Q_j(x_\H)}\leq\mu\epsilon (n-r+1).
\end{equation}
Therefore, 
\begin{align}
    \norm{\nabla Q_i(x_\H)}=&\norm{\sum_{j\in T}\nabla Q_j(x_\H)-\sum_{j\in S}\nabla Q_j(x_\H)} \leq\norm{\sum_{j\in T}\nabla Q_j(x_\H)}+\norm{\sum_{j\in S}\nabla Q_j(x_\H)} \nonumber \\
        \leq&(n-r)\mu\epsilon+(n-r+1)\mu\epsilon=(2n-2r+1)\mu\epsilon. \label{eqn:apprx_gradient_bnd}
\end{align}
Note that the inequality \eqref{eqn:apprx_gradient_bnd} can be applied to any $i\in[n]$ with a suitable choice of $S$ above.\\

On the other hand, by Assumption~\ref{assum:lipschitz}, for any $x\in\mathbb{R}^d$,
\begin{equation}
    \norm{\nabla Q_i(x)-\nabla Q_i(x_\H)}\leq\mu\norm{x-x_\H}.
\end{equation}
By triangle inequality,
\begin{equation}
    \norm{\nabla Q_i(x)}\leq\norm{\nabla Q_i(x_\H)}+\mu\norm{x-x_\H}.
\end{equation}
Combining above and \eqref{eqn:apprx_gradient_bnd},
\begin{equation}
    \label{eqn:apprx_gradient_bnd_2}
    \norm{\nabla Q_i(x)}\leq(2n-2r+1)\mu\epsilon+\mu\norm{x-x_\H}\leq2n\mu\epsilon+\mu\norm{x-x_\H}.
\end{equation}
Recall that $x^t\in\W$ for all $t$, where $\W$ is a compact set. Also recall that we defined in \eqref{eqn:def-gamma} that there exists a $\Gamma=\max_{x\in\W}\norm{x-x_\H}<\infty$, such that $\norm{x^t-x_\H}\leq\Gamma$ for all $t$. 
Therefore, for all $t$, 
\begin{align}
    \label{eqn:apprx_phi_t_bnd}
    \norm{\sum_{j\in S^t}\nabla Q_j(x^t)}&\leq\sum_{j\in S^t}\norm{\nabla Q_j(x^t)} \leq\mnorm{S^t}\cdot\left(2n\mu\epsilon+\mu\norm{x^t-x_\H}\right) \nonumber \\
    &\leq(n-r)\left(2n\mu\epsilon+\mu\Gamma\right)<\infty.
\end{align}

\textbf{Second}, consider the following term $\displaystyle\phi_t=\iprod{x^t-x_\H}{\sum_{j\in S^t}\nabla Q_j(x^t)}$.
We have
\begin{align}
    \label{eqn:apprx_phi_t_1}
    \phi_t=&\iprod{x^t-x_\H}{\sum_{j\in S^t}\nabla Q_j(x^t)} \nonumber \\
        =&\iprod{x^t-x_\H}{\sum_{j\in S^t}\nabla Q_j(x^t)+\sum_{k\in [n]\backslash S^t}\nabla Q_k(x^t)-\sum_{k\in [n]\backslash S^t}\nabla Q_k(x^t)} \nonumber \\
        =&\iprod{x^t-x_\H}{\sum_{j\in [n]}\nabla Q_j(x^t)}-\iprod{x^t-x_\H}{\sum_{k\in [n]\backslash S^t}\nabla Q_k(x^t)}.
\end{align}

For the first term in \eqref{eqn:apprx_phi_t_1}, recall that $Q_S(x)=\frac{1}{\mnorm{S}}\sum_{j\in S}Q_j(x)$ with $\mnorm{S}=n$, i.e, $S=[n]$. Also note that by the definition of $x_\H$, $\sum_{j\in[n]}\nabla Q_j(x_\H)=0$. With Assumption~\ref{assum:strongly-convex-ft}, 
\begin{align}
    \label{eqn:apprx_phi_t_1_1}
    &\iprod{x^t-x_\H}{\sum_{j\in [n]}\nabla Q_j(x^t)}=\iprod{x^t-x_\H}{\sum_{j\in [n]}\nabla Q_j(x^t)-\sum_{j\in [n]}\nabla Q_j(x_\H)} \nonumber \\
        =&n\cdot\iprod{x^t-x_\H}{\nabla Q_{[n]}(x^t)-\nabla Q_{[n]}(x_\H)}\geq n\gamma\norm{x^t-x_\H}^2.
\end{align}

For the second term in \eqref{eqn:apprx_phi_t_1}, by Cauchy-Schwartz inequality,
\begin{align}
    \label{eqn:apprx_phi_t_1_2}
    \iprod{x^t-x_\H}{\sum_{k\in [n]\backslash S^t}\nabla Q_k(x^t)}&=\sum_{k\in [n]\backslash S^t}\iprod{x^t-x_\H}{\nabla Q_k(x^t)}\nonumber \\
        &\leq\sum_{k\in [n]\backslash S^t}\norm{x^t-x_\H}\cdot\norm{\nabla Q_k(x^t)}.
\end{align}

Combining \eqref{eqn:apprx_phi_t_1}, \eqref{eqn:apprx_phi_t_1_1}, and \eqref{eqn:apprx_phi_t_1_2},
\begin{equation}
    \label{eqn:apprx_phi_t_2}
    \phi_t\geq n\gamma\norm{x^t-x_\H}^2-\sum_{k\in[n]\backslash S^t}\norm{x^t-x_\H}\norm{\nabla Q_k(x^t)}.
\end{equation}
Substituting from \eqref{eqn:apprx_gradient_bnd_2} in above, note that $\mnorm{[n]\backslash S^t}=r$,
\begin{align}
    \label{eqn:apprx_phi_t_3}
    \phi_t&\geq n\gamma\norm{x^t-x_\H}^2-\sum_{k\in[n]\backslash S^t}\norm{x^t-x_\H}(2n\mu\epsilon+\mu\norm{x^t-x_\H}) \nonumber \\
        &=n\gamma\norm{x^t-x_\H}^2-r\norm{x^t-x_\H}(2n\mu\epsilon+\mu\norm{x^t-x_\H}) \nonumber \\
        &=n\gamma\left(1-\dfrac{r}{n}\cdot\dfrac{\mu}{\gamma}\right)\norm{x^t-x_\H}\left(\norm{x^t-x_\H}-\dfrac{2r\mu\epsilon}{\gamma\left(1-\dfrac{r}{n}\cdot\dfrac{\mu}{\gamma}\right)}\right).
\end{align}
Recall that we defined $\displaystyle\alpha\triangleq1-\dfrac{r}{n}\cdot\dfrac{\mu}{\gamma}>0$.
We have
\begin{equation}
    \phi_t\geq\alpha n\gamma\norm{x^t-x_\H}\left(\norm{x^t-x_\H}-\dfrac{2r\mu}{\alpha\gamma}\epsilon\right).
    \label{eqn:phi_t_final}
\end{equation}
\eqref{eqn:phi_t_final} implies for any $\D^*>\D\triangleq\dfrac{2r\mu}{\alpha\gamma}\epsilon$,
\begin{equation}
    \phi_t\geq\alpha n\gamma\D^*(\D^*-\D)>0~\textrm{ when }~\norm{x^t-x_\H}\geq\D^*. 
\end{equation}
Combining above with Theorem~\ref{thm:async-fault-toler}, we have  
\begin{equation}
    \lim_{t\rightarrow\infty}\norm{x^t-x_\H}\leq\D^*.
\end{equation}
\end{proof}

\hrule 

\subsection{Derivation of \eqref{eqn:async-parameters-stale} with stale gradients}
\label{appdx:a-4}

In Section~\ref{sub:fg-special}, we also discussed an asynchronous special case utilizing stale gradients. Substituting the parameters \eqref{eqn:async-parameters} in Theorem~\ref{thm:async-fault-toler}, we write the result in its full form as follows, and provide its proof. Note that in this case $f=0$ and $r\geq0$, we use the GAR \eqref{eqn:update-straggler}:
\begin{equation}
    \ga \left(g_j^t|j\in S^t;n,f,r\right)=\sum_{i=0}^\tau\sum_{j\in T^{t;t-i}}\nabla Q_j(x^{t-i}).
    \label{eqn:update-straggler-appdx}
\end{equation}
The definition of $T^{t;t-i}$ and its relationship with $S^t$ can be found in Section~\ref{sub:fg-special}. 

Also recall the definition of $\phi_t$ in \eqref{eqn:phi-def}: 
\begin{align*}
    \phi_t\triangleq\iprod{x^t-x_\H}{\ga\left(g_j^t|\,j\in S^t;n,f,r\right)}.
\end{align*}
Recall that we defined in Section~\ref{sec:full-grad} that for any non-faulty agents $\H$ with $\mnorm{\H}=n-f$ in an execution, $x_\H\in\W$ is a unique minimum point of the aggregate cost function of agents in $\H$. Note that $\H=[n]$, so $x_\H=\arg\min_{x\in\W}\sum_{j\in[n]}Q_j(x)$ is the unique minimum of the aggregate cost functions of all agents, and the vector $g_j^t=\nabla Q_j(x^t)$ is the gradient of $Q_j$ at $x^t$. 

\begin{mdframed}[everyline=true]
    \textsc{Theorem~\ref{thm:async-fault-toler}-Async-Stale.}
    \textit{Let $\H=[n]$ be all agents in the system. Let $x_\H = \arg\min_{x\in\R^d}\sum_{j\in\H}Q_j(x)$; $x_\H\in\W$ is unique. Suppose that $f=0$, $r\geq0$, Assumption~\ref{assum:lipschitz} and \ref{assum:strongly-convex-ft} hold true, and the cost functions of all agents satisfies $(0,r;\epsilon)$-redundancy. 
     Assume that $\eta_t$ satisfy $\sum_{t=0}^\infty\eta_t=\infty$, $\sum_{t=0}^\infty\eta_t^2<\infty$, and $\eta_t\geq\eta_{t+1}$ for all $t$.
    Suppose the GAR in Algorithm~\ref{alg} is \eqref{eqn:update-straggler-appdx}, and there exists a $\tau\geq0$ such that $\mnorm{T^t}\geq n-r$ for all $t$. Assume that 
    \begin{align*}
       \alpha\triangleq1-\frac{r}{n}\cdot\frac{\mu}{\gamma}>0. 
    \end{align*}
    It can be shown that
    \begin{align*}
        \phi_t \geq \alpha n \gamma \delta \D^* > 0 \textrm{ when } \norm{x^t-x_{\H}}\geq\D^*,\\
        \textrm{ for all }\D^*>\D\triangleq\dfrac{\mu\left(2r + \tau\eta_0G\right)}{\alpha\gamma}\epsilon,
    \end{align*}
    where $G=n\mu(2n\epsilon+\Gamma)$, $\Gamma = \max_{x\in\W}\norm{x-x_\H}$, and $\tau$ is the parameter in \eqref{eqn:update-straggler-appdx}.
    Then Algorithm~\ref{alg} with the GAR \eqref{eqn:update-straggler-appdx} converges with
    \begin{equation}
        \lim_{t\rightarrow\infty}\norm{x^t-x_\H}\leq\D^*. 
    \end{equation}}
\end{mdframed}

For the convenience of our argument, we let $T^{t;t-k}\triangleq\varnothing$ for all $k>t$, meaning the set of agents from which the server receives gradients in a negative iteration number is empty, by a slight abuse of notation. Similarly, we also define $\eta_t=\eta_0$ for $-\tau\leq t\leq0$. Note that we  have  $\eta_t\geq\eta_{t+1}$ and $\eta_0\geq\eta_t$, for $t\geq -\tau$. 

\begin{proof}
\textbf{First}, we show that $\norm{\sum_{i=0}^\tau\sum_{j\in T^{t;t-i}}\nabla Q_j(x^{t-i})}$ is bounded for all $t$. By Assumption~\ref{assum:lipschitz} and $(0,r;\epsilon)$-redundancy, following the same argument in the proof of Theorem~\ref{thm:async-fault-toler}-Async in~\ref{appdx:a-3}, we obtain that for all $j\in[n]$, (cf. \eqref{eqn:apprx_gradient_bnd})
\begin{equation}
    \norm{\nabla Q_j(x_\H)}\leq(2n-2r+1)\mu\epsilon.
\end{equation}
Furthermore, by Assumption~\ref{assum:lipschitz}, for all $x\in\R^d$, (cf. \eqref{eqn:apprx_gradient_bnd_2})
\begin{equation}
    \norm{\nabla Q_j(x)}\leq 2n\mu\epsilon+\mu\norm{x-x_\H}.
    \label{eqn:apprx_gradient_bnd_2-stale-gd}
\end{equation}
Recall that $x^t\in\W$ for all $t$, where $\W$ is a compact set. There exists a $\Gamma=\max_{x\in\W}\norm{x-x_\H}<\infty$, such that $\norm{x^t-x_\H}\leq\Gamma$ for all $t$. 

Recall that we defined $T^{t;t-k}\triangleq\varnothing$ for all $k>t$. Therefore, for all $t$,
\begin{align}
    &\norm{\sum_{i=0}^\tau\sum_{j\in T^{t;t-i}}\nabla Q_j(x^{t-i})}\leq \sum_{i=0}^\tau\sum_{j\in T^{t;t-i}}\norm{\nabla Q_j(x^{t-i})} \nonumber \\
    \leq& \sum_{i=0}^\tau\mnorm{T^{t;t-i}}\cdot(2n\mu\epsilon+\mu\norm{x^{t-i}-x_\H})\leq \sum_{i=0}^\tau\mnorm{T^{t;t-i}}\cdot(2n\mu\epsilon+\mu\Gamma)
\end{align}
Recall that by the definition of $T^{t;t-i}$, we have $T^{t;t-i_1}\cap T^{t;t-i_2}=\varnothing$ for all $0\leq i_1, i_2\leq\tau$, $i_1\neq i_2$. Therefore, $\sum_{i=0}^\tau\mnorm{T^{t;t-i}}\leq n$. Thus,
\begin{align}
    \label{eqn:update-bnd-generalized}
    &\norm{\sum_{i=0}^\tau\sum_{j\in T^{t;t-i}}\nabla Q_j(x^{t-i})}\leq n(2n\mu\epsilon+\mu\Gamma)<\infty.
\end{align}
Let us define
\begin{align}
    G = n\mu(2n\epsilon+\Gamma).
    \label{eqn:def-G}
\end{align}
We have $\norm{\sum_{i=0}^\tau\sum_{j\in T^{t;t-i}}\nabla Q_j(x^{t-i})}\leq G<\infty$.

\textbf{Second}, consider the term 
\begin{equation}
    \label{eqn:phi_t-gen}
    \phi_t=\iprod{x^t-x_\H}{\sum_{i=0}^\tau\sum_{j\in T^{t;t-i}}\nabla Q_j(x^{t-i})}.
\end{equation}
Note that with the non-expansion property of Euclidean projection onto a closed convex set we have
\begin{align}
    \label{eqn:origin-of-phi-t}
    \norm{x^{t+1}-x_\H}^2&= \norm{\left[x^t-\eta_t\sum_{i=0}^\tau\sum_{j\in T^{t;t-i}}\nabla Q_j(x^{t-i})\right]_\W-x_\H}^2 \nonumber \\
    &\leq \norm{x^t-\eta_t\sum_{i=0}^\tau\sum_{j\in T^{t;t-i}}\nabla Q_j(x^{t-i})-x_\H}^2 \nonumber \\
    &\leq\norm{x^t-x_\H}^2-2\eta_t\phi_t+\eta_t^2\norm{\sum_{i=0}^\tau\sum_{j\in T^{t;t-i}}\nabla Q_j(x^{t-i})}^2.
\end{align}

Let $T^t=\bigcup_{i=0}^\tau T^{t;t-i}$ be the set of agents whose gradients are used for the update at iteration $t$.
\begin{align}
    \sum_{i=0}^\tau\sum_{j\in T^{t;t-i}}\nabla Q_j(x^{t-i})=&\sum_{i=0}^\tau\sum_{j\in T^{t;t-i}}\left(\nabla Q_j(x^t) - \nabla Q_j(x^t) + \nabla Q_j(x^{t-i}) \right) \nonumber \\
    =&\sum_{i=0}^\tau\sum_{j\in T^{t;t-i}}\nabla Q_j(x^t) - \sum_{i=0}^\tau\sum_{j\in T^{t;t-i}}\left(\nabla Q_j(x^{t}) - \nabla Q_j(x^{t-i})\right) \nonumber \\
    =&\sum_{j\in T^t}\nabla Q_j(x^t) - \sum_{i=0}^\tau\sum_{j\in T^{t;t-i}}\left(\nabla Q_j(x^{t}) - \nabla Q_j(x^{t-i})\right).
\end{align}
Therefore, we have
\begin{align}
    \label{eqn:phi_t-generalized}
    \phi_t=&\iprod{x^t-x_\H}{\sum_{i=0}^\tau\sum_{j\in T^{t;t-i}}\nabla Q_j(x^{t-i})} \nonumber \\
    =&\iprod{x^t-x_\H}{\sum_{j\in T^{t}}\nabla Q_j(x^{t})} - \iprod{x^t-x_\H}{\sum_{i=0}^\tau\sum_{j\in T^{t;t-i}}\left(\nabla Q_j(x^t)-\nabla Q_j(x^{t-i})\right)}.
\end{align}
Let us denote
\begin{align}
    \phi_{t;1} =& \iprod{x^t-x_\H}{\sum_{j\in T^{t}}\nabla Q_j(x^{t})}, \\
    \phi_{t;2} =& \iprod{x^t-x_\H}{\sum_{i=0}^\tau\sum_{j\in T^{t;t-i}}\left(\nabla Q_j(x^t)-\nabla Q_j(x^{t-i})\right)}.
\end{align}
From \eqref{eqn:phi_t-generalized} we have $\phi_t = \phi_{t;1} - \phi_{t;2}$.

For the first term in \eqref{eqn:phi_t-generalized}, $\phi_{t;1}$, we have
\begin{align}
    \label{eqn:phi_t1}
    \phi_{t;1}=&\iprod{x^t-x_\H}{\sum_{j\in T^{t}}\nabla Q_j(x^{t})} \nonumber \\
    =&\iprod{x^t-x_\H}{\sum_{j\in T^{t}}\nabla Q_j(x^{t})+\sum_{k\in[n]\backslash T^t}\nabla Q_k(x^t)-\sum_{k\in[n]\backslash T^t}\nabla Q_k(x^t)} \nonumber \\
    =&\iprod{x^t-x_\H}{\sum_{j\in[n]}\nabla Q_j(x^{t})}-\iprod{x^t-x_\H}{\sum_{k\in[n]\backslash T^t}\nabla Q_k(x^t)}.
\end{align}
Recall \eqref{eqn:apprx_phi_t_1_1} and \eqref{eqn:apprx_phi_t_1_2}, we have the same result as we have in \eqref{eqn:apprx_phi_t_2}:
\begin{equation}
    \phi_{t;1}\geq n\gamma\norm{x^t-x_\H}^2-\sum_{k\in[n]\backslash T^t}\norm{x^t-x_\H}\norm{\nabla Q_k(x^t)}.
\end{equation}
Note that $\mnorm{T^t}\geq n-r$, or $\mnorm{[n]\backslash T^t}\leq r$. Combining above and \eqref{eqn:apprx_gradient_bnd_2-stale-gd},
\begin{align}
    \label{eqn:phi_t1_bnd}
    \phi_{t;1}\geq& n\gamma\norm{x^t-x_\H}^2-r\norm{x^t-x_\H}\left(2n\mu\epsilon+\mu\norm{x^t-x_\H}\right).
\end{align}

For the second term in \eqref{eqn:phi_t-generalized}, $\phi_{t;2}$, by Cauchy-Schwartz inequality, we have
\begin{align}
    \label{eqn:phi_t2}
    \phi_{t;2}=&\iprod{x^t-x_\H}{\sum_{i=0}^\tau\sum_{j\in T^{t;t-i}}\left(\nabla Q_j(x^{t}) - \nabla Q_j(x^{t-i})\right)} \nonumber \\
    \leq& \norm{x^t-x_\H}\cdot\norm{\sum_{i=0}^\tau\sum_{j\in T^{t;t-i}}\left(\nabla Q_j(x^{t}) - \nabla Q_j(x^{t-i})\right)}.
\end{align}
Consider the factor $\norm{\sum_{i=0}^t\sum_{j\in T^{t;t-i}}\left(\nabla Q_j(x^{t}) - \nabla Q_j(x^{t-i})\right)}$. By triangle inequality,
\begin{align}
    \label{eqn:phi_t2_step_a}
    \norm{\sum_{i=0}^\tau\sum_{j\in T^{t;t-i}}\left(\nabla Q_j(x^{t}) - \nabla Q_j(x^{t-i})\right)}\leq&\sum_{i=0}^\tau\sum_{j\in T^{t;t-i}}\norm{\nabla Q_j(x^{t}) - \nabla Q_j(x^{t-i})}.
\end{align}
By Assumption~\ref{assum:lipschitz} and $x^t\in\W$ for all $t$, 
\begin{equation}
    \label{eqn:phi_t2_step_b}
    \norm{\nabla Q_j(x^{t}) - \nabla Q_j(x^{t-i})}\leq\mu\norm{x^t-x^{t-i}}.
\end{equation}
Recall that we defined $T^{t;t-k}\triangleq\varnothing$, $\forall k>t$. By repeatedly applying the update rule \eqref{eqn:update-straggler}, we have
\begin{align}
    & \norm{x^{t}-x^{t-i}}= \norm{\left[x^{t-1}-\eta_{t-1}\sum_{h=0}^\tau\sum_{j\in T^{t-1;t-1-h}}\nabla Q_j(x^{t-1-h})\right]_\W-x^{t-i}} \nonumber \\
        &\leq\norm{x^{t-1}-\eta_{t-1}\sum_{h=0}^\tau\sum_{j\in T^{t-1;t-1-h}}\nabla Q_j(x^{t-1-h})-x^{t-i}} \nonumber \\
        &=\norm{\left[x^{t-2}-\eta_{t-2}\sum_{h=0}^\tau\sum_{j\in T^{t-2;t-2-h}}\nabla Q_j(x^{t-2-h})\right]_\W-\eta_{t-1}\sum_{h=0}^\tau\sum_{j\in T^{t-1;t-1-h}}\nabla Q_j(x^{t-1-h})-x^{t-i}} \nonumber \\
        &\leq\norm{x^{t-2}-\eta_{t-2}\sum_{h=0}^\tau\sum_{j\in T^{t-2;t-2-h}}\nabla Q_j(x^{t-2-h})-\eta_{t-1}\sum_{h=0}^\tau\sum_{j\in T^{t-1;t-1-h}}\nabla Q_j(x^{t-1-h})-x^{t-i}} \nonumber \\
        &=\norm{x^{t-2}-\sum_{k=1}^2\eta_{t-k}\sum_{h=0}^\tau\sum_{j\in T^{t-k;t-k-h}}\nabla Q_j(x^{t-k-h}) - x^{t-i}} \nonumber \\
        &=\cdots \nonumber \\
        &\leq\norm{x^{t-i}-\sum_{k=1}^i\eta_{t-k}\sum_{h=0}^\tau\sum_{j\in T^{t-k;t-k-h}}\nabla Q_j(x^{t-k-h})-x^{t-i}}.
\end{align}
Therefore,
\begin{align}
    \norm{x^t-x^{t-i}}\leq&\norm{\sum_{k=1}^i\eta_{t-k}\sum_{h=0}^\tau\sum_{j\in T^{t-k;t-k-h}}\nabla Q_j(x^{t-k-h})} \nonumber \\
        \leq&\sum_{k=1}^i\eta_{t-k}\norm{\sum_{h=0}^\tau\sum_{j\in T^{t-k;t-k-h}}\nabla Q_j(x^{t-k-h})}.
\end{align}
Recall that we defined $G=n\mu(2n\epsilon+\Gamma)$ in $\eqref{eqn:def-G}$. Also note that for all $t$, $\eta_{t}\geq\eta_{t+1}$. Substitute from \eqref{eqn:update-bnd-generalized} in above, we have 
\begin{equation}
    \norm{x^t-x^{t-i}}\leq\sum_{k=1}^i\eta_{t-k} G\leq i\cdot \eta_{t-i} G. 
\end{equation}
Combining \eqref{eqn:phi_t2_step_a}, \eqref{eqn:phi_t2_step_b} and above, we have
\begin{align}
    &\norm{\sum_{i=0}^\tau\sum_{j\in T^{t;t-i}}\left(\nabla Q_j(x^{t}) - \nabla Q_j(x^{t-i})\right)}\leq\sum_{i=0}^\tau\sum_{j\in T^{t;t-i}}\norm{\nabla Q_j(x^{t}) - \nabla Q_j(x^{t-i})} \nonumber \\
        \leq&\sum_{i=0}^\tau\sum_{j\in T^{t;t-i}}\mu i\cdot \eta_{t-i} G.
\end{align}
Recall that we defined $T^{t;t-i}=\varnothing$ if $i>t$, and $\eta_t=\eta_0$ for all $-\tau\leq t\leq0$. Therefore, 
we have
\begin{align}
    \sum_{i=0}^\tau\sum_{j\in T^{t;t-i}}\mu i\cdot \eta_{t-i} G\leq G\cdot\sum_{i=0}^\tau\sum_{j\in T^{t;t-i}}\mu\tau\eta_{t-\tau}=\mu\tau G\cdot\sum_{j\in T^{t}}\eta_{t-\tau} 
        \leq&\mu\tau\eta_{t-\tau} G\cdot n.
\end{align}
Combining with \eqref{eqn:phi_t2}, from above we have
\begin{equation}
    \label{eqn:phi_t2_bnd}
    \phi_{t;2}\leq\mu\tau\eta_{t-\tau}n G\norm{x^t-x_\H}.
\end{equation}
Recall that $\phi_t=\phi_{t;1}-\phi_{t;2}$. Also, note that $\eta_t\geq\eta_{t+1}$, thus $\eta_0\geq\eta_t\,(\forall t)$.
Combining \eqref{eqn:phi_t1_bnd} and \eqref{eqn:phi_t2_bnd}, we have
\begin{align}
    \phi_t =& \phi_{t;1} - \phi_{t;2} \nonumber \\
        \geq& n\gamma\norm{x^t-x_\H}^2-r\norm{x^t-x_\H}\left(2n\mu\epsilon+\mu\norm{x^t-x_\H}\right) -\mu\tau\eta_{t-\tau}n G\norm{x^t-x_\H} \nonumber \\
        \geq& \left(n\gamma - r\mu\right)\norm{x^t-x_\H}^2 - (2rn\mu\epsilon+\mu\tau\eta_0 nG)\norm{x^t-x_\H} \nonumber \\
        =&n\gamma\left(1-\dfrac{r}{n}\cdot\dfrac{\mu}{\gamma}\right)\norm{x^t-x_\H}\left(\norm{x^t-x_\H}-\dfrac{2r\mu\epsilon + \mu\tau\eta_0G}{\gamma\left(1-\dfrac{r}{n}\cdot\dfrac{\mu}{\gamma}\right)}\right).
\end{align}
Recall that we assume $\displaystyle\alpha=1-\dfrac{r}{n}\cdot\dfrac{\mu}{\gamma}>0$.
We have
\begin{equation}
    \phi_t\geq\alpha n\gamma\norm{x^t-x_\H}\left(\norm{x^t-x_\H}-\dfrac{\mu\left(2r + \tau\eta_0G\right)}{\alpha\gamma}\epsilon\right).
    \label{eqn:phi_t_final-stale}
\end{equation}
\eqref{eqn:phi_t_final-stale} implies for any $\D^*>\D\triangleq\dfrac{\mu\left(2r + \tau\eta_0G\right)}{\alpha\gamma}\epsilon$, 
\begin{equation}
    \label{eqn:phi_t_final_final}
    \phi_t\geq\alpha n\gamma\D^*(\D^*-\D)>0~\textrm{ when }~\norm{x^t-x_\H}\geq\D^*. 
\end{equation}
Combining above with Theorem~\ref{thm:async-fault-toler}, we have
\begin{align}
    \lim_{t\rightarrow\infty}\norm{x^t-x_\H}\leq\D^*.
\end{align}

\end{proof}

\section{Proofs of theorems in Section~\ref{sec:stochastic}}
\label{part:3}

In this section, we present the detailed proof of the convergence rate results presented in Section~\ref{sec:stochastic} of the main paper. We substitute the parameters under different problem settings and state the theorems separately, then proceed with the proofs.

\subsection{Observations and lemmas from the assumptions}

Before the proof, we first introduce and prove some results derived from the assumptions we made. 

\subsubsection{Preliminaries}
We defined the following notations in Section~\ref{sec:stochastic}:


For a stochastic iterative algorithm, we denote $\boldsymbol{z}_i^t\triangleq\{z_{i_1}^t,...,z_{i_k}^t\}$ to be the set of $k$ data samples sampled i.i.d. by agent $i$ at iteration $t$. For each agent $i$ and iteration $t$, we define a random variable 
\begin{equation}
    \zeta_i^t\triangleq\begin{cases}
        \z_i^t, & \textrm{ agent $i$ is non-faulty}, \\
        g_i^t, & \textrm{ agent $i$ is faulty}
    \end{cases}
\end{equation}
representing the source of randomness introduced by agent $i$ at iteration $t$. We further denote $\zeta^t=\left\{\zeta_i^t,\,i=1,...,n\right\}$ to be the source of randomness introduced in iteration $t$, and let $\E_t$ be a shorthand for the expectation with respect to the collective random variables $\zeta^0,...,\zeta^t$, given the initial estimate $x^0$. Specifically, 
\begin{equation}
    \E_t(\cdot)\triangleq\E_{\zeta^0,...,\zeta^t}(\cdot), ~\forall t\geq0.
\end{equation}
Let us further define the notation $\E_{\zeta^t}(\cdot)$ to be the expectation with respect to the random variable $\zeta^t$ and given $x^t$, and $\E_{\zeta_i^t}(\cdot)$ to be the expectation with respect to the random variable $\zeta_i^t$ and given $x^t$.

\subsubsection{Observations}

Consider an arbitrary iteration $t$ of the stochastic version of Algorithm~\ref{alg}. Given the current model estimate $x^t$, recall from definition~\eqref{eqn:def-g-i-t} that for each non-faulty agent $i$ we have
\begin{align*}
    g_i^t = \frac{1}{k} \sum_{j=1}^k \nabla\ell(x^t,z_{i_j}^t) ,
\end{align*}
where the gradient of loss function $\ell(\cdot,\cdot)$ is with respect to its first argument $x^t$ and $z_{i_1}^t,...,z_{i_k}^t$ denotes i.i.d.~data samples from distribution $\mathcal{D}_i$. Therefore, from the definition of $\zeta^t$, generally, for each non-faulty agent $i$ and a deterministic mapping $\Psi$, we have
\begin{equation}
    \E_{\zeta^t} \Psi\left(g_i^t\right)=\E_{\z_i^t}\Psi\left(g_i^t\right), 
    \label{eqn:expectation-zeta-cge}
\end{equation}
where recall that $\boldsymbol{z}_i^t\triangleq\{z_{i_1}^t,...,z_{i_k}^t\}$ for all non-faulty agent $i$, and $\E_{\z_i^t}(\cdot)$ stands for the expectation with respect to the randomness in $\z_i^t$ and given $x^t$.
Accordingly, given $x^t$, we obtain that 
\begin{equation}
\label{eqn:def-expectation-zeta-g-cge}
    \E_{\zeta^t}\left[g_i^t\right]=\E_{\z_i^t}\left[g_i^t\right] = \frac{1}{k}\cdot\E_{\z_i^t}\sum_{j=1}^k\left(\nabla\ell(x^t,z_{i_j}^t)\right) .
\end{equation}
As $\z_i^t$ is a collection of $k$ elements $z_{i_1}^t,...,z_{i_k}^t$ drawn independently from $\mathcal{D}_i$, given $x^t$, from~\eqref{eqn:def-expectation-zeta-g-cge} we obtain that
\begin{equation}
    \E_{\zeta^t}\left[g_i^t\right]=\frac{1}{k}\cdot\sum_{j=1}^k\E_{z_{i_j}^t\sim\mathcal{D}_i}\nabla\ell(x^t,z_{i_j}^t).
    \label{eqn:expectation-zeta-g-cge}
\end{equation}
Note that
\begin{equation}
    \nabla Q_i(x)=\E_{z\sim\mathcal{D}_i}\nabla \ell(w,z),~\forall w\in\R^d.
    \label{eqn:qi-expectation-cge}
\end{equation}
Substituting from \eqref{eqn:qi-expectation-cge} in \eqref{eqn:expectation-zeta-g-cge}, we obtain that for any non-faulty agent $i$,
\begin{equation}
    \E_{\zeta^t}\left[g_i^t\right] = \frac{1}{k}\sum_{j=1}^k\nabla Q_i(x^t)=\nabla Q_i(x^t).
    \label{eqn:expectation-g-i-t-cge}
\end{equation}

\subsubsection{Lemma~\ref{lemma:1-cge} on expectations related to gradients}
~

\begin{mdframed}
\begin{lemma}
    \label{lemma:1-cge}
    For any iteration $t$, if Assumption~\ref{assum:bound-grad} holds true, for each non-faulty agent $i$, the two inequalities hold true:
    \begin{equation}
        \E_{\zeta_i^t}\norm{g_i^t-\E_{\zeta_i^t}\left[g_i^t\right]}\leq\sigma.
        \label{eqn:assum-bound-grad-cge}
    \end{equation}
    \begin{equation}
        \E_{\zeta^t}\norm{g_i^t}^2\leq\sigma^2+\norm{\nabla Q_i(x^t)}^2.
        \label{eqn:lemma-1-cge}
    \end{equation}
\end{lemma}
\end{mdframed}
\begin{proof}
    For the \textbf{first} inequality: Note that $\left(\cdot\right)^2$ is a convex function. By Jensen's inequality, 
    \begin{equation}
        \left(\E_{\zeta_i^t}\norm{g_i^t-\E_{\zeta_i^t}\left[g_i^t\right]}\right)^2\leq\E_{\zeta_i^t}\norm{g_i^t-\E_{\zeta_i^t}\left[g_i^t\right]}^2.
    \end{equation}
    Substitute above in Assumption~\ref{assum:bound-grad}, we have
    \begin{equation}
        \left(\E_{\zeta_i^t}\norm{g_i^t-\E_{\zeta_i^t}\left[g_i^t\right]}\right)^2\leq\sigma^2.
    \end{equation}
    Since both sides are non-negative, taking square roots on both sides we have \eqref{eqn:assum-bound-grad-cge} that
    \begin{equation*}
        \E_{\zeta_i^t}\norm{g_i^t-\E_{\zeta_i^t}\left[g_i^t\right]}\leq\sigma.
    \end{equation*}

    For the \textbf{second} inequality: Let $i$ be an arbitrary non-faulty agent. Using the definition of Euclidean norm, for any $t$,
    \begin{equation}
        \norm{g_i^t-\E_{\zeta^t}\left[g_i^t\right]}^2 = \norm{g_i^t}^2-2\iprod{g_i^t}{\E_{\zeta^t}\left[g_i^t\right]} + \norm{\E_{\zeta^t}\left[g_i^t\right]}^2.
    \end{equation}
    Upon taking expectations on both sides, we obtain that 
    \begin{align}
        \E_{\zeta^t}\norm{g_i^t-\E_{\zeta^t}\left[g_i^t\right]}^2 &= \E_{\zeta^t}\norm{g_i^t}^2-\E_{\zeta^t}2\iprod{g_i^t}{\E_{\zeta^t}\left[g_i^t\right]} + \E_{\zeta^t}\norm{\E_{\zeta^t}\left[g_i^t\right]}^2 \nonumber \\\
        &= \E_{\zeta^t}\norm{g_i^t}^2-2\E_{\zeta^t}\left[g_i^t\right]^T\cdot\E_{\zeta^t}\left[g_i^t\right] + \norm{\E_{\zeta^t}\left[g_i^t\right]}^2 \nonumber \\
        &= \E_{\zeta^t}\norm{g_i^t}^2- \norm{\E_{\zeta^t}\left[g_i^t\right]}^2,
    \end{align}
    where $x^T$ stands for the transpose of a vector $x$.
    Note that from \eqref{eqn:expectation-zeta-cge}, $\E_{\zeta^t}\norm{g_i^t-\E_{\zeta^t}\left[g_i^t\right]}^2=\E_{\z^t}\norm{g_i^t-\E_{\z^t}\left[g_i^t\right]}^2$, and $\E_{\zeta^t}\left[g_i^t\right] = \E_{\z^t}\left[g_i^t\right]$. therefore
    \begin{equation}
        \E_{\z^t}\norm{g_i^t-\E_{\z^t}\left[g_i^t\right]}^2= \E_{\zeta^t}\norm{g_i^t}^2- \norm{\E_{\z^t}\left[g_i^t\right]}^2.
        \label{eqn:expectation-difference-cge}
    \end{equation}
    Combining \eqref{eqn:def-expectation-zeta-g-cge} and \eqref{eqn:expectation-g-i-t-cge} we have $\E_{\z_i^t}\left[g_i^t\right]=\E_{\zeta_i^t}\left[g_i^t\right]=\nabla Q_i(x^t)$. Substituting this in \eqref{eqn:expectation-difference-cge}, 
    \begin{equation}
        \E_{\z^t}\norm{g_i^t-\E_{\z^t}\left[g_i^t\right]}^2= \E_{\zeta^t}\norm{g_i^t}^2-\norm{\nabla Q_i(x^t)}^2.
        \label{eqn:expectation-difference-cge-2}
    \end{equation}
    Combining Assumption~\ref{assum:bound-grad} and \eqref{eqn:expectation-zeta-cge}, we have $\E_{\z^t}\norm{g_i^t-\E_{\z^t}\left[g_i^t\right]}^2\leq\sigma^2$. Substituting this in \eqref{eqn:expectation-difference-cge-2}, we obtain \eqref{eqn:lemma-1-cge} that
    \begin{equation*}
        \E_{\zeta^t}\norm{g_i^t}^2\leq\sigma^2+\norm{\nabla Q_i(x^t)}^2.
    \end{equation*}
\end{proof}

\hrule

\subsubsection{Lemma~\ref{lemma:bound-exp-largest-gradient} on the bound of the largest norm of the gradients among honest agents}

We first recall below a general order-statistics bound to obtain an upper bound on the largest gradient norm among honest agents. \\

\begin{mdframed}
    \begin{lemma}
        \label{lemma:bound-largest-rv}
        Consider $n$ random variables $x_1,...,x_n$. Let $\widehat\mu \triangleq \max_{i} \E \left[ x_i \right]$, and $\widehat\sigma \triangleq \max_{i} \left[ \sqrt{ \E \left[ \left( x_i - \E \left[ x_i \right] \right)^2 \right]} \right]$. For $\widehat x \triangleq \max_{i} x_i$, the following holds true.  
        \begin{align}
            \E{\left[\widehat x\right]} \leq \widehat\sigma\sqrt{n-1} + \widehat\mu.
            \label{eqn:bound-largest-rv}
        \end{align}
    \end{lemma}
\end{mdframed}
\begin{proof}
    Let $y_i \triangleq x_i-\mu_i$, where $\mu_i$ denotes the expectation $x_i$, i.e., $\mu_i \triangleq \E[x_i]$. Note that, for all $i$, $\E\left[y_i\right]=0$, and $\E \left[ \left(y_i - \E\left[y_i\right] \right)^2\right] \leq \widehat{\sigma}^2$.

    Let $y_{k:n}$ denote the $k$-th order of $y_i,...,y_n$. That is, given the values of the random variables $y_i,...,y_n$, the value of $y_{k:n}$ is given by the $k$-largest value in $\{y_i,...,y_n\}$. From results on order statistics (specifically, (2) and~(4) in~\citep{arnold1979bounds}), 
    \begin{align}
        \E\left[y_{k:n}\right] \leq \widehat{\sigma} \sqrt{\frac{k-1}{n-k+1}}.
        \label{eqn:bound-exp-ykn}
    \end{align}
    Let $\widehat{y} \triangleq y_{n:n}$. Therefore,~\eqref{eqn:bound-exp-ykn} implies that
    \begin{align}
        \E{\left[\widehat y\right]} \leq \widehat{\sigma} \, \sqrt{n-1}. \label{eqn:bnd_max_exp_y}
    \end{align}
    
    Now, recall that 
    $\widehat \mu \triangleq \max_{i} \mu_i$. Therefore, 
    \begin{align}
        \widehat{y} = \max_i \{x_i-\mu_i\} \geq \max_i \{x_i - \max_i\mu_i\}  = \widehat x-\widehat\mu,
        \label{eqn:bound-yhat}
    \end{align}
    where $\widehat x \triangleq \max_{i} x_i$. From~\eqref{eqn:bound-yhat}, we obtain that
    \begin{align*}
        \E{\left[\widehat y\right]} \geq  \E[\widehat x] - \widehat\mu .
    \end{align*}
    Substituting from~\eqref{eqn:bnd_max_exp_y} in the above concludes the proof, i.e., we obtain that
    \begin{align*}
        \E{\left[\widehat x\right]} \leq \widehat\sigma\sqrt{n-1} + \widehat\mu.
    \end{align*}
\end{proof}

\begin{mdframed}
    \begin{lemma}
        \label{lemma:bound-exp-largest-gradient}
        Let $\H$ be a set of $n-f$ honest agents in any given execution.
        Let $v_t$ be the agent that sends the stochastic gradient of the largest norm among agents in $\H$ in iteration $t$ (and break tie in favor of the smallest index). Specifically, 
        $v_t := \min \{i \in \H ~ ; ~ \norm{g_{i}^t} \geq \norm{g_j^t} \forall j \in \H \} $. If Assumption~\ref{assum:bound-grad} holds true, we have
        \begin{equation}
            \E_{\zeta^{t}}{\norm{g_{v_t}^t}}\leq \sigma(\sqrt{n-f-1}+1) + \max_{i\in\H}\norm{\nabla Q_i(x^t)}.
            \label{eqn:lemma-bound-exp-largest-gradient}
        \end{equation}
    \end{lemma}
\end{mdframed}
\begin{proof}
    By Jensen's inequality, since the square function is convex, for all $i \in \H$, 
    \begin{align}
        \E_{\zeta^t}\norm{g_i^t}\leq\sqrt{\E_{\zeta^t}\norm{g_i^t}^2}.
        \label{eqn:bound-exp-gradient-norm-0}
    \end{align}
    Also, recall, from \eqref{eqn:lemma-1-cge} in Lemma~\ref{lemma:1-cge}, that 
    \begin{equation*}
        \E_{\zeta^t}\norm{g_i^t}^2\leq \sigma^2+\norm{\nabla Q_i(x^t)}^2.
    \end{equation*}
    Therefore, substituting \eqref{eqn:lemma-1-cge} in \eqref{eqn:bound-exp-gradient-norm-0}, we have
    \begin{align}
        \E_{\zeta^t}\norm{g_i^t}\leq\sqrt{\sigma^2+\norm{\nabla Q_i(x^t)}^2}.
        \label{eqn:bound-exp-gradient-norm}
    \end{align}

    Also, consider the variance of $\norm{g_i^t}$:
    \begin{align}
        \E_{\zeta^t}\left[\left(\norm{g_i^t}-\E_{\zeta^t}{\norm{g_i^t}}\right)^2\right]=\E_{\zeta^t}{\norm{g_i^t}^2} - \left(\E_{\zeta^t}{\norm{g_i^t}}\right)^2.
    \end{align}
    Since the Euclidean norm is convex, by Jensen's inequality, 
    \begin{align}
        \E_{\zeta^t}{\norm{g_i^t}} \geq \norm{\E_{\zeta^t}{\left[g_i^t\right]}}.
    \end{align}
    Combining with \eqref{eqn:expectation-g-i-t-cge} that $\E_{\zeta^t}{\left[g_i^t\right]} = \nabla Q_i(x^t)$, we have
    \begin{align}
        \E_{\zeta^t}{\norm{g_i^t}}\geq \norm{\nabla Q_i(x^t)}.
    \end{align}
    Therefore, 
    \begin{align}
        \E_{\zeta^t}\left[\left(\norm{g_i^t}-\E_{\zeta^t}{\norm{g_i^t}}\right)^2\right] &=\E_{\zeta^t}{\norm{g_i^t}^2} - \left(\E_{\zeta^t}{\norm{g_i^t}}\right)^2 \nonumber \\
        &\leq \E_{\zeta^t}\norm{g_i^t}^2 - \norm{\nabla Q_i(x^t)}^2.
    \end{align}
    Combining above with \eqref{eqn:lemma-1-cge}, i.e., $\E_{\zeta^t}\norm{g_i^t}^2 - \norm{\nabla Q_i(x^t)}^2 \leq \sigma^2$, we have
    \begin{align}
        \E_{\zeta^t}\left[\left(\norm{g_i^t}-\E_{\zeta^t}{\norm{g_i^t}}\right)^2\right] &\leq \E_{\zeta^t}\norm{g_i^t}^2 - \norm{\nabla Q_i(x^t)}^2 \leq \sigma^2.
        \label{eqn:bound-gradient-variance}
    \end{align}

    Now recall \eqref{eqn:bound-largest-rv} from Lemma~\ref{lemma:bound-largest-rv}. In our case, we have a group of random variables $\norm{g_i^t}$'s, where $i\in\H$, $\mnorm{\H}=n-f$. Therefore,
    \begin{align}
        \E_{\zeta^t}{\norm{g_{v_t}^t}}\leq \left(\max_{i\in\H}\sqrt{\E_{\zeta^t}\left[\left(\norm{g_i^t}-\E_{\zeta^t}\norm{g_i^t}\right)^2\right]}\right)\cdot\sqrt{\mnorm{\H}-1} + \max_{i\in\H}\left(\E_{\zeta^t}\norm{g_i^t}\right).
    \end{align}
    Substituting from \eqref{eqn:bound-exp-gradient-norm} and \eqref{eqn:bound-gradient-variance} in above, we have
    \begin{align*}
        \E_{\zeta^t}{\norm{g_{v_t}^t}} &\leq \max_{i\in\H}\sqrt{\sigma^2 + \norm{\nabla Q_i(x^t)}^2} + \sigma\cdot\sqrt{\mnorm{\H}-1} \nonumber \\
        &= \sqrt{\sigma^2 + \max_{i\in\H}\norm{\nabla Q_i(x^t)}^2} + \sigma\cdot\sqrt{n-f-1} \nonumber \\
        &\leq \sigma + \max_{i\in\H}\norm{\nabla Q_i(x^t)} + \sigma\sqrt{n-f-1} \nonumber \\
        &=\sigma(1+\sqrt{n-f-1}) + \max_{i\in\H}\norm{\nabla Q_i(x^t)}.
    \end{align*}
\end{proof}

\hrule 

\subsubsection{Lemmas on $\gamma$ and $\mu$}

\begin{mdframed}
\begin{lemma}
    \label{lemma:gamma-mu}
    If assumptions~\ref{assum:lipschitz} and \ref{assum:strongly-convex-ft} hold true simultaneously then 
    \begin{equation}
        \gamma \leq \mu.
    \end{equation}
\end{lemma}
\end{mdframed}

The proof of this lemma can be found in Appendix C of \cite{liu2021approximate-full}.


\subsubsection{Lemma~\ref{lemma:converge-sgd} on expected convergence of estimates}

Recall that we defined in Section~\ref{sec:full-grad} that for any non-faulty agents $\H$ with $\mnorm{\H}=n-f$ in an execution, $x_\H\in\W$ is a unique minimum point of the aggregate cost function of agents in $\H$.

\begin{mdframed}
\begin{lemma}
    \label{lemma:converge-sgd}
    Consider the general iterative update rule \eqref{eqn:update}. Suppose $\M\geq0$ and $\rho\in[0,1)$ are two real values. If 
    \begin{align}
        \E_{\zeta^t}\norm{x^{t+1}-x_\H}^2\leq\rho\norm{x^t-x_\H}^2 + \M.
        \label{eqn:expectation-w-t-square-sgd-cge-5}
    \end{align}
    we have
    \begin{align}
        \E_{t}\norm{x^{t+1}-x_\H}^2&\leq\rho^{t+1}\norm{x^0-x_\H}^2 + \left(\frac{1-\rho^{t+1}}{1-\rho}\right)\M.
    \end{align}
\end{lemma}
\end{mdframed}

The proof of this lemma can be found in the appendices of \cite{gupta2021byzantine}, and is inspired by the analysis presented in~\cite[Section 4]{bottou2018optimization}.

\subsection{Proof of Theorem~\ref{thm:cge} - Derivation of \eqref{eqn:stochastic-params-a}}
\label{appdx:b-2}

Recall that we defined the following parameters in Section~\ref{sec:stochastic}:
\begin{itemize}[nosep]
    \item The \textit{resilience margin}
    \begin{align}
        \alpha=1-\frac{f-r}{n-r}-\frac{f+r}{n-r}\cdot\frac{2\mu}{\gamma}.
        \label{eqn:def-alpha-cge}
    \end{align}
    \item The parameter that determines the step size
    \begin{align}
        \overline{\eta} = \cfrac{2(n-r)\gamma\alpha}{(n-f)^2\mu^2+2(n-r)^2\mu^2}.
        \label{eqn:def-eta-bar-cge}
    \end{align}
\end{itemize}

Recall the definition of $x_\H$ in Section~\ref{sec:full-grad}: for any non-faulty agents $\H$ with $\mnorm{\H}=n-f$ in an execution, $x_\H\in\W$ is a unique minimum point of the aggregate cost function of agents in $\H$. The results in \eqref{eqn:stochastic-params-a} can be described in full as follows:

\begin{mdframed}
    \textsc{Theorem~\ref{thm:cge}-CGE. }
    \textit{Consider Algorithm~\ref{alg} with stochastic updates, and the GAR in use is the CGE gradient filter \eqref{eqn:cge-footnote}. 
    Suppose Assumptions~\ref{assum:lipschitz}, \ref{assum:strongly-convex-ft}, and \ref{assum:bound-grad} hold true, the expected cost functions of the agents in the system satisfy $(f, r;\epsilon)$-redundancy, $\alpha>0$, step size in \eqref{eqn:update} $\eta_t=\eta>0$ for all $t$, and $n\geq 2f+3r$. Let
    \begin{align}
        \M =& 4\left(\left(2(f+r)+(n-f)^2\eta\mu\right)^2 +m^2(n-f)^2\eta^2\mu^2\right)\epsilon^2 \nonumber \\
        &\,+\left(4\left(\frac{f+r}{m\mu}\right)^2\left(\sqrt{n-f-1}+1\right)^2 + (n-f)^2\eta^2\right)\sigma^2.
        \label{eqn:def-m-cge}
    \end{align}
    If $\eta<\overline{\eta}$, the following holds true:
    \begin{itemize}
        \item The value of 
        \begin{align}
            \rho = 1-2(n-f)\eta\gamma+4(f+r)\eta\mu+(n-f)^2\eta^2\mu^2 + 2(n-r)^2\eta^2\mu^2
            \label{eqn:def-rho-cge}
        \end{align}
        satisfies $0<\rho<1$, and 
        \item Given the initial estimate $x^0$ arbitrarily chosen from $\R^d$, for all $t\geq0$,
        \begin{align}
            \E_{t}\norm{x^{t+1}-x_\H}^2&\leq\rho^{t+1}\norm{x^0-x_\H}^2 + \left(\frac{1-\rho^{t+1}}{1-\rho}\right)\M.
            \label{eqn:expectation-bound-1-cge}
        \end{align}
    \end{itemize}}
\end{mdframed}

\begin{proof}
    \textbf{First}, we show a recursive bound over the expected value of $\norm{x^t-x_\H}^2$. We use $m=n-r$ as a short hand throughout this proof. Following the \textbf{first} part of the proof of Theorem~\ref{thm:async-fault-toler}-CGE in Appendix~\ref{appdx:a-2}, we know that by Assumption~\ref{assum:lipschitz} and $(f,r;\epsilon)$-redundancy, for every non-faulty agent $i\in[n]$, we obtain \eqref{eqn:honest-bound-everywhere} that
    \begin{equation}
        \norm{\nabla Q_i(x)}\leq(2m-4f+1)\mu\epsilon+\mu\norm{x-x_{\H}}\leq2m\mu\epsilon+\mu\norm{x-x_{\H}}.
        \label{eqn:honest-bound-everywhere-copy-stochastics}
    \end{equation}
    
    Let us denote $\g^t$ the output of the gradient filter applied, i.e.
    \begin{equation}
        \g^t=\ga\left(g_j^t|j\in S^t;m,f\right). 
        \label{eqn:gothic-g-t-async-cge}
    \end{equation}
    With $\eta_t=\eta$ and above, for each iteration $t$ we have
    \begin{equation}
        x^{t+1}=\left[x^t-\eta\g^t\right]_\W.
        \label{eqn:update-gothic-g-t-async-cge}
    \end{equation}
    Subtracting $x_\H$ and taking norm on both sides. Using the non-expansion property of Euclidean projection onto a closed convex set,
    \begin{equation}
        \norm{x^{t+1}-x_\H}= \norm{\left[x^t-\eta_t\g^t\right]_\W-x_\H} \leq\norm{x^t-\eta_t\g^t-x_\H}.
    \end{equation}
    Taking square on both sides, we have
    \begin{align}
        \norm{x^{t+1}-x_\H}^2\leq \norm{x^t-x_\H}^2-2\eta\iprod{x^t-x_\H}{\g^t}+\eta^2\norm{\g^t}^2.
        \label{eqn:iterate-general-cge}
    \end{align}
    
    Consider the CGE gradient filter, where 
    \begin{equation}
        \g^t=\sum_{l=1}^{m-f}g_{i_l}^t,
        \label{eqn:cge-filter}
    \end{equation}
    where the gradients are sorted by their norms as 
    \begin{equation}
        \norm{g_{i_1}^t}\leq...\leq\norm{g_{i_{m-f-1}}^t}\leq\norm{g_{i_{m-f}}^t}\leq\norm{g_{i_{m-f+1}}^t}\leq...\leq\norm{g_{i_m}^t}.
    \end{equation}
    As there are at most $f$ Byzantine agents, in each iteration $t$, for each $i_l\notin\H$ and $l\in[m-f]$, there exists a unique $k_l\in\H$ and $k_l\notin\{i_1,...,i_{m-f}\}$, such that
    \begin{align}
        \norm{g_{i_l}^t}\leq \norm{g_{k_l}^t}.
    \end{align}
    The above implies that
    \begin{align}
        \norm{\g^t}\leq\sum_{l=1}^{m-f}\norm{g_{i_l}^t}\leq 
        \sum_{j\in\H}\norm{g_j^t}.
    \end{align}
    By triangle inequality and AM-QM inequality (i.e., $\frac{1}{n}\sum_{j=1}^nx_j\leq\sqrt{\frac{1}{n}\sum_{j=1}^nx_j^2}$ for any $n$ positive real $x_j$'s),
    \begin{align}
        \norm{\g^t}^2 \leq \left(\sum_{j\in\H}\norm{g_j^t}\right)^2\leq\mnorm{\H}\sum_{j\in\H}\norm{g_j^t}^2=(n-f)\sum_{j\in \H}\norm{g_j^t}^2.
        \label{eqn:cge-square-bound-cge}
    \end{align}
    Now, consider the term $\iprod{x^t-x_\H}{\g^t}$. let $\H^t=\{i_1,...,i_{m-f}\}\cap\H$, and let $\B^t=\{i_1,...,i_{m-f}\}\backslash\H^t$. Note that 
    \begin{equation}
        \mnorm{\H^t}\geq m-2f, ~\textrm{ and }~\mnorm{\B^t}\leq f.
    \end{equation}
    From \eqref{eqn:cge-filter}, 
    \begin{equation}
        \g^t=\sum_{j\in\H^t}g_j^t+\sum_{j\in\B^t}g_j^t.
    \end{equation}
    Therefore, 
    \begin{equation}
        \iprod{x^t-x_\H}{\g^t} = \iprod{x^t-x_\H}{\sum_{j\in\H^t}g_j^t} + \iprod{x^t-x_\H}{\sum_{j\in\B^t}g_j^t}.
    \end{equation}
    We define 
    \begin{equation}
        \phi_t=\iprod{x^t-x_\H}{\g^t}=\sum_{j\in\H^t}\iprod{x^t-x_\H}{g_j^t} + \sum_{j\in\B^t}\iprod{x^t-x_\H}{g_j^t}.
        \label{eqn:phi-t-sgd-cge}
    \end{equation}
    Substituting above and \eqref{eqn:cge-square-bound-cge} in \eqref{eqn:iterate-general-cge}, we obtain that
    \begin{equation}
        \norm{x^{t+1}-x_\H}^2\leq\norm{x^t-x_\H}^2-2\eta\phi_t+(n-f)\eta^2\sum_{i\in\H}\norm{g_i^t}^2.
    \end{equation}
    Recall that $\zeta^t=\left\{\zeta_1^t,...,\zeta_n^t\right\}$, and $x^{t+1}$ is a function of the set of random variables $\zeta^t$. Also note that $\E_{\zeta^t}\norm{x^t-x_\H}^2=\norm{x^t-x_\H}^2$, since $x^t$ does not depend on $\zeta^t$. Taking expectation $\E_{\zeta^t}$ on both sides, we have
    \begin{equation}
        \E_{\zeta^t}\norm{x^{t+1}-x_\H}^2\leq\norm{x^t-x_\H}^2-2\eta\E_{\zeta^t}\left[\phi_t\right]+(n-f)\eta^2\sum_{i\in\H}\E_{\zeta^t}\norm{g_i^t}^2. 
        \label{eqn:expectation-w-t-square-cge}
    \end{equation}
    
    From \eqref{eqn:phi-t-sgd-cge} the definition of $\phi_t$,
    \begin{align}
        \E_{\zeta^t}\left[\phi_t\right]& = \E_{\zeta^t}\sum_{j\in\H^t}\iprod{x^t-x_\H}{g_j^t} + \E_{\zeta^t}\sum_{j\in\B^t}\iprod{x^t-x_\H}{g_j^t} \nonumber \\
        & = \sum_{j\in\H^t}\iprod{x^t-x_\H}{\E_{\zeta^t}\left[g_j^t\right]} +\E_{\zeta^t}\sum_{j\in\B^t}\iprod{x^t-x_\H}{g_j^t}.
        \label{eqn:exp-phi-t-sgd-cge}
    \end{align}
    The first term of \eqref{eqn:exp-phi-t-sgd-cge} becomes
    \begin{align}
        \sum_{j\in\H^t}\iprod{x^t-x_\H}{\E_{\zeta^t}\left[g_j^t\right]} &= \iprod{x^t-x_\H}{\sum_{j\in\H}\E_{\zeta^t}\left[g_j^t\right]} - \iprod{x^t-x_\H}{\sum_{j\in\H\backslash\H^t}\E_{\zeta^t}\left[g_j^t\right]}
    \end{align}
    Recall from \eqref{eqn:expectation-g-i-t-cge} that for any $j\in\H$, $\E_{\zeta^t}\left[g_j^t\right] = \nabla Q_j(x^t)$. By Assumption~\ref{assum:strongly-convex-ft} and the fact that $\nabla\sum_{j\in\H}Q_j(x_\H)=0$,
    \begin{align}
        \iprod{x^t-x_\H}{\sum_{j\in\H}\E_{\zeta^t}\left[g_j^t\right]} &= \iprod{x^t-x_\H}{\sum_{j\in\H}\nabla Q_j(x^t)} \nonumber \\
        &= \iprod{x^t-x_\H}{\sum_{j\in\H}\nabla Q_j(x^t)-\sum_{j\in\H}\nabla Q_j(x_\H)}\geq \mnorm{\H}\gamma\norm{x^t-x_\H}^2.
        \label{eqn:exp-phit-part1A-sgd-cge}
    \end{align}
    By Cauchy-Schwartz inequality,
    \begin{equation}
        \iprod{x^t-x_\H}{\sum_{j\in\H\backslash\H^t}\E_{\zeta^t}\left[g_j^t\right]}= \sum_{j\in\H\backslash\H^t}\iprod{x^t-x_\H}{\E_{\zeta^t}\left[g_j^t\right]} \leq \sum_{j\in\H\backslash\H^t}\norm{x^t-x_\H}\E_{\zeta^t}\norm{g_j^t}.
        \label{eqn:exp-phit-part1B-sgd-cge}
    \end{equation}
    Therefore, by combining \eqref{eqn:exp-phit-part1A-sgd-cge} and \eqref{eqn:exp-phit-part1B-sgd-cge}, we have
    \begin{align}
        \sum_{j\in\H^t}\iprod{x^t-x_\H}{\E_{\zeta^t}\left[g_j^t\right]} &\geq \mnorm{\H}\gamma\norm{x^t-x_\H}^2 - \sum_{j\in\H\backslash\H^t}\norm{x^t-x_\H}\E_{\zeta^t}\norm{g_j^t}.
        \label{eqn:exp-phit-part1-sgd-cge}
    \end{align}
    Also, by Cauchy-Schwartz inequality, for any $j\in[n]$, 
    \begin{equation}
        \iprod{x^t-x_\H}{g_j^t}\geq -\norm{x^t-x_\H}\norm{g_j^t}.
    \end{equation}
    Recall the sorting of vectors $\{g_j^t\}_{j=1}^n$. For an arbitrary $j\in\B^t$ and an arbitrary $j'\in\H\backslash\H^t$, 
    \begin{equation}
        \norm{g_j^t}\leq\norm{g_{j'}^t}.
        \label{eqn:bound-B-t}
    \end{equation}
    Recall that $\B^t=\left\{i_1,...,i_{m-f}\right\}\backslash\H^t$. Thus, $\mnorm{\B^t}=m-f-\mnorm{\H^t}$. Also, as $\mnorm{\H}=n-f$, $\mnorm{\H\backslash\H^t}=n-f-\mnorm{\H^t}$. That is, $\mnorm{\B^t}\leq\mnorm{\H\backslash\H^t}$. Therefore,  we have
    \begin{align}
       \sum_{j\in\B^t}\iprod{x^t-x_\H}{g_j^t}&\geq-\sum_{j\in\B^t}\norm{x^t-x_\H}\norm{g_j^t}\geq-\sum_{j\in\H\backslash\H^t}\norm{x^t-x_\H}\norm{g_j^t}.
       \label{eqn:B-t-sgd-cge}
    \end{align}
    Taking expectation on both sides, for the second term of \eqref{eqn:exp-phi-t-sgd-cge} we have
    \begin{align}
       \E_{\zeta^t}\sum_{j\in\B^t}\iprod{x^t-x_\H}{g_j^t}\geq\E_{\zeta^t}\left[-\sum_{j\in\H\backslash\H^t}\norm{x^t-x_\H}\norm{g_j^t}\right]=-\sum_{j\in\H\backslash\H^t}\norm{x^t-x_\H}\E_{\zeta^t}\norm{g_j^t}.
       \label{eqn:exp-B-t-sgd-cge}
    \end{align}
    Combining \eqref{eqn:exp-phi-t-sgd-cge}, \eqref{eqn:exp-phit-part1-sgd-cge}, and \eqref{eqn:exp-B-t-sgd-cge}, we have
    \begin{align}
        \E_{\zeta^t}\left[\phi_t\right] &\geq \mnorm{\H}\gamma\norm{x^t-x_\H}^2-2\sum_{j\in\H\backslash\H^t}\norm{x^t-x_\H}\E_{\zeta^t}\norm{g_j^t}.
        \label{eqn:exp-phit-sgd-cge-intermediate}
    \end{align}

    Note that in each iteration $t$, there exists a $v_t\in\H$ such that $\norm{g_j^t}\leq\norm{g_{v_t}^t}$ for all $j\in\H$, i.e., the gradient sent by agent $v_t$ has the largest norm among all gradients from non-faulty agents in $\H$ in iteration $t$. By Lemma~\ref{lemma:bound-exp-largest-gradient},
    \begin{align}
        &\sum_{j\in\H\backslash\H^t}\norm{x^t-x_\H}\E_{\zeta^t}\norm{g_j^t} \leq \mnorm{\H\backslash\H^t}\norm{x^t-x_\H}\E_{\zeta^t}\norm{g_{v_t}^t}
        \nonumber \\
        \leq& \mnorm{\H\backslash\H^t}\norm{x^t-x_\H}\left(\sigma\left(\sqrt{n-f-1}+1\right)+\max_{i\in\H}\norm{\nabla Q_i(x^t)}\right) \nonumber \\
        =& \sigma\left(\sqrt{n-f-1}+1\right)\mnorm{\H\backslash\H^t}\norm{x^t-x_\H}+\mnorm{\H\backslash\H^t}\norm{x^t-x_\H}\left(\max_{i\in\H}\norm{\nabla Q_i(x^t)}\right).
        \label{eqn:exp-phit-part2-sgd-cge}
    \end{align}
    Combining \eqref{eqn:exp-phit-sgd-cge-intermediate} and \eqref{eqn:exp-phit-part2-sgd-cge}, we have
    \begin{align}
        \E_{\zeta^t}\left[\phi_t\right] \geq \mnorm{\H}\gamma\norm{x^t-x_\H}^2 &-2\mnorm{\H\backslash\H^t}\norm{x^t-x_\H}\left(\max_{i\in\H}\norm{\nabla Q_i(x^t)}\right) \nonumber \\
        &-2\sigma\left(\sqrt{n-f-1}+1\right)\mnorm{\H\backslash\H^t}\norm{x^t-x_\H}. 
    \end{align}
    Substituting \eqref{eqn:honest-bound-everywhere-copy-stochastics} in above, we have
    \begin{align}
        \E_{\zeta^t}\left[\phi_t\right] \geq \mnorm{\H}\gamma\norm{x^t-x_\H}^2&-2\mnorm{\H\backslash\H^t}\norm{x^t-x_\H}\left(2m\mu\epsilon + \mu\norm{x^t-x_\H}\right) \nonumber \\
        &-2\sigma\left(\sqrt{n-f-1}+1\right)\mnorm{\H\backslash\H^t}\norm{x^t-x_\H}. 
    \end{align}
    As $\mnorm{\H}=n-f$, $\mnorm{\H^t}\geq m-2f$, and $\mnorm{\H\backslash\H^t}=n-f-\mnorm{H^t}\leq f+r$, the above implies that
    \begin{align}
        \E_{\zeta^t}\left[\phi_t\right]&\geq (n-f)\gamma\norm{x^t-x_\H}^2-2(f+r)\norm{x^t-x_\H}\left(2m\mu\epsilon + \mu\norm{x^t-x_\H}\right) \nonumber \\
        &\qquad\qquad\qquad\qquad\qquad\,-2(f+r)\sigma\left(\sqrt{n-f-1}+1\right)\norm{x^t-x_\H} \nonumber \\
        &\geq \left((n-f)\gamma - 2(f+r)\mu\right)\norm{x^t-x_\H}^2 \nonumber \\
        &\quad\quad - 2\left(2m\mu\epsilon+\sigma\left(\sqrt{n-f-1}+1\right)\right)(f+r)\norm{x^t-x_\H}.
        \label{eqn:expectation-second-term-sgd-cge}
    \end{align}
    
    Now consider the third term $(n-f)\eta^2\sum_{i\in\H}\E_{\zeta^t}\norm{g_i^t}^2$ in \eqref{eqn:expectation-w-t-square-cge}. Recall from Lemma~\ref{lemma:1-cge} that for any non-faulty agent $j$, we have \eqref{eqn:lemma-1-cge}
    \begin{equation*}
        \E_{\zeta^t}\norm{g_j^t}^2\leq\sigma^2+\norm{\nabla Q_j(x^t)}^2.
    \end{equation*}
    Substituting \eqref{eqn:honest-bound-everywhere} in above, we have
    \begin{align}
        \E_{\zeta^t}\norm{g_j^t}^2\leq \sigma^2+\norm{\nabla Q_j(x^t)}^2= \sigma^2+\left(2m\mu\epsilon + \mu\norm{x^t-x_\H}\right)^2.
    \end{align}
    Therefore, 
    \begin{align}
        (n-f)\eta^2\sum_{i\in\H}\E_{\zeta^t}\norm{g_i^t}^2 &\leq (n-f)\mnorm{\H}\eta^2\left(\sigma^2+\left(2m\mu\epsilon + \mu\norm{x^t-x_\H}\right)^2\right) \nonumber \\
        &= (n-f)^2\eta^2\left(\sigma^2+\left(2m\mu\epsilon + \mu\norm{x^t-x_\H}\right)^2\right).
        \label{eqn:expectation-third-term-sgd-cge}
    \end{align}
    
    Substituting \eqref{eqn:expectation-second-term-sgd-cge} and \eqref{eqn:expectation-third-term-sgd-cge} in \eqref{eqn:expectation-w-t-square-cge}, we have
    \begin{align}
        &\E_{\zeta^t}\norm{x^{t+1}-x_\H}^2\leq\norm{x^t-x_\H}^2-2\eta\E_{\zeta^t}\left[\phi_t\right]+ (n-f)\eta^2\sum_{i\in\H}\E_{\zeta^t}\norm{g_i^t}^2 \nonumber \\
        \leq&\norm{x^t-x_\H}^2 \nonumber \\
            &\, -2\eta\left[\left((n-f)\gamma - 2(f+r)\mu\right)\norm{x^t-x_\H}^2 - 2\left(2m\mu\epsilon+\sigma\left(\sqrt{n-f-1}+1\right)\right)(f+r)\norm{x^t-x_\H}\right] \nonumber \\
            &\, +(n-f)^2\eta^2\left(2m\mu\epsilon + \mu\norm{x^t-x_\H}\right)^2 + (n-f)^2\eta^2\sigma^2 \nonumber \\
        =&\left(1-2(n-f)\eta\gamma+4(f+r)\eta\mu+(n-f)^2\eta^2\mu^2\right)\norm{x^t-x_\H}^2 \nonumber \\
        &\,+\left(4m\eta\mu\epsilon\left(2(f+r)+(n-f)^2\eta\mu\right)+4(f+r)\left(\sqrt{n-f-1}+1\right)\eta\sigma\right)\norm{x^t-x_\H} \nonumber \\
        &\,+4m^2(n-f)^2\eta^2\mu^2\epsilon^2+ (n-f)^2\eta^2\sigma^2.
        \label{eqn:expectation-w-t-square-sgd-cge-2}
    \end{align}
    Consider the second term in \eqref{eqn:expectation-w-t-square-sgd-cge-2}. Notice that for any two real values $a$ and $b$, we have $2ab\leq a^2+b^2$. Thus we have
    \begin{align}
        &\left(4m\eta\mu\epsilon\left(2(f+r)+(n-f)^2\eta\mu\right)+4(f+r)\left(\sqrt{n-f-1}+1\right)\eta\sigma\right)\norm{x^t-x_\H} \nonumber \\
        =&2\left(2m\eta\mu\epsilon\left(2(f+r)+(n-f)^2\eta\mu\right)\norm{x^t-x_\H}\right)+2\left(2(f+r)\left(\sqrt{n-f-1}+1\right)\eta\sigma)\norm{x^t-x_\H}\right) \nonumber \\
        \leq& m^2\eta^2\mu^2\norm{x^t-x_\H}^2 + 4\left(2(f+r)+(n-f)^2\eta\mu\right)^2\epsilon^2 \nonumber \\
        &+m^2\eta^2\mu^2\norm{x^t-x_\H}^2+4\left(\frac{f+r}{m\mu}\right)^2\left(\sqrt{n-f-1}+1\right)^2\sigma^2 \nonumber \\
        =& 2m^2\eta^2\mu^2\norm{x^t-x_\H}^2 + 4\left(2(f+r)+(n-f)^2\eta\mu\right)^2\epsilon^2 +4\left(\frac{f+r}{m\mu}\right)^2\left(\sqrt{n-f-1}+1\right)^2\sigma^2.
        \label{eqn:expectation-w-t-square-sgd-cge-2-2}
    \end{align}
    Substituting above in \eqref{eqn:expectation-w-t-square-sgd-cge-2}, we have
    \begin{align}
        &\E_{\zeta^t}\norm{x^{t+1}-x_\H}^2\leq\norm{x^t-x_\H}^2-2\eta\E_{\zeta^t}\left[\phi_t\right]+ (n-f)\eta^2\sum_{i\in\H}\E_{\zeta^t}\norm{g_i^t}^2 \nonumber \\
        \leq&\left(1-2(n-f)\eta\gamma+4(f+r)\eta\mu+(n-f)^2\eta^2\mu^2\right)\norm{x^t-x_\H}^2 \nonumber \\
        &\,+2m^2\eta^2\mu^2\norm{x^t-x_\H}^2 + 4\left(2(f+r)+(n-f)^2\eta\mu\right)^2\epsilon^2 +4\left(\frac{f+r}{m\mu}\right)^2\left(\sqrt{n-f-1}+1\right)^2\sigma^2 \nonumber \\
        &\,+4m^2(n-f)^2\eta^2\mu^2\epsilon^2+ (n-f)^2\eta^2\sigma^2 \nonumber \\
        =&\left(1-2(n-f)\eta\gamma+4(f+r)\eta\mu+(n-f)^2\eta^2\mu^2+2m^2\eta^2\mu^2\right)\norm{x^t-x_\H}^2 \nonumber \\
        &\, + 4\left(\left(2(f+r)+(n-f)^2\eta\mu\right)^2 +m^2(n-f)^2\eta^2\mu^2\right)\epsilon^2 \nonumber \\
        &\,+\left(4\left(\frac{f+r}{m\mu}\right)^2\left(\sqrt{n-f-1}+1\right)^2 + (n-f)^2\eta^2\right)\sigma^2.
        \label{eqn:expectation-w-t-square-sgd-cge-2-new}
    \end{align}
    
    Let $B\triangleq4\left(\left(2(f+r)+(n-f)^2\eta\mu\right)^2 +m^2(n-f)^2\eta^2\mu^2\right)$ and $C\triangleq\linebreak4\left(\frac{f+r}{m\mu}\right)^2\left(\sqrt{n-f-1}+1\right) + (n-f)^2\eta^2$. 
    Thus, from \eqref{eqn:expectation-w-t-square-sgd-cge-2-new} we obtain
    \begin{align}
        \label{eqn:expectation-w-t-square-sgd-cge-3}
        \E_{\zeta^t}\norm{x^{t+1}-x_\H}^2&\leq \left(1-2(n-f)\eta\gamma+4(f+r)\eta\mu+(n-f)^2\eta^2\mu^2+2m^2\eta^2\mu^2\right)\norm{x^t-x_\H}^2 \nonumber \\
        &\qquad+B\epsilon^2+C\sigma^2.
    \end{align}
    Recalling \eqref{eqn:def-rho-cge} the definition of $\rho$, we have
    \begin{align}
        \label{eqn:expectation-w-t-square-sgd-cge-4}
        \E_{\zeta^t}\norm{x^{t+1}-x_\H}^2&\leq \rho\norm{x^t-x_\H}^2+B\epsilon^2+C\sigma^2.
    \end{align}
    Note that $B,C\geq0$. 

    \textbf{Second}, we show $0<\rho<1$. Recall that in \eqref{eqn:def-alpha-cge} we defined 
    \begin{equation*}
        \alpha=1-\frac{f-r}{m}-\frac{f+r}{m}\cdot\frac{2\mu}{\gamma},
    \end{equation*} 
    where $m=n-r$.
    We have
    \begin{equation}
        (n-f)\gamma-2(f+r)\mu = m\gamma\alpha.
        \label{eqn:replace-alpha-sgd-cge}
    \end{equation}
    So $\rho$ can be written as
    \begin{align}
        \rho&=1-2m\eta\gamma\alpha+(n-f)^2\eta^2\mu^2+2m^2\eta^2\mu^2 \nonumber \\
        &=1-\left(2m\eta\gamma\alpha-\left((n-f)^2\mu^2+2m^2\mu^2\right)\eta^2\right) \nonumber \\
        &= 1- \left((n-f)^2\mu^2+2m^2\mu^2\right)\eta\left(\frac{2m\gamma\alpha}{(n-f)^2\mu^2+2m^2\mu^2}-\eta\right).
        \label{eqn:rho-sgd-cge-2}
    \end{align}
    Recall \eqref{eqn:def-eta-bar-cge} that $\overline{\eta}=\cfrac{2m\gamma\alpha}{(n-f)^2\mu^2+2m^2\mu^2}$. From above we obtain that
    \begin{align}
        \rho = 1-\left((n-f)^2\mu^2+2m^2\mu^2\right)\eta(\overline{\eta}-\eta).
    \end{align}
    Note that 
    \begin{align}
        \eta(\overline{\eta}-\eta)=\left(\frac{\overline{\eta}}{2}\right)^2-\left(\eta-\frac{\overline{\eta}}{2}\right)^2.
    \end{align}
    Therefore,
    \begin{align}
        \rho &= 1-\left((n-f)^2\mu^2+2m^2\mu^2\right)\left[\left(\frac{\overline{\eta}}{2}\right)^2-\left(\eta-\frac{\overline{\eta}}{2}\right)^2\right] \nonumber \\
        &= \left((n-f)^2\mu^2+2m^2\mu^2\right)\left(\eta-\frac{\overline{\eta}}{2}\right)^2 +  1-\left((n-f)^2\mu^2+2m^2\mu^2\right)\left(\frac{\overline{\eta}}{2}\right)^2.
    \end{align}
    Since $\eta\in(0,\overline{\eta})$, the minimum value of $\rho$ can be obtained when $\eta=\overline{\eta}/2$,
    \begin{align}
        \min_\eta\rho=1-\left((n-f)^2\mu^2+2m^2\mu^2\right)\left(\frac{\overline{\eta}}{2}\right)^2.
    \end{align}
    On the other hand, since $\eta\in(0,\overline{\eta})$, from \eqref{eqn:rho-sgd-cge-2}, $\rho<1$. Thus,
    \begin{align}
        1-\left((n-f)^2\mu^2+2m^2\mu^2\right)\left(\frac{\overline{\eta}}{2}\right)^2\leq\rho<1.
    \end{align}
    Substituting \eqref{eqn:def-eta-bar-cge} in above implies that $\rho\in\left[1-\cfrac{(m\gamma\alpha)^2}{(n-f)^2\mu^2+2m^2\mu^2},1\right)$. Note that since $\alpha>0$, due to \eqref{eqn:replace-alpha-sgd-cge} we have $(n-f)\gamma-2(f+r)\mu>0$. Recall from Lemma~\ref{lemma:gamma-mu} that $\gamma\leq\mu$. Also note that for two non-negative values $a$ and $b$, $(a-b)^2=a^2-2ab+b^2\leq a^2+b^2$. Thus
    \begin{equation}
        0< ((n-f)\gamma-2(f+r)\mu)^2\leq(n-f)^2\gamma^2+4(f+r)^2\mu^2\leq \left((n-f)^2+4(f+r)^2\right)\mu^2.
        \label{eqn:gamma-mu-related-ineq-sgd-cge-1}
    \end{equation}
    Recall from \eqref{eqn:replace-alpha-sgd-cge} that $(n-f)\gamma-2(f+r)\mu=m\gamma\alpha$ and that we defined $m=n-r$, we have 
    \begin{align}
        1-\cfrac{(m\gamma\alpha)^2}{(n-f)^2\mu^2+2m^2\mu^2}\geq&1-\frac{\left((n-f)^2+4(f+r)^2\right)\mu^2}{(n-f)^2\mu^2+2m^2\mu^2} \nonumber \\
        =&\frac{2(n-r)^2-4(f+r)^2}{(n-f)^2+2(n-r)^2} \nonumber \\
        =&\frac{2\left(n+\sqrt{2}f+(\sqrt{2}-1)r\right)\left(n-\sqrt{2}f-(1+\sqrt{2})r\right)}{(n-f)^2+2(n-r)^2}.
    \end{align}
    Note that we assume $n\geq 2f+3r$ (cf. Theorem~\ref{thm:cge}-CGE description). Thus, the right-hand side of the above is positive. 
    Therefore, $0<\rho<1$. 
    
    \textbf{Third}, we show the convergence property. 
    Recall that in \eqref{eqn:def-m-cge} we defined
    \begin{align*}
        \M 
        =& B\epsilon^2+C\sigma^2.
    \end{align*}
    From \eqref{eqn:expectation-w-t-square-sgd-cge-4}, definition of $\rho$ in \eqref{eqn:def-rho-cge}, and definition of $\M$ above, we have
    \begin{align}
        \E_{\zeta^t}\norm{x^{t+1}-x_\H}^2\leq&\rho\norm{x^t-x_\H}^2 + \M.
    \end{align}
    By Lemma~\ref{lemma:converge-sgd}, since $\M\geq0$ and $\rho\in[0,1)$, we have
    \begin{align*}
        \E_{t}\norm{x^{t+1}-x_\H}^2&\leq\rho^{t+1}\norm{x^0-x_\H}^2 + \left(\frac{1-\rho^{t+1}}{1-\rho}\right)\M. 
    \end{align*}
\end{proof}

\hrule

\subsection{Derivation of \eqref{eqn:stochastic-params-c}}
\label{appdx:b-4}

For the special case of asynchronous stochastic optimization discussed in Section~\ref{sub:stochastic-special}, there are 0 Byzantine faulty agents, and up to $r$ stragglers. In \eqref{eqn:stochastic-params-c} there are the following parameters:
\begin{itemize}[nosep]
    \item The \textit{resilience margin}
    \begin{align}
        \alpha = 1-\frac{r}{n}\cdot\frac{\mu}{\gamma},
        \label{eqn:def-alpha}
    \end{align}
    \item The parameter that determines the step size
    \begin{align}
        \overline{\eta} = \frac{2n\gamma\alpha}{3n^2\mu^2},
        \label{eqn:def-eta-bar}
    \end{align}
\end{itemize}

Recall the definition of $x_\H$ in Section~\ref{sec:full-grad}: for any non-faulty agents $\H$ with $\mnorm{\H}=n-f$ in an execution, $x_\H\in\W$ is a unique minimum point of the aggregate cost function of agents in $\H$. When $f=0$ there is only one set $\H=[n]$ and one corresponding $x_\H$. The results in \eqref{eqn:stochastic-params-c} can be described in full as follows:

\begin{mdframed}
    \textsc{Theorem~\ref{thm:cge}-Async. }
    \textit{Consider Algorithm~\ref{alg} with stochastic updates, and the GAR in use is as follows:
    \begin{align*}
        \mathsf{GradAgg}\left(g_j^t|j\in S^t;n,0,r\right)=\sum_{j\in{S^t}}g_j^t,
    \end{align*}
    where $g_j^t$ is the stochastic gradient of agent $j$ at iteration $i$.
    Suppose Assumptions~\ref{assum:lipschitz}, \ref{assum:strongly-convex-ft}, and \ref{assum:bound-grad} hold true, the expected cost functions of the agents in the system satisfy $(0,r;\epsilon)$-redundancy, $\alpha>0$ and step size in \eqref{eqn:update} $\eta_t=\eta>0$ for all $t$. Let
    \begin{align}
        \M &=  4\left(\left(r +n^2\eta\mu\right)^2+n^4\eta^2\mu^2\right)\epsilon^2 +\left(n^2\eta^2+\left(\frac{r}{n\mu}\right)^2\left(\sqrt{n-1}+1\right)^2 \right)\sigma^2.
        \label{eqn:def-m}
    \end{align}
    If $\eta<\overline{\eta}$, the following holds true:
    \begin{itemize}
        \item The value of 
        \begin{align}
            \rho = 1-2(n\gamma-r\mu)\eta + 3n^2\eta^2\mu^2,
            \label{eqn:def-rho-async}
        \end{align}
        satisfies $0<\rho<1$, and \\
        \comment{Lowerbound of $\rho$ is indeed $<0$. (changed later in the proof as well)\\TODO: check this for the proof above in Appdx C.2 as well}
        \item Given the initial estimate $x^0$ arbitrarily chosen from $\W$, for all $t\geq0$,
        \begin{align}
            \E_{t}\norm{x^{t+1}-x_\H}^2&\leq\rho^{t+1}\norm{x^0-x_\H}^2 + \left(\frac{1-\rho^{t+1}}{1-\rho}\right)\M.
            \label{eqn:expectation-bound-1}
        \end{align}
    \end{itemize}}
\end{mdframed}

\begin{proof}
    \textbf{First}, we show a recursive bound over the expected value of $\norm{x^t-x_\H}^2$. Following the \textbf{first} part of the proof of Theorem~\ref{thm:async-fault-toler}-Async in Appendix~\ref{appdx:a-3}, we know that by Assumption~\ref{assum:lipschitz} and $(0,r;\epsilon)$-redundancy, for all $i\in[n]$, we obtain the same result as in \eqref{eqn:apprx_gradient_bnd_2} that
    \begin{equation}
        \norm{\nabla Q_i(x)}\leq2n\mu\epsilon+\mu\norm{x-x_\H}.
        \label{eqn:apprx_gradient_bnd_2-apdx-b4}
    \end{equation}
    
    Define $\g^t=\sum_{j\in S^t}g_j^t$. Recall our iterative update \eqref{eqn:update}. Using the non-expansion property of Euclidean projection onto a closed convex set\footnote{$\norm{x-x_\H}\geq\norm{[x]_\W-x_\H},~\forall w\in\mathbb{R}^d$.}, with $\eta_t=\eta$ for all $t$, we have
    \begin{equation}
        \label{eqn:apprx-one-step-sgd}
        \norm{x^{t+1}-x_\H}\leq\norm{x^t-\eta\g^t-x_\H}.
    \end{equation}
    Taking square on both sides, we have
    \begin{align}
        \label{eqn:rate-sgd-0}
        \norm{x^{t+1}-x_\H}^2\leq&\norm{x^t-x_\H}^2-2\eta\iprod{x^t-x_\H}{\g^t}+\eta^2\norm{\g^t}^2
    \end{align}
    By triangle inequality and AM-QM inequality (i.e., $\frac{1}{n}\sum_{j=1}^nx_j\leq\sqrt{\frac{1}{n}\sum_{j=1}^nx_j^2}$ for any $n$ positive real $x_j$'s),
    \begin{align}
        \norm{\g^t}^2 \leq\left(\sum_{j\in S^t}\norm{g_j^t}\right)^2\leq \left(\sum_{j\in [n]}\norm{g_j^t}\right)^2\leq\mnorm{[n]}\sum_{j\in[n]}\norm{g_j^t}^2=n\sum_{j\in[n]}\norm{g_j^t}^2.
    \end{align}
    Substitute above in \eqref{eqn:rate-sgd-0}, we have
    \begin{align}
        \label{eqn:rate-sgd-1}
        \norm{x^{t+1}-x_\H}^2\leq&\norm{x^t-x_\H}^2-2\eta\iprod{x^t-x_\H}{\g^t}+n\eta^2\sum_{j\in[n]}\norm{g_j^t}^2.
    \end{align}
    We define
    \begin{align}
        \phi_t=\iprod{x^t-x_\H}{\g^t} = \sum_{j\in S^t}\iprod{x^t-x_\H}{g_j^t}.
    \end{align}
    Substituting above in \eqref{eqn:rate-sgd-1}, we have
    \begin{align}
        \label{eqn:rate-sgd-2}
        \norm{x^{t+1}-x_\H}^2\leq&\norm{x^t-x_\H}^2-2\eta\phi_t+n\eta^2\sum_{j\in[n]}\norm{g_j^t}^2.
    \end{align}
    Recall that $\zeta^t=\{\zeta_1^t,...,\zeta_n^t\}$, and $x^{t+1}$ is a function of the set of random variables $\zeta^t$. Also note that $\E_{\zeta^t}\norm{x^t-x_\H}^2=\norm{x^t-x_\H}^2$. Taking expectation $\E_{\zeta^t}$ on both sides, we have
    \begin{align}
        \label{eqn:rate-sgd-3}
        \E_{\zeta^t}\norm{x^{t+1}-x_\H}^2\leq&\norm{x^t-x_\H}^2-2\eta\E_{\zeta^t}\left[\phi_t\right]+n\eta^2\sum_{j\in [n]}\E_{\zeta^t}\norm{g_j^t}^2.
    \end{align}
    
    Consider $\E_{\zeta^t}\left[\phi_t\right]$. With the definition of $\g^t$ we have
    \begin{align}
        \label{eqn:phi_t_bnd-sgd}
        \E_{\zeta^t}\left[\phi_t\right] = \E_{\zeta^t}\iprod{x^t-x_\H}{\g^t} = \E_{\zeta^t}\iprod{x^t-x_\H}{\sum_{j\in S^t}g_j^t}.
    \end{align}
    Recall from \eqref{eqn:expectation-g-i-t-cge} that for any $j\in[n]$, $\E_{\zeta^t}\left[g_j^t\right]=\nabla Q_j(x^t)$. Substituting this in \eqref{eqn:phi_t_bnd-sgd}, we have
    \begin{align}
        \label{eqn:phi_t_bnd-sgd-1}
        \E_{\zeta^t}\left[\phi_t\right] &= \E_{\zeta^t}\iprod{x^t-x_\H}{\sum_{j\in S^t}g_j^t} \nonumber \\
        & = \E_{\zeta^t}\iprod{x^t-x_\H}{\sum_{j\in S^t}g_j^t + \sum_{j\in [n]\backslash S^t}g_j^t -\sum_{j\in[n]\backslash S^t}g_j^t} \nonumber \\
        & = \E_{\zeta^t}\iprod{x^t-x_\H}{\sum_{j\in [n]}g_j^t} - \iprod{x^t-x_\H}{\sum_{j\in[n]\backslash S^t}g_j^t} \nonumber \\
        & = \iprod{x^t-x_\H}{\sum_{j\in [n]}\E_{\zeta^t}\left[g_j^t\right]} - \E_{\zeta^t}\iprod{x^t-x_\H}{\sum_{j\in[n]\backslash S^t}g_j^t} \nonumber \\
        & = \iprod{x^t-x_\H}{\sum_{j\in [n]}\nabla Q_j(x^t)} - \E_{\zeta^t}\iprod{x^t-x_\H}{\sum_{j\in[n]\backslash S^t}g_j^t}.
    \end{align}
    For the first term in \eqref{eqn:phi_t_bnd-sgd-1}, recall that $x_\H$ is the minimum of  $\sum_{j\in[n]}Q_j(x)$, i.e., $\sum_{j\in[n]}\nabla Q_j(x_\H)=0$. By Assumption~\ref{assum:strongly-convex-ft}, we have
    \begin{align}
        \label{eqn:phi_t_bnd-sgd-2-1}
        \iprod{x^t-x_\H}{\sum_{j\in [n]}\nabla Q_j(x^t)} = \iprod{x^t-x_\H}{\sum_{j\in[n]}\nabla Q_j(x^t)-\sum_{j\in[n]}\nabla Q_j(x_\H)} \geq n\gamma\norm{x^t-x_\H}^2.
    \end{align}
    For the second term in \eqref{eqn:phi_t_bnd-sgd-1}, by Cauchy-Schwartz inequality,
    \begin{align}
        \E_{\zeta^t}\iprod{x^t-x_\H}{\sum_{j\in[n]\backslash S^t}g_j^t} &\geq - \E_{\zeta^t}\left[\sum_{j\in[n]\backslash S^t}\norm{x^t-x_\H}\norm{g_j^t}\right] = - \sum_{j\in[n]\backslash S^t}\norm{x^t-x_\H}\E_{\zeta^t}\norm{g_j^t}.
        \label{eqn:phi_t_bnd-sgd-2-1-0}
    \end{align}
    Note that in each iteration $t$, there exists a $v_t\in[n]$ such that $\norm{g_j^t}\leq\norm{g_{v_t}^t}$ for all $j\in[n]$, i.e., the gradient sent by agent $v_t$ has the largest norm among all agents in $[n]$ in iteration $t$. Now, 
    \begin{align}
        &\sum_{j\in[n]\backslash S^t}\norm{x^t-x_\H}\E_{\zeta^t}\norm{g_j^t} \leq \mnorm{[n]\backslash S^t}\norm{x^t-x_\H}\E_{\zeta^t}\norm{g_{v_t}^t} \nonumber \\
        \leq& \mnorm{[n]\backslash S^t}\norm{x^t-x_\H}\left(\sigma\left(\sqrt{n-1}+1\right) + \max_{i\in[n]}\norm{\nabla Q_i(x^t)}\right) \nonumber \tag{by Lemma~\ref{lemma:bound-exp-largest-gradient}} \\
        \leq & \sigma\left(\sqrt{n-1}+1\right)\mnorm{[n]\backslash S^t}\norm{x^t-x_\H} + \mnorm{[n]\backslash S^t}\norm{x^t-x_\H}\left(\max_{i\in[n]}\norm{\nabla Q_i(x^t)}\right).
        \label{eqn:phi_t_bnd-sgd-2-1-1}
    \end{align}
    Combining \eqref{eqn:phi_t_bnd-sgd-1}, \eqref{eqn:phi_t_bnd-sgd-2-1}, \eqref{eqn:phi_t_bnd-sgd-2-1-0}, and \eqref{eqn:phi_t_bnd-sgd-2-1-1}, we have
    \begin{align}
        \E_{\zeta^t}[\phi_t] \geq n\gamma\norm{x^t-x_\H}^2 &- \sigma\left(\sqrt{n-1}+1\right)\mnorm{[n]\backslash S^t}\norm{x^t-x_\H} \nonumber \\
        &- \mnorm{[n]\backslash S^t}\norm{x^t-x_\H}\left(\max_{i\in[n]}\norm{\nabla Q_i(x^t)}\right).
    \end{align}
    Note that $\mnorm{[n]\backslash S^t}=r$. Substituting \eqref{eqn:apprx_gradient_bnd_2-apdx-b4} in above, we have
    \begin{align}
        \label{eqn:phi_t_bnd-sgd-2-2}
        \E_{\zeta^t}[\phi_t] \geq n\gamma\norm{x^t-x_\H}^2 &- r\sigma\left(\sqrt{n-1}+1\right)\norm{x^t-x_\H} \nonumber \\
        &- r\norm{x^t-x_\H}(2n\mu\epsilon +\mu\norm{x^t-x_\H}).
    \end{align}
    
    Now consider $\sum_{j\in[n]}\E_{\zeta^t}\norm{g_j^t}^2$. Recall \eqref{eqn:lemma-1-cge} from Lemma~\ref{lemma:1-cge} that
    \begin{equation*}
        \E_{\zeta^t}\norm{g_i^t}^2\leq\sigma^2+\norm{\nabla Q_i(x^t)}^2.
    \end{equation*}
    Combining above and \eqref{eqn:apprx_gradient_bnd_2-apdx-b4}, we have
    \begin{equation}
        \label{eqn:expectation-second-term-sgd}
        \sum_{j\in [n]}\E_{\zeta^t}\norm{g_j^t}^2\leq n\left(\sigma^2+\left(2n\mu\epsilon+\mu\norm{x^t-x_\H}\right)^2\right).
    \end{equation}
    
    Substituting \eqref{eqn:phi_t_bnd-sgd-2-2} and \eqref{eqn:expectation-second-term-sgd} in \eqref{eqn:rate-sgd-3},
    \begin{align}
        \label{eqn:rate-sgd-4}
        &\E_{\zeta^t}\norm{x^{t+1}-x_\H}^2 \nonumber \\
        \leq&\norm{x^t-x_\H}^2 \nonumber \\ 
        &\qquad- 2\eta\left[(n\gamma - r\mu)\norm{x^t-x_\H}^2 -\left(2n\mu\epsilon+\sigma\left(\sqrt{n-1}+1\right)\right)r\norm{x^t-x_\H}\right] \nonumber \\
        &\qquad+n^2\eta^2\left(\sigma^2+\left(2n\mu\epsilon+\mu\norm{x^t-x_\H}\right)^2\right) \nonumber \\
        \leq& (1-2(n\gamma-r\mu)\eta + n^2\eta^2\mu^2)\norm{x^t-x_\H}^2 \nonumber \\
        &\qquad+\left(4n\eta\mu\epsilon\left(r + n^2\eta\mu\right) + 2r\left(\sqrt{n-1}+1\right)\eta\sigma\right)\norm{x^t-x_\H} \nonumber \\
        &\qquad+n^2\eta^2(\sigma^2+4n^2\mu^2\epsilon^2).
    \end{align}
    Consider the second term in \eqref{eqn:rate-sgd-3}. Notice that for any two real values $a$ and $b$, we have $2ab\leq a^2+b^2$. Thus, we have
    \begin{align}
        &\left(4n\eta\mu\epsilon\left(r + n^2\eta\mu\right) + 2r\left(\sqrt{n-1}+1\right)\eta\sigma\right)\norm{x^t-x_\H} \nonumber \\
        \leq& 2\left(2n\eta\mu\epsilon\left(r + n^2\eta\mu\right)\norm{x^t-x_\H}\right) + 2\left(r\left(\sqrt{n-1}+1\right)\eta\sigma\norm{x^t-x_\H}\right) \nonumber \\
        \leq& n^2\eta^2\mu^2\norm{x^t-x_\H}^2 + 4\left(r + n^2\eta\mu\right)^2\epsilon^2 \nonumber \\
        &\quad+ n^2\eta^2\mu^2\norm{x^t-x_\H}^2 + \left(\frac{r}{n\mu}\right)^2\left(\sqrt{n-1}+1\right)^2\sigma^2 \nonumber\\
        \leq& 2n^2\eta^2\mu^2\norm{x^t-x_\H}^2 + 4\left(r + n^2\eta\mu\right)^2\epsilon^2 + \left(\frac{r}{n\mu}\right)^2\left(\sqrt{n-1}+1\right)^2\sigma^2.
    \end{align}
    Substituting above in \eqref{eqn:rate-sgd-4}, we have
    \begin{align}
        \label{eqn:rate-sgd-4-new}
        \E_{\zeta^t}\norm{x^{t+1}-x_\H}^2
        &\leq (1-2(n\gamma-r\mu)\eta + n^2\eta^2\mu^2)\norm{x^t-x_\H}^2 \nonumber \\
        &\qquad+2n^2\eta^2\mu^2\norm{x^t-x_\H}^2 + 4\left(r + n^2\eta\mu\right)^2\epsilon^2 + \left(\frac{r}{n\mu}\right)^2\left(\sqrt{n-1}+1\right)^2\sigma^2 \nonumber \\
        &\qquad+n^2\eta^2(\sigma^2+4n^2\mu^2\epsilon^2) \nonumber \\
        &\leq (1-2(n\gamma-r\mu)\eta + 3n^2\eta^2\mu^2)\norm{x^t-x_\H}^2 \nonumber \\
        &\qquad + 4\left(\left(r +n^2\eta\mu\right)^2+n^4\eta^2\mu^2\right)\epsilon^2 +\left(n^2\eta^2+\left(\frac{r}{n\mu}\right)^2\left(\sqrt{n-1}+1\right)^2 \right)\sigma^2.
    \end{align}
    
    Let $B=4\left(\left(r +n^2\eta\mu\right)^2+n^4\eta^2\mu^2\right)$ and $C=\left(n^2\eta^2+\left(\cfrac{r}{n\mu}\right)^2\left(\sqrt{n-1}+1\right)^2 \right)$. 
    Thus, from \eqref{eqn:rate-sgd-4-new} we obtain
    \begin{align}
        \label{eqn:rate-sgd-5}
        \E_{\zeta^t}\norm{x^{t+1}-x_\H}^2&\leq (1-2(n\gamma-r\mu)\eta + 3n^2\eta^2\mu^2)\norm{x^t-x_\H}^2+B\epsilon^2+C\sigma^2.
    \end{align}
    Recalling \eqref{eqn:def-rho-async},  the definition of $\rho$, we have
    \begin{align}
        \label{eqn:rate-sgd-6}
        \E_{\zeta^t}\norm{x^{t+1}-x_\H}^2&\leq \rho\norm{x^t-x_\H}^2+B\epsilon^2+C\sigma^2.
    \end{align}
    Note that $B,C\geq0$. 

    \textbf{Second}, we show $0<\rho<1$. Recall that in \eqref{eqn:def-alpha} we defined 
    \begin{align*}
        \alpha = 1-\frac{r}{n}\cdot\frac{\mu}{\gamma}.
    \end{align*}
    We have
    \begin{align}
    n\gamma-r\mu=n\gamma\alpha. 
    \end{align}
    Recall from \eqref{eqn:def-rho-async} that $\rho=1-2(n\gamma-r\mu)\eta + 3n^2\eta^2\mu^2$. So $\rho$ can be written as 
    \begin{align}
        \rho =& 1-2n\eta\gamma\alpha + 3n^2\eta^2\mu^2 \nonumber \\
        =& 1- \left(3n^2\mu^2\right)\eta\left(\frac{2n\gamma\alpha}{3n^2\mu^2}-\eta\right).
    \end{align}
    Recalling \eqref{eqn:def-eta-bar} that $\overline{\eta}=\cfrac{2n\gamma\alpha}{3n^2\mu^2}$, from above we obtain
    \begin{align}
        \rho = 1-\left(3n^2\mu^2\right)\eta(\overline{\eta}-\eta).
        \label{eqn:rho-sgd-2}
    \end{align}
    Note that 
    \begin{align}
        \eta(\overline{\eta}-\eta)=\left(\frac{\overline{\eta}}{2}\right)^2-\left(\eta-\frac{\overline{\eta}}{2}\right)^2.
    \end{align}
    Therefore,
    \begin{align}
        \rho &= 1-\left(3n^2\mu^2\right)\left[\left(\frac{\overline{\eta}}{2}\right)^2-\left(\eta-\frac{\overline{\eta}}{2}\right)^2\right] \nonumber \\
        &= \left(3n^2\mu^2\right)\left(\eta-\frac{\overline{\eta}}{2}\right)^2 +  1-\left(3n^2\mu^2\right)\left(\frac{\overline{\eta}}{2}\right)^2.
    \end{align}
    Since $\eta\in(0,\overline{\eta})$, the minimum value of $\rho$ can be obtained when $\eta=\overline{\eta}/2$,
    \begin{align}
        \min_\eta\rho=1-\left(3n^2\mu^2\right)\left(\frac{\overline{\eta}}{2}\right)^2.
    \end{align}
    On the other hand, since $\eta\in(0,\overline{\eta})$, from \eqref{eqn:rho-sgd-2}, $\rho<1$. Thus,
    \begin{align}
        1-\left(3n^2\mu^2\right)\left(\frac{\overline{\eta}}{2}\right)^2\leq\rho<1.
    \end{align}
    Substituting \eqref{eqn:def-eta-bar} in above implies that $\rho\in\left[1-\cfrac{\left(n\gamma\alpha\right)^2}{3n^2\mu^2},1\right)$. Note that since $\alpha>0$, we have $n\gamma-r\mu>0$. Thus, with $\gamma\leq\mu$ (cf. Lemma~\ref{lemma:gamma-mu}),
    \begin{align}
        (n\gamma-r\mu)^2\leq (n-r)^2\mu^2.
    \end{align}
    Recalling that $n\gamma-r\mu=n\gamma\alpha$, we have 
    \begin{align}
        \cfrac{(n\gamma\alpha)^2}{3n^2\mu^2}\leq \cfrac{(n-r)^2\mu^2}{3n^2\mu^2}\leq\frac{1}{3}.
    \end{align}
    Therefore the lower bound of $\rho\geq1-1/3>0$. So $0<\rho<1$.

    \textbf{Third}, we show the convergence property. 
    Recall that in \eqref{eqn:def-m} we defined
    \begin{align*}
        \M = B\epsilon^2+C\sigma^2.
    \end{align*}
    From \eqref{eqn:rate-sgd-6}, definition of $\rho$ in \eqref{eqn:def-rho-async}, and definition of $\M$ above, we have
    \begin{align}
        \label{eqn:rate-sgd-7}
        \E_{\zeta^t}\norm{x^{t+1}-x_\H}^2&\leq \rho\norm{x^t-x_\H}^2 + \M.
    \end{align}
    By Lemma~\ref{lemma:converge-sgd}, since $\M\geq0$ and $\rho\in[0,1)$, we have
    \begin{align*}
        \E_{t}\norm{x^{t+1}-x_\H}^2&\leq\rho^{t+1}\norm{x^0-x_\H}^2 + \left(\frac{1-\rho^{t+1}}{1-\rho}\right)\M. 
    \end{align*}
\end{proof}



\end{document}